%% file: main-sosym.tex
\documentclass[pdflatex,sn-mathphys-num,iicol]{sn-jnl}

\usepackage{graphicx}%
\usepackage{multirow}%
\usepackage{amsmath,amssymb,amsfonts}%
\usepackage{amsthm}%
\usepackage{mathrsfs}%
\usepackage[title]{appendix}%
\usepackage{xcolor}%
\usepackage{textcomp}%
\usepackage{manyfoot}%
\usepackage{booktabs}%
\usepackage{listings}%

\theoremstyle{thmstyleone}%
\theoremstyle{thmstyletwo}%

\theoremstyle{thmstylethree}%

\usepackage{amsmath,amssymb,amsfonts}
\usepackage{graphicx}
\usepackage{textcomp}
\usepackage{xcolor}
\usepackage{booktabs}
\usepackage{tabularx}
\usepackage{subcaption}
\usepackage{hyperref}
\def\BibTeX{{\rm B\kern-.05em{\sc i\kern-.025em b}\kern-.08em
    T\kern-.1667em\lower.7ex\hbox{E}\kern-.125emX}}
 \newcommand{\boxedtext}[1]{\fbox{\scriptsize\bfseries\textsf{#1}}}
\newcommand{\nota}[2]{
	\boxedtext{#1}
		{\small$\blacktriangleright$\emph{\textsl{#2}}$\blacktriangleleft$}
}
\newcommand{\todo}[1]{{\color{red}\nota{TODO}{#1}}} 

\newcolumntype{P}[1]{>{\centering\arraybackslash}p{#1}}
\newcolumntype{Y}{>{\centering\arraybackslash}X}

\usepackage{enumitem}

\newlist{questions}{enumerate}{2}
\newlength{\questionsubindent}
\setlist[questions,1]{wide=0pt, labelsep=0.5em, label=\textbf{RQ\arabic*.}, ref=RQ\arabic*}
\setlist[questions,2]{wide=0pt, leftmargin=\questionsubindent,
    labelindent=\questionsubindent, labelsep=0.5em,
    label=\textbf{RQ\arabic{questionsi}.\arabic*.},
    ref=RQ\arabic{questionsi}.\arabic*}

\usepackage{AMMALanguages}
\usepackage[linesnumbered,plainruled,vlined,noend,commentsnumbered]{algorithm2e}

\usepackage{float}

\usepackage{multirow}

\usepackage{listings}
\usepackage{parcolumns}
\usepackage[framemethod=TikZ]{mdframed}
\usepackage{pgfplots}
\mdfdefinestyle{rqanswer}{backgroundcolor=gray!15,outerlinewidth=1}

\input{preamble/commands}
\input{preamble/tables}
\input{preamble/artifact-styles}

\begin{document}

\title[LLM-based Generation of Diverse and Realistic Domain Model Instances]{LLM-based Generation of Diverse and Realistic Domain Model Instances}


\author*[1]{\fnm{Andrei} \sur{Coman}}\email{andreialexcoman@uma.es}

\author[1]{\fnm{Lola} \sur{Burgueño}}\email{lolaburgueno@uma.es}

\author[2]{\fnm{Dominik} \sur{Bork}}\email{dominik.bork@tuwien.ac.at}

\author[3]{\fnm{Manuel} \sur{Wimmer}}\email{manuel.wimmer@jku.at}

\affil*[1]{\orgdiv{ITIS Software}, \orgname{University of Málaga}, \orgaddress{\city{Málaga}, \country{Spain}}}

\affil[2]{\orgname{TU Wien}, \orgaddress{\city{Vienna}, \country{Austria}}}

\affil[3]{\orgname{Johannes Kepler University Linz}, \orgaddress{\city{Linz}, \country{Austria}}}


\input{sections/Abstract
}

\maketitle

\input{sections/Introduction}

\input{sections/BackgroundMotivation}

\input{sections/Approach}

\input{sections/Evaluation}

\input{sections/Results}

\input{sections/Threats}

\input{sections/RelatedWork}

\input{sections/Conclusion}

\begin{appendices}
\input{sections/Appendix}
\end{appendices}

\bibliography{literature}

\end{document}

%% file: preamble/commands.tex
\newcommand{\secref}[1]{Sec.~\ref{#1}}

\newcommand{\assignedto}[1]{%
    \ifthenelse{\boolean{showannotations}}%
    {\textbf{\noindent\ding{46}\textcolor{white}{~\colorbox{\assignementcolor}{Assigned to:}}~\textcolor{\assignementcolor}{#1}\\}%
    }
    {}
}

\newcommand{\nb}[4]{
    \fcolorbox{gray}{#2}{\bfseries\sffamily\scriptsize{#1}}%
	{\sf\small$\blacktriangleright$\textcolor{#4}{\textit{#3}}$\blacktriangleleft$}%
}

\newcommand\personmarker[2]{\noindent\nb{#1}{yellow}{#2}{violet}}

\newcommand\manuel[1]{\noindent\personmarker{Manuel}{#1}}

\newcommand\dominik[1]{\noindent\personmarker{Dominik}{#1}}
\newcommand\andrei[1]{\noindent\personmarker{Andrei}{#1}}

\newcommand{\rem}[1]{%
    \ifthenelse{\boolean{showannotations}}%
    {\textcolor{oldtextcolor}{\st{#1}}}%
    {}%
}

\newcommand\add[1]{%
    \ifthenelse{\boolean{showannotations}}%
    {\textcolor{newtextcolor}{{#1}}}%
    {#1}%
}

\newcommand\rep[2]{%
    \ifthenelse{\boolean{showannotations}}%
    {\rem{#1}~\add{#2}}%
    {#2}%
}

\newcommand{\viewerurl}{https://a-coman.github.io/llm-viewer/gpt_5_2}
\newcommand{\ilinst}[3]{\href{\viewerurl/#1/gen#2/}{#3}}
\newcommand{\cotinst}[4]{\href{\viewerurl/#1/gen#2/?type=cot\string&category=#3}{#4}}

\renewcommand\todo[1]{}
\renewcommand\personmarker[2]{}

%% file: preamble/tables.tex
\usepackage{collcell}

\newcommand{\thd}[1]{\multicolumn{1}{c}{#1}}

\makeatletter
\newlength{\tbl@w}
\newcommand{\tbl@r}[2]{\settowidth{\tbl@w}{#1}\makebox[\tbl@w][r]{#2}}
\newcommand{\tbl@l}[2]{\settowidth{\tbl@w}{#1}\makebox[\tbl@w][l]{#2}}
\newcommand{\tbl@c}[2]{\settowidth{\tbl@w}{#1}\makebox[\tbl@w][c]{#2}}
\def\tbl@dash{--}

\newcommand{\tbl@num}[1]{%
  \def\tbl@cell{#1}%
  \ifx\tbl@cell\tbl@dash\tbl@c{\tbl@tmpl}{--}\else\tbl@r{\tbl@tmpl}{#1\unskip}\fi}
\newcolumntype{N}[1]{>{\def\tbl@tmpl{#1}\collectcell\tbl@num}c<{\endcollectcell}}

\def\tbl@fracsplit#1/#2\relax{%
  \tbl@r{\tbl@tmpl}{#1\unskip}/\tbl@l{\tbl@tmpl}{\ignorespaces#2\unskip}}
\newcommand{\tbl@frac}[1]{%
  \in@{/}{#1}%
  \ifin@\tbl@fracsplit#1\relax
  \else\tbl@c{\tbl@tmpl/\tbl@tmpl}{#1\unskip}\fi}
\newcolumntype{F}[1]{>{\def\tbl@tmpl{#1}\collectcell\tbl@frac}c<{\endcollectcell}}

\def\tbl@pmsplit#1$\pm$#2\relax{%
  \tbl@r{\tbl@tmplm}{#1\unskip}${}\pm{}$\tbl@l{\tbl@tmpls}{\ignorespaces#2\unskip}}
\newcommand{\tbl@pm}[1]{%
  \def\tbl@cell{#1}%
  \ifx\tbl@cell\tbl@dash\tbl@c{\tbl@tmplm${}\pm{}$\tbl@tmpls}{--}%
  \else
    \in@{$\pm$}{#1}%
    \ifin@\tbl@pmsplit#1\relax
    \else\tbl@c{\tbl@tmplm${}\pm{}$\tbl@tmpls}{#1\unskip}\fi
  \fi}
\newcolumntype{M}[2]{>{\def\tbl@tmplm{#1}\def\tbl@tmpls{#2}\collectcell\tbl@pm}c<{\endcollectcell}}
\makeatother

%% file: preamble/artifact-styles.tex
\usepackage{placeins}
\usepackage{needspace}

\newcommand{\sosymListingNumber}[1]{%
    \ifnum\value{lstnumber}=0\relax
    \else{\ttfamily\tiny #1}%
    \fi}

\newif\ifsosymPromptHeadingNumberHidden
\newcommand{\sosymPromptListingNumber}[2]{%
    \ifnum\value{lstnumber}=0\relax
    \else
        \ifsosymPromptHeadingNumberHidden
            {\ttfamily\tiny\number\numexpr#2-1\relax}%
        \else
            \ifnum\value{lstnumber}=#1\relax
                \global\sosymPromptHeadingNumberHiddentrue
            \else
                {\ttfamily\tiny #2}%
            \fi
        \fi
    \fi}

\newlength{\sosymWrapIndent}

\lstdefinestyle{sosym-listing}{
    basicstyle=\sosymCodeRegular\fontsize{8}{9}\selectfont,
    identifierstyle=\sosymCodeRegular,
    keywordstyle=\sosymCodeBold,
    emphstyle=\sosymCodeItalic,
    commentstyle=\sosymCodeItalic,
    stringstyle=\sosymCodeRegular,
    columns=fullflexible,
    keepspaces=true,
    numberblanklines=false,
    numbers=left,
    numberstyle=\sosymListingNumber,
    stepnumber=1,
    numbersep=6pt,
    showspaces=false,
    showstringspaces=false,
    showtabs=false,
    tabsize=2,
    breaklines=true,
    breakatwhitespace=true,
    breakindent=\sosymWrapIndent,
    frame=tb,
    framerule=0.4pt,
    framesep=3pt,
    rulecolor=\color{black},
    xleftmargin=1.5em,
    xrightmargin=0pt,
    captionpos=b,
    floatplacement=tbp,
    float=true,
    aboveskip=0pt,
    belowskip=0pt,
    abovecaptionskip=0.5\baselineskip
}

\makeatletter
\let\lst@beginfloat\@float
\let\lst@endfloat\end@float
\makeatother
\newcommand{\sosymPromptHeadingRule}{%
    \rule[-0.85\baselineskip]{0pt}{0.85\baselineskip}%
    \makebox[0pt][l]{%
        \raisebox{-0.55\baselineskip}[0pt][0pt]{%
            \rule{\linewidth}{0.4pt}}}}

\newcommand{\sosymListingHeading}[1]{%
    \global\sosymPromptHeadingNumberHiddenfalse
    \sosymPromptHeadingRule{\sosymCodeBold #1}\vadjust{\nobreak}}

\newcommand{\sosymPromptUserRules}{%
    \rule{0pt}{1.55\baselineskip}%
    \rule[-0.85\baselineskip]{0pt}{0.85\baselineskip}%
    \makebox[0pt][l]{%
        \raisebox{1.02\baselineskip}[0pt][0pt]{%
            \rule{\linewidth}{0.4pt}}}%
    \makebox[0pt][l]{%
        \raisebox{-0.55\baselineskip}[0pt][0pt]{%
            \rule{\linewidth}{0.4pt}}}}

\newcommand{\sosymPromptUserHeading}[1]{%
    \sosymPromptUserRules{\sosymCodeBold #1}\vadjust{\nobreak}}

\SetAlFnt{\sosymCodeRegular\fontsize{8}{9}\selectfont}
\SetAlCapFnt{\small}
\SetAlCapNameFnt{\small}

\SetNlSty{sosymAlgoNumber}{}{}
\SetAlgoNlRelativeSize{0}
\SetAlgoCaptionSeparator{:}
\SetAlgoSkip{}
\SetAlCapSkip{2.65pt}
\SetInd{0.5em}{0.8em}
\SetAlgoHangIndent{\sosymWrapIndent}
\makeatletter
\def\@algocf@pre@plainruled{%
    \moveright\algomargin\vbox{\hrule width\dimexpr\algocf@ruledwidth-\algomargin\relax
        height\algoheightrule depth0pt}%
    \kern\interspacetitleruled}
\renewcommand{\algocf@makecaption@plainruled}[2]{%
    \global\sbox\algocf@capbox{%
        \parbox[t]{\columnwidth}{\raggedright\algocf@captiontext{#1}{#2}}}}
\def\@algocf@post@plainruled{%
    \kern\interspacealgoruled
    \moveright\algomargin\vbox{\hrule width\dimexpr\algocf@ruledwidth-\algomargin\relax
        height\algoheightrule depth0pt}}
\makeatother
\makeatletter
\newcommand{\SetKwInOutFit}[2]{%
  \sbox\algocf@inoutbox{\AlFnt\KwSty{#2\algocf@typo:}}%
  \ifdim\wd\algocf@inoutbox>\inoutsize
    \setlength{\inoutsize}{\wd\algocf@inoutbox}%
    \sbox\algocf@inoutbox{\AlFnt\parbox[t]{\inoutsize}{\KwSty{#2\algocf@typo\hfill:}}~}\setlength{\inoutindent}{\wd\algocf@inoutbox}%
  \fi
  \algocf@newcommand{#1}[1]{%
    \ifthenelse{\boolean{algocf@hanginginout}}{\relax}{\algocf@seteveryparhanging{\relax}}%
    \ifthenelse{\boolean{algocf@inoutnumbered}}{\relax}{\algocf@seteveryparnl{\relax}}%
    {\let\\\algocf@newinout\hangindent=\inoutindent\hangafter=1\parbox[t]{\inoutsize}{\KwSty{#2\algocf@typo\hfill:}}~##1\par}%
    \algocf@linesnumbered
    \ifthenelse{\boolean{algocf@hanginginout}}{\relax}{\algocf@reseteveryparhanging}%
  }}
\makeatother
\SetKwInOutFit{Input}{Input}
\SetKwInOutFit{Output}{Output}
\newlength{\sosymIORuleAbove}
\newlength{\sosymIORuleBelow}
\newcommand{\sosymAlgoIORule}{%
    \par\kern\sosymIORuleAbove\hrule height0.4pt\kern\sosymIORuleBelow}
\newcommand{\algparam}[2]{#1 {\sosymCodeItalic #2}}
\DontPrintSemicolon
\SetKwProg{Fn}{def}{:}{}

\SetKwSty{sosymAlgoKw}
\SetFuncSty{sosymAlgoCode}
\SetArgSty{sosymAlgoCode}

\SetDataSty{sosymAlgoCode}
\SetCommentSty{sosymAlgoComment}
\newcommand{\algvar}[1]{#1}
\newcommand{\algstr}[1]{"#1"}
\SetKw{KwNot}{not}
\SetKw{KwAnd}{and}
\SetKw{KwElem}{in}

%% file: sections/Abstract.tex
\abstract{Large Language Models (LLMs) have recently been proposed to support domain modeling tasks, mostly for the completion of partial models. However, another important task is the instantiation of domain models into concrete objects, which has received little attention. Existing generators focus on structural validity, but leave aside whether instances are realistic and diverse, although only such instances are understandable for humans and useful, e.g., for teaching modeling, for model-based testing, or as training data for machine learning.

To address this gap, this paper presents LIMoGen (\textbf{L}LM-based \textbf{I}nstance \textbf{Mo}del \textbf{Gen}eration), an approach that generates UML object diagrams from UML class diagrams with LLMs that use feedback from the USE model validation tool to repair malformed or non-conforming instances. We compare two prompting strategies, a single instruction-based prompt and a Chain-of-Thought strategy that steers the generation toward specific unusual scenarios, such as boundary, edge, and over-constraint cases. We applied LIMoGen to ten domain models and evaluated the generated instances for validity, coverage, diversity, and realism, the latter with domain-specific checks and an LLM-as-a-judge audited by human experts. The results show that all instances are well-formed, conforming, and diverse in values and structure. Most realism problems are inconsistencies between values, which the judge largely detects, as confirmed by the experts. Furthermore, we show that these findings help improve the domain models, as uninstantiated classes suggest abstract classes, over-constraint scenarios reveal multiplicities that are too strict, and realism problems uncover implicit domain knowledge that can be encoded as additional invariants.
\manuel{todo: manuel : review this}
\andrei{Note: I think max abstract is 250 words}
}

\keywords{UML, Model Instantiation, Object Diagram, Class Diagram, Generative AI, LLM, USE}

%% file: sections/Introduction.tex
\section{Introduction}
\label{sec:introduction}
In recent years, we have seen several applications of AI for software engineering in general~\cite{belzner2023large,hou2023large} and for software modeling in particular~\cite{CMAI-SMS.23,Raedler.MDE-AI.24}. Most of the presented approaches operate at the type level of structural models, e.g., UML class diagrams being probably the most prominent notation, by proposing (semi)-automated model completions (see, e.g.,~\cite{ChaabenBS23,ChenYCLMV23}), or the provision of synthetically generated~\cite{He.2026} datasets of structural models. 

However, one aspect that has received little attention is the instance level; e.g., consider UML object diagrams for AI-based automation support in software modeling of concrete instantiations~\cite{BurguenoRSW25}. 
Having support for this level would be of major importance, as the type-level descriptions must be tested at the instance level in terms of completeness and correctness~\cite{BurguenoCCG19}. Only then can the type level serve as a reliable reference for validating the instance level. 

Previous studies followed different approaches to automate the synthetic instantiation of structural models by using, e.g., formal methods such as constraint solvers~\cite{CabotCR07,KuhlmannG12,SemerathNV18,Oszkar2019,BurguenoCCG19} and graph transformations~\cite{EhrigKT09,Taentzer12,RadkeABHT18,NassarKKT20} or heuristic-based approaches such as search-based techniques~\cite{AliIAB13,SoltanaSB20,KARIMI2024}.
The work by Semeráth et al. also showed that \textit{``Real instance models, created by humans [\dots] provide low mutation score''}~\cite{SemerathFBV20}, i.e., human-made models are realistic but structurally homogeneous, while synthetic generators optimize structure and ignore realism. Our goal is to obtain both. 
Some recent approaches, e.g.,~\cite{LopezC23}, aim to automatically mutate existing real-world models. However, using these approaches requires a large number of high-quality instance models as a prerequisite, which can be challenging in itself. 

Only recently have the first LLM-based instance generators appeared~\cite{PanPZWK25,PoloGambinRH26,HummelRHCKR26}, and they mainly focus on the structure of the instances or on their fidelity to a description of an existing system (see \secref{sec:relatedwork}). Thus, one remaining major challenge is generating instance models that go beyond structural validity or fidelity to an existing system, are realistic, and, at the same time, are diverse enough to exercise the type-level description under different scenarios. We define the notion of \textit{realistic} in this paper in the sense, that models are realistic if their instance property values relate to phenomena that can be observed in the real-world e.g., \textit{meaningful values} (e.g., contextually meaningful names like a street name in Spain could have a term `calle'), \textit{plausible values} (e.g., age of a human being between 0 and 120), values \textit{conforming to some norm or standard} (e.g., SWIFT code, an email address, or a phone number) etc. Furthermore, realistic values should be consistent and coherent within a single model to describe a realistic scenario coherently (e.g., in a banking model, the country code of an account's IBAN, the bank's BIC, and its country attribute should be coherent).
Note that we differentiate realistic from real values. The former can be considered representative of statistical distributions manifest in reality, while the latter would further expect values that themselves refer to traceable entities of the real world. 

\begin{figure*}[ht!]
\centerline{\includegraphics[width=.8\linewidth]{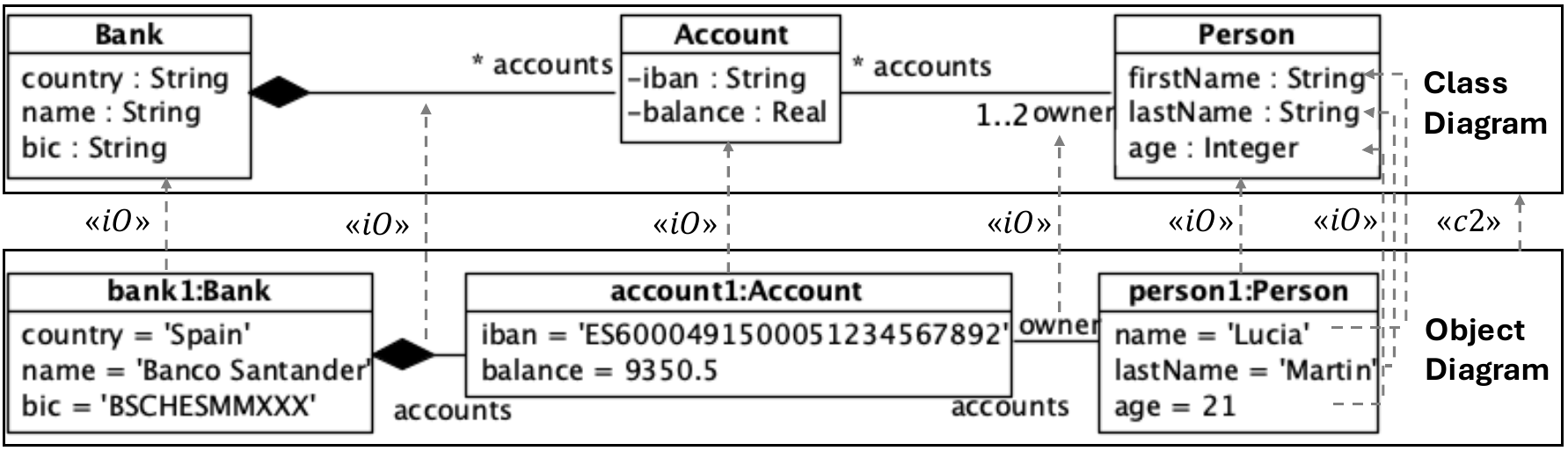}}
\caption{Class diagram example (top) with example instantiation as object diagram (bottom).}
\label{fig:modeling-at-a-glance}
\end{figure*}

To tackle the aforementioned challenge, this work investigates the use of LLMs to automatically generate diverse and realistic UML object diagrams from given UML class diagrams. We follow a design science approach~\cite{Wieringa14}, where LIMoGen is the artifact, evaluated through experiments. In particular, we make these contributions.
\begin{enumerate}[label=\textbf{C\arabic*.}, ref=C\arabic*]
    \item \textbf{Approach.} We introduce LIMoGen (\textbf{L}LM-based \textbf{I}nstance \textbf{Mo}del \textbf{Gen}eration), an approach, together with its supporting toolchain, that instantiates class diagrams as realistic and diverse object diagrams by combining two LLM prompting strategies with the USE~\cite{GogollaBR07} model validation tool, which provides feedback to repair invalid instances.

    \item \textbf{Evaluation.} We define an evaluation methodology for generated instance models and apply it to evaluate LIMoGen experimentally on ten domain models. It measures the validity of the generated instances using USE's well-formedness and conformance checks; their coverage and diversity with model coverage metrics, exact-match and normalized Levenshtein distances, Shannon evenness, and graph edit distance; and their realism with automated domain checks and an LLM-as-a-judge whose verdicts are audited by human experts. We also show how these results can be used to improve the underlying domain models by making uncovered classes abstract, relaxing overly restrictive constraints, or encoding additional OCL invariants to enhance realism. Finally, we publish all generated instances, along with the domain models, prompts, judge verdicts, expert audit, and derived OCL invariants, as a replication package~\cite{repo}.




\end{enumerate}

The remainder of this paper is structured as follows. In \secref{sec:motivation}, we establish the background of this work and provide a motivating example. In \secref{sec:approach}, we present our LIMoGen approach for LLM-based domain model instantiation. We discuss the comprehensive evaluation of our approach in \secref{sec:validation}. In \secref{sec:relatedwork}, we discuss related approaches before we conclude the paper with an outlook on future work in \secref{sec:conclusion}. 

%% file: sections/BackgroundMotivation.tex
\section{Background and Motivation}
\label{sec:motivation}

Class diagrams are among the most widely used structural modeling languages in software engineering~\cite{BudgenBBKP11,HUTCHINSON2014144}. They allow modeling of domains using classes to represent different entity types, associations to represent relationships between entity types, and attributes to model the intrinsic properties of those types. For instance, see the upper part of Fig.~\ref{fig:modeling-at-a-glance}, which describes the relationships between the concepts of a financial domain: \textit{Bank}, \textit{Account}, and \textit{Person}. 
In addition, several constraints can be defined for the different elements to capture also these aspects of the domain, such as multiplicities of the association ends (e.g., an account can have one or two Persons as owners), data types, or even more complex constraints in terms of using a constraint language such as the Object Constraint Language (OCL).

While class diagrams focus on defining the types of a domain, object diagrams represent concrete situations. For instance, the lower part of Fig.~\ref{fig:modeling-at-a-glance} assigns the person \textit{`Lucia Martin'} to the bank account with the IBAN \textit{`ES600...'} of the \textit{`Banco Santander'} by instantiating the elements of the corresponding class diagram. While this is often a manual process for describing a concrete domain setting, the question arises as to whether automation support can be provided at this level, as it is currently emerging at the class diagram level~\cite{ChaabenBS23,ChenYCLMV23}. 

To realize LLM-based class diagram instantiation, an LLM would first need to consider the following aspects. First, as a basis, the syntax of object diagrams must be known (i.e., what objects are, the links between objects, and the slots of objects for storing values). Second, the relationships between the type level and the instance level must be known. This comprises the \textit{instanceOf} (iO) relationships between objects and their class, as well as the \textit{conformsTo} (c2) relationships between the object diagrams and their class diagram. The mechanisms underlying these relationships must also be known to the LLMs to reason about valid instantiations, i.e., how to ensure consistency between the class and object diagrams. Third, domain knowledge, captured in the provided class diagrams, is required to build meaningful, human-understandable instantiations that yield realistic models. Finally, to generate diverse models, e.g., to reason about domain coverage, deep knowledge of the first three points needs to be combined.

These requirements map to the three dimensions we use throughout the paper. Well-formedness and conformance together form the \textit{validity} dimension of an instance: an object diagram is \emph{well-formed} if it is a valid object graph in the concrete syntax (here SOIL), and \emph{conforming} if it respects the multiplicities and OCL constraints of its class diagram. The second dimension, \emph{diversity}, applies to a set of instances, and also to the values inside one instance. It has two facets: \emph{value diversity}, which is variation in attribute values, and \emph{structural diversity}, which is variation in the numbers and combinations of objects and links. Diversity is complemented by \emph{coverage}, i.e., whether the instances use all elements of the class diagram (its classes, attributes, and relationships), since a set of instances can vary a lot while never touching some parts of the model. The third dimension is \emph{realism}: an object diagram is \emph{realistic} if its values and their combinations relate to phenomena that can be observed in the real world.

The availability of realistic models would be helpful in many scenarios, such as explaining the expressiveness and solution space of type-level structural models using concrete examples. This may also be very helpful for teaching abstractions to students. Another scenario is the generation of test models for model-processing tools (e.g., code generators or model transformations) and model-based software systems (e.g., through novel approaches to test amplification). Furthermore, it may be beneficial in the future to explore type-level domain abstractions by reasoning about certain instantiations, which may help complete or improve the type-level descriptions.

Another scenario would be to further use this approach to generate repositories of models for training machine learning models. Currently, there is increasing interest in such model repositories~\cite{Hamid17,LopezIC22-ModelSet,Glaser25-EAModelSet,GlaserBB26,VerbruggenNGSHCRSB25,He.2026}, yet few provide large numbers of high-quality models. In particular, the qualitative aspect is intriguing, as many models found in public GitHub repositories are of low quality (e.g., providing only template or dummy models)~\cite{Laue.24,DelicAVNB25}.

%% file: sections/Approach.tex
\section{Approach}
\label{sec:approach}
\begin{figure*}[t]
    \centering
    \includegraphics[width=0.8\linewidth]{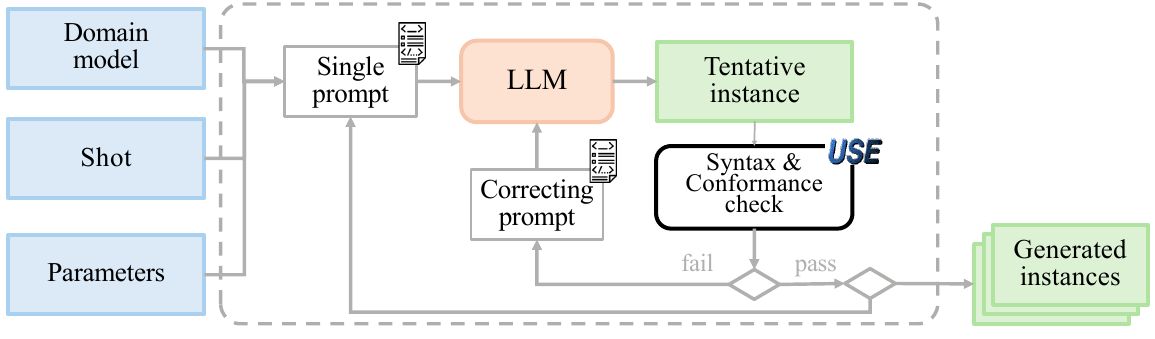}
    \caption{Overview of our Instruction Learning model instance generation strategy}
    \label{fig:learningstrategies-IL}
\end{figure*}

In the following, we first introduce the preliminaries of LIMoGen (\secref{sec:approach:prelim}) before detailing the two different strategies we implemented for LLM-based domain model instantiation (cf.\ Figs.~\ref{fig:learningstrategies-IL} and~\ref{fig:learningstrategies-CoT}). 
The first strategy uses a single prompt with a combination of \textit{instruction learning} and \textit{one-shot learning} (Fig.~\ref{fig:learningstrategies-IL}; introduced in \secref{sec:approach:instruct-learn}), while the second strategy is inspired by \textit{Chain-of-Thought (CoT) learning} with three prompts (Fig.~\ref{fig:learningstrategies-CoT}; introduced in \secref{sec:approach:CoT-learn}). 
The rationale for exploring these two strategies is that the former is more efficient and sustainable because it uses fewer prompts, whereas the latter has the potential to yield more diverse instances by asking the LLM to generate specific instance categories (see Section~\ref{sec:approach:CoT-learn}).\footnote{In our evaluation (\secref{sec:generation-protocol}), generating 300 IL instances took 321 LLM calls with 10.1M input and 0.9M output tokens (about US\$31), whereas generating 300 CoT instances took 642 calls with 19.2M input and 1.8M output tokens (about US\$59), both at official API costs at the time of generation.}

\subsection{Preliminaries}
\label{sec:approach:prelim}

To present class diagrams and object diagrams in a compact, comprehensible format to LLMs, we have chosen the UML-based Specification Environment (USE) tool~\cite{GogollaBR07}. USE provides a textual syntax for representing domain models and includes the SOIL language for defining instances.
Listings~\ref{lst:domainModelUSE} and~\ref{lst:instanceModelUSE} show excerpts of the domain model and instance model of Fig.~\ref{fig:modeling-at-a-glance}, respectively. Note that the domain model in USE can be enriched with OCL constraints.

\begin{lstlisting}[language=USE,firstnumber=0,escapeinside={(*@}{@*)},caption=Excerpt of the domain model represented as a class diagram in USE (cf.\ Fig.~\ref{fig:modeling-at-a-glance}),label=lst:domainModelUSE]
(*@\sosymListingHeading{Domain model:}@*)
model BankAccount
class Account
attributes
    iban: String
    balance : Real
end
class Person
attributes
    firstName:String
    lastName:String
    age : Integer
end [...]
association Ownership between
    Person [1..2] role owner
    Account [*] role accounts
end [...]
constraints
context Account inv AdultOwners:
    self.owner->forAll(p | p.age >= 18) [...]
\end{lstlisting}

\begin{lstlisting}[language=USE,firstnumber=0,escapeinside={(*@}{@*)},caption=Excerpt of an instance model in SOIL (cf.\ Fig.~\ref{fig:modeling-at-a-glance}),label=lst:instanceModelUSE]
(*@\sosymListingHeading{Instance model:}@*)
!new Account('account1')
!account1.iban := 'ES6000491500051234567892'
!account1.balance := 9350.5
!new Person('person1')
!person1.age := 21
!person1.firstName := 'Lucia'
!person1.lastName := 'Martin'
!insert (person1, account1) into Ownership
\end{lstlisting}

\subsection{IL Instance Generation}
\label{sec:approach:instruct-learn}

Our first strategy, Instruction Learning (IL)\footnote{Also referred to as \textit{Simple} in the repo~\cite{repo}}, uses a single prompt that employs instruction learning. In this prompt, we provide detailed instructions about the task we ask the LLM to perform.
It has been noted that LLMs generally perform poorly on low-resource languages (languages with limited training data) and domain-specific languages~\cite{joel_survey_2024}.
Specific text-based modeling languages such as SOIL fall into this category.
Hence, we combine instruction learning with one-shot learning. One-shot learning is a prompting technique in which the LLM is provided with a single example (shot) of what it is expected to produce. We provide this single shot in the system prompt (cf.~Listing~\ref{lst:promptInstructionLearning}). With this shot, our goal is to help the LLM learn about the particularities of the languages we are using, in our case, USE and SOIL, helping the LLM to learn how the instances need to be generated, especially from a syntactic point of view. Therefore, we provide the LLM with an example (shot) of how to instantiate a domain model in a concrete scenario. Fig.~\ref{fig:learningstrategies-IL} graphically depicts the steps to generate instances, and a pseudo-code presentation is given by Algorithm~\ref{alg:algorithmSimple}. 


\begin{algorithm}[!tp]
\Input{\algparam{dm}{domain model}\\
\algparam{shot}{SOIL syntax example}\\
\algparam{numInstances}{instance count}\\
\algparam{maxChecks}{checks per instance}}
\Output{\algparam{instances}{set of instances}}
\sosymAlgoIORule
\Fn{run(\algvar{dm}, \algvar{shot}, \algvar{numInstances}, \algvar{maxChecks})}{
    \algvar{instances} $\leftarrow$ \{\}\;
    openNewChat()\;
    \algvar{tInst} $\leftarrow$ prompt(\algvar{dm}, \algvar{shot})\;
    \algvar{inst} $\leftarrow$ checkConformance(\algvar{tInst})\;
    add(\algvar{inst}, \algvar{instances})\;
    \For{\algvar{n} $\leftarrow$ 2 \KwTo \algvar{numInstances}\nllabel{alg:il:loop}}{
        \algvar{tInst} $\leftarrow$ prompt(\algstr{nextInstance})\;
        \algvar{inst} $\leftarrow$ checkConformance(\algvar{tInst})\;
        add(\algvar{inst}, \algvar{instances})\nllabel{alg:il:loopEnd}\;
    }
    \KwRet{\algvar{instances}}\;
}
\BlankLine
\Fn{checkConformance(\algvar{tInst})}{
    \tcp{uses \algvar{dm} and \algvar{maxChecks} from the input}
    \algvar{i} $\leftarrow$ 0\;
    \algvar{sPass}, \algvar{err} $\leftarrow$ syntaxCheckUSE(\algvar{tInst})\nllabel{alg:il:syn}\;
    \While{\KwNot \algvar{sPass} \KwAnd \algvar{i} < \algvar{maxChecks}\nllabel{alg:il:synLoop}}{
        \algvar{tInst} $\leftarrow$ prompt(\algstr{correct}, \algvar{err})\;
        \algvar{sPass}, \algvar{err} $\leftarrow$ syntaxCheckUSE(\algvar{tInst})\;
        \algvar{i} $\leftarrow$ \algvar{i} + 1\nllabel{alg:il:synLoopEnd}\;
    }
    \If{\algvar{sPass}}{
        \algvar{i} $\leftarrow$ 0\;
        \algvar{cPass}, \algvar{err} $\leftarrow$ conformanceCheckUSE(\algvar{tInst})\nllabel{alg:il:conf}\;
        \While{\KwNot \algvar{cPass} \KwAnd \algvar{i} < \algvar{maxChecks}\nllabel{alg:il:confLoop}}{
            \algvar{tInst} $\leftarrow$ prompt(\algstr{correct}, \algvar{err})\;
            \algvar{cPass}, \algvar{err} $\leftarrow$ conformanceCheckUSE(\algvar{tInst})\;
            \algvar{i} $\leftarrow$ \algvar{i} + 1\nllabel{alg:il:confLoopEnd}\;
        }
    }
    \KwRet{\algvar{tInst}}\;
}
\caption{Instruction Learning instance generation}
\label{alg:algorithmSimple}
\end{algorithm}

Our approach uses the domain model\footnote{Also referred to as \textit{conceptual model} in the prompts} in textual USE format, a shot illustrating the instance model syntax in SOIL, and two input parameters: the number of instances to be created and the maximum number of times the LLM is asked to correct an erroneously generated instance. 

Our approach starts by creating a single prompt following the template that Listing~\ref{lst:promptInstructionLearning} shows, where 
\{\{modelUML\}\} in line 14 is to be replaced by a class diagram in the textual USE format (cf.\ Listing~\ref{lst:domainModelUSE}); and \{\{syntaxExample\}\} in line 10 by the syntax example shot in SOIL format. 

We have designed the syntax example shot to cover a good number of instance elements, such as comments, primitive data types, and the instantiation of objects and relations in USE. 
For space reasons, we cannot provide all prompt templates here, but they are available in our online supplementary material repository~\cite{repo} in \texttt{docs/ai-prompts}. 

\begin{lstlisting}[language=Prompt,firstnumber=0,numberstyle={\sosymPromptListingNumber{12}},caption=Prompt for instruction learning,label=lst:promptInstructionLearning]
(*@\sosymListingHeading{System prompt:}@*)
Given a conceptual model expressed in the UML-based Specification Environment (USE), your task is to generate valid and realistic instances that conform to the provided model.
<requirements>
- Instances must be syntactically correct according to the USE syntax_reference.
- Avoid unnecessary comments and output the instance in plain text (i.e., not markdown).
- Make sure instances fulfill all the model's constraints, and that multiplicities, relationships, and attributes are valid and realistic.
- Provide multiple instances with diverse data values and structure.
</requirements>
<syntax_reference>
Here there is a snippet showing how to create objects and set values in the specific .soil language required:
{{syntaxExample}}
</syntax_reference>
(*@\sosymPromptUserHeading{User prompt:}@*)
Create an instance for the following domain model:
<domain_model>
{{modelUML}}
</domain_model>
\end{lstlisting}


To ensure diversity among the generated model instances, we leverage the LLM's \textit{``memory''} and request additional instances within the same \textit{``chat''}. After the LLM generated one instance, we use the prompt described in Listing~\ref{lst:prompt-diverse} to ask the LLM to generate another instance instead of asking for all of them at once. We repeat this process as many times as the user has specified in the input parameter \texttt{numInstances} (see lines~\ref{alg:il:loop}--\ref{alg:il:loopEnd} in Algorithm~\ref{alg:algorithmSimple}). 

\begin{lstlisting}[language=Prompt,firstnumber=0,caption=Prompt for generating another instance,label=lst:prompt-diverse]
(*@\sosymListingHeading{User prompt:}@*)
Please generate another instance that is structurally and semantically different from the previous ones.
\end{lstlisting} 

For each tentative instance generated by the LLM, we use USE to check the conformance to the class diagram (syntax and constraints) (lines~\ref{alg:il:syn} and~\ref{alg:il:conf} in Algorithm~\ref{alg:algorithmSimple}).
When a check fails, we use the prompt in Listing~\ref{lst:prompt-feedback}, which includes the error that USE throws to help the LLM fix its mistake. We repeat this process until the instance passes the test or up to the maximum number of times that the user specified in the parameter \texttt{maxChecks} (lines~\ref{alg:il:synLoop}--\ref{alg:il:synLoopEnd} and~\ref{alg:il:confLoop}--\ref{alg:il:confLoopEnd} in Algorithm~\ref{alg:algorithmSimple}).

\begin{lstlisting}[language=Prompt,firstnumber=0,caption=Prompt for correcting a tentative model instance,label=lst:prompt-feedback]
(*@\sosymListingHeading{User prompt:}@*)
The last output is partially incorrect:
{{errors}}
Please provide the complete output corrected
\end{lstlisting}

\subsection{CoT Instance Generation}
\label{sec:approach:CoT-learn}
\begin{figure*}[t]
    \centering
    \includegraphics[width=0.8\linewidth]{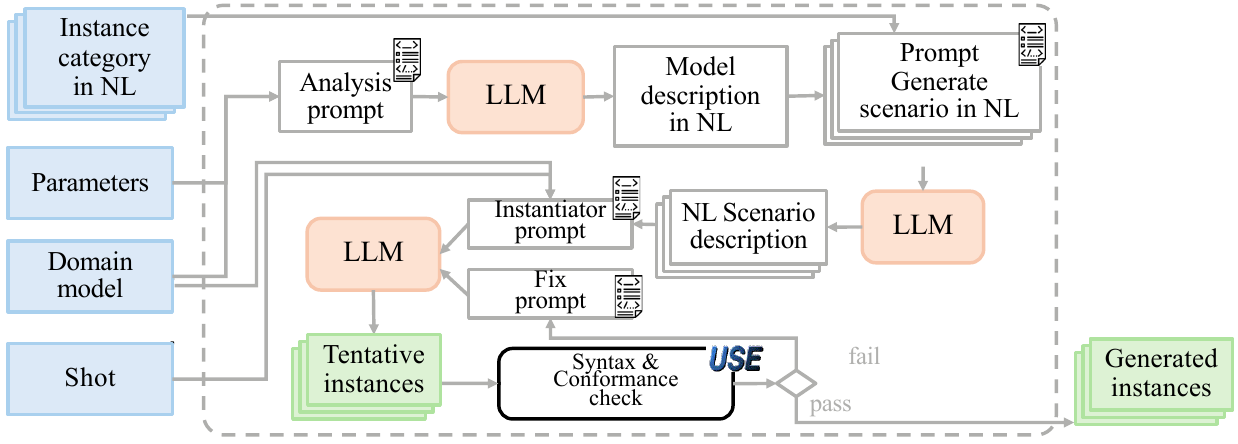}
    \caption{Overview of our Chain of Thought Learning generation strategy}
    \label{fig:learningstrategies-CoT}
\end{figure*}

Inspired by the Chain-of-Thought strategy~\cite{Wei2022CoT}, we implement \textit{``CoT''}, a prompt-chaining approach combined with one-shot learning, where the natural-language output of each step is passed as input to the next prompt to further improve the LLM-based model instance generation. 
We provide a visual representation of our algorithm in Fig.~\ref{fig:learningstrategies-CoT} and its formalization in Algorithm~\ref{alg:algorithmCoT}.

In addition to the inputs of the Instruction Learning strategy, the CoT strategy receives a list describing the categories of instances that the LLM is expected to generate. 
Our approach starts by providing the LLM with the domain model as input and asking it to generate a natural-language description of it. This step aims to help improve the LLM's understanding of the domain to be modeled. 
The prompt we have designed is described in Listing~\ref{lst:prompt-analysis}.

\begin{algorithm}[!t]
\Input{\algparam{dm}{domain model}\\
\algparam{shot}{SOIL syntax example}\\
\algparam{categories}{instance categories}\\
\algparam{numInstances}{count per category}\\
\algparam{maxChecks}{checks per instance}}
\Output{\algparam{instances}{set of instances}}
\sosymAlgoIORule
\Fn{run(\algvar{dm}, \algvar{categories}, \algvar{shot}, \algvar{numInstances}, \algvar{maxChecks})}{
    \algvar{instances} $\leftarrow$ \{\}\;
    openNewChat()\tcp*{model description}
    \algvar{modelDesc} $\leftarrow$ prompt(\algstr{analyze}, \algvar{dm})\nllabel{alg:cot:analyze}\;
    \ForEach{\algvar{c} \KwElem \algvar{categories}\nllabel{alg:cot:catLoop}}{
        \tcp{the categories run concurrently}
        genScenarios(\algvar{c}, \algvar{modelDesc}, \algvar{scenarios})\nllabel{alg:cot:catLoopEnd}\;
    }
    openNewChat()\nllabel{alg:cot:instChat}\tcp*{soil instantiation}
    \ForEach{(\algvar{c}, \algvar{s}) \KwElem \algvar{scenarios}\nllabel{alg:cot:instLoop}}{
        \tcp{in the order the scenarios arrive}
        \eIf{\algvar{instances} = \{\}}{
            \algvar{tInst} $\leftarrow$ prompt(\algvar{dm}, \algvar{shot}, \algvar{s})\;
        }{
            \algvar{tInst} $\leftarrow$ prompt(\algstr{continue}, \algvar{s})\;
        }
        \algvar{inst} $\leftarrow$ checkConformance(\algvar{tInst}, \algvar{c})\;
        add(\algvar{inst}, \algvar{instances})\nllabel{alg:cot:instLoopEnd}\;
    }
    \KwRet{\algvar{instances}}\;
}
\BlankLine
\Fn{genScenarios(\algvar{c}, \algvar{modelDesc}, \algvar{scenarios})}{
    openNewChat()\tcp*{for the scenarios of \algvar{c}}
    add((\algvar{c}, prompt(\algvar{c}, \algvar{modelDesc})), \algvar{scenarios})\;
    \For{\algvar{n} $\leftarrow$ 2 \KwTo \algvar{numInstances}}{
        add((\algvar{c}, prompt(\algstr{nextScenario})), \algvar{scenarios})\;
    }
}
\BlankLine
\Fn{checkConformance(\algvar{tInst}, \algvar{c})}{
    \tcp{As in Algorithm~\ref{alg:algorithmSimple}, but only the syntax check runs when \algvar{c} is over-constraint}
}
\caption{CoT instance generation}
\label{alg:algorithmCoT}
\end{algorithm}

\begin{lstlisting}[language=Prompt,firstnumber=0,numberstyle={\sosymPromptListingNumber{15}},caption=Prompt for analyzing a domain model,label=lst:prompt-analysis]
(*@\sosymListingHeading{System prompt:}@*)
You are tasked with analyzing conceptual models represented as class diagrams and expressed in the UML-based specification environment using its native syntax.
You must adhere to the following requirements:
- Use very clear language.
- Do not overexplain, be concise.
- Multiplicities must be very clear and easy to understand.
You should follow the structure and requirements below:
## Description
Start by explaining the overall structure and purpose of the model.
### Components
Break down the components of the model (i.e., classes and attributes), describing each, their type and purpose.
## Relationships
Describe the relationships between the components of the model, dependencies and multiplicities (i.e., minimum and maximum number of instances of one class that can be associated with instances of another class). Describe the multiplicities at both ends of each association.
## Invariants
Define the invariants that apply to the model (i.e., those constraints that must be fulfilled).
(*@\sosymPromptUserHeading{User prompt:}@*)
Analyze the following UML class diagram:
{{modelUML}}
\end{lstlisting}

\begin{figure*}[t]
    \centering
    \includegraphics[width=0.8\linewidth]{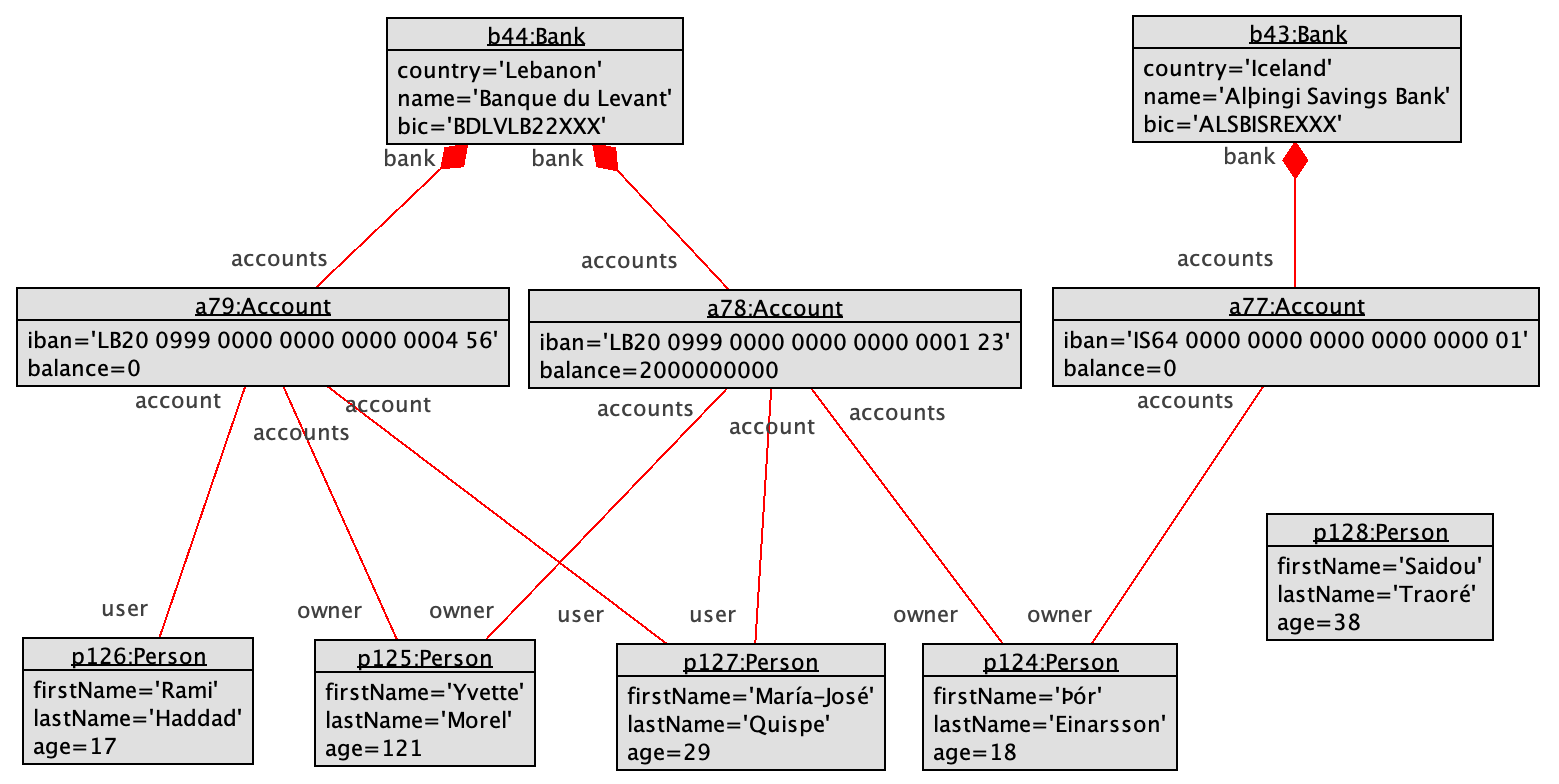}
    \caption{Boundary instance \cotinst{bank}{4}{boundary}{\texttt{Bank boundary4}} generated for the Bank model. It has an owner aged exactly 18 (the minimum allowed by invariant \texttt{AdultOwners}), two accounts with a balance of exactly 0, and accounts with one and two owners (the bounds of multiplicity \texttt{1..2}). Where the model defines no bound, the instance explores the limits of plausible values, such as an owner aged 121 or a balance of 2,000,000,000.}
    \label{fig:boundary-bank}
\end{figure*}

Then, for each \textit{category}, we ask the LLM to generate \texttt{numInstances} characteristic natural-language model instances (scenarios) for the provided domain model that fall into that category (lines~\ref{alg:cot:catLoop}--\ref{alg:cot:catLoopEnd} in Algorithm~\ref{alg:algorithmCoT}). Each category has its own chat, and the categories run concurrently. As in the IL strategy, the first scenario of a category is requested with the prompt in Listing~\ref{lst:prompt-system}, and each further one with a follow-up prompt in the same chat that asks for another scenario of the same category that is structurally and semantically different from the previous ones. The categories are inspired by classical black-box test-design techniques~\cite{OstrandB88,Myers2011}, which partition the input space into categories and select representative values from each. Listing~\ref{lst:prompt-system} shows the prompt we use.
\begin{itemize}
    \item \emph{Baseline scenario:} a standard, typical scenario in which each class and relationship of the model is present at least once, if applicable. The object diagram in Fig.~\ref{fig:modeling-at-a-glance} illustrates a small instance of this kind.
    \item \emph{Complex scenario:} a scenario that contains multiple interrelated entities and/or entities involved in several constraints at once. For example, in the Bank model, several people who jointly own some accounts and use others, held at banks in different countries.
    \item \emph{Boundary scenario:} a scenario at the structural and value limits of the model, i.e., at the ends of the ranges it defines, such as the bounds of multiplicities and the limits of attribute values. Fig.~\ref{fig:boundary-bank} shows an example for the Bank model.
    \item \emph{Edge scenario:} a scenario that is valid and possible in real life, but unusual or unlikely, not necessarily at the end of any defined range. While the boundary scenario targets the syntactic limits of the model, the edge scenario targets the limits of realism. For example, a bank in Greenland without any accounts, or an account used by a 16-year-old, which the model allows because only owners must be adults. Fig.~\ref{fig:edge-bank} shows an instance generated for this category. (Note that as we discuss in \secref{sec:rq3-realism} (RQ3.1), it also contains an incoherent combination of values, which shows that the LLM does not always keep such scenarios realistic)
    \item \emph{Over-constraint\footnote{Also referred as \textit{Invalid} in the repo~\cite{repo}} scenario:} a scenario that represents a plausible real-life situation but intentionally violates the model's multiplicities or constraints, exposing restrictions that are too strict. For example, a joint account owned by three siblings violates the multiplicity \texttt{1..2} of role \texttt{owner} in the Bank model. Fig.~\ref{fig:ovc-example-hotel} shows a generated example for the Hotel model.
\end{itemize}

\begin{lstlisting}[language=Prompt,firstnumber=0,numberstyle={\sosymPromptListingNumber{8}},caption=Prompt for generating an instance in plain English,label=lst:prompt-system]
(*@\sosymListingHeading{System prompt:}@*)
Your task is to generate a complete and diverse instance, in plain English, for a given category and based on a provided conceptual model description. The instance must adhere to these requirements:
- Be self-contained: Include all required attributes, relationships, and related entities in full detail.
- Conform to the model: Fulfill the constraints, multiplicities, relationships, and attributes defined in the class diagram model.
- Understand the context: Ensure that its attributes and relationships are relevant.
- Avoid duplication of instances: Take into consideration those instances previously built to avoid redundancy.
- Semantic diversity: From a semantic point of view, incorporate varied scenarios, including regional, linguistic, or cultural differences.
- Structural diversity: Include instances with different numbers of elements, different numbers of relationships and complexity, and create varied examples by changing entity attributes.
(*@\sosymPromptUserHeading{User prompt:}@*)
# Conceptual model description:
{{modelDescription}}
{{categoryPrompt}}
\end{lstlisting}

For illustration, Listing~\ref{lst:prompt-boundaryInstances} shows the prompt we have used for the \emph{boundary scenario} category.

\begin{lstlisting}[language=Prompt,firstnumber=0,caption=Prompt for the boundary scenario,label=lst:prompt-boundaryInstances]
(*@\sosymListingHeading{Category prompt:}@*)
# Category: Boundary Instances
Create a boundary case instance. This is an instance that focuses on the extreme upper or lower limits of valid input ranges. For example:
- Upper or lower limits of multiplicities.
- For numbers in a range, the minimum and maximum valid values.
- Empty collections when possible, i.e., when they do not violate the semantics of the model or its constraints.
\end{lstlisting}

Each scenario description is then turned into an instance in SOIL syntax (lines~\ref{alg:cot:instChat}--\ref{alg:cot:instLoopEnd} in Algorithm~\ref{alg:algorithmCoT}). All scenarios of all categories share a single instantiation chat and are processed in the order in which the category chats deliver them. The first scenario is sent with the prompt in Listing~\ref{lst:prompt-instantiator}, together with the class diagram and the syntax example shot in SOIL format. Each further scenario is sent in the same chat with a short follow-up prompt that asks the LLM to continue with the new description, following the syntax example and the specification. In this way, the class diagram, the shot, and the previously created instances remain in the context, which helps the LLM avoid duplicate names and variables.

\begin{lstlisting}[language=Prompt,firstnumber=0,numberstyle={\sosymPromptListingNumber{10}},caption=Prompt for the generation of an instance,label=lst:prompt-instantiator]
(*@\sosymListingHeading{System prompt:}@*)
You are tasked with creating instances of a conceptual model in the UML-based Specification Environment (USE). You will receive:
1. The UML class diagram that the instance follows.
2. A sample syntax of instances creation.
3. A description of the instance that needs to be created.
Your goal is to generate these instances based on the provided description, adhering strictly to these requirements:
- The output must be in plain text, with no additional comments, descriptions, or explanations.
- Ensure that the created instance adheres to the provided description.
- Follow the syntax sample provided, without deviation.
- Take into account previously created instances to avoid using duplicate naming.
(*@\sosymPromptUserHeading{User prompt:}@*)
# UML class diagram:
{{modelUML}}
# Syntax example of instances creation:
{{syntaxExample}}
# Instance description:
Please create the instance according to this specification:
{{list}}
\end{lstlisting}

Like for the IL strategy, we use USE to perform conformance checks and corrections (when needed) before adding the generated instance to the output set of instances. The correction prompts are sent in the same instantiation chat. Over-constraint instances intentionally violate the model, so for them we only check and correct the syntax.

\subsection{Tool Support}
\label{sec:approach:tools}
We implemented LIMoGen as a Java-based tool leveraging LangChain4j\footnote{\url{https://docs.langchain4j.dev/}} and the UML-based Specification Environment Tool (USE)~\cite{GogollaBR07} to check for syntactic and conformance errors. All LLMs are accessed through OpenRouter\footnote{\url{https://openrouter.ai/}}, a service that exposes models from different vendors behind a single API. Throughout our preliminary experiments, we evaluated multiple LLMs, including GPT-4o, DeepSeek V3, and Gemini 2.0. Ultimately, we selected GPT-5.2 as our model of choice for the instance generation due to its state-of-the-art status at testing time, its reasonable speed, and its initial performance using a variant of the banking domain model\footnote{Results of this initial assessment of different LLMs are provided in~\cite{repo} \texttt{docs/preliminary}}.

It should be noted that, although we used the described tools, LIMoGen remains agnostic to the programming language, the selected libraries, and the LLM, which is crucial given the rapid evolution of LLMs. LangChain4j's model abstractions and OpenRouter make it straightforward to switch to other LLMs in the future, requiring only a change to the model identifier string in the tool configuration.

%% file: sections/Evaluation.tex
\section{Evaluation}
\label{sec:validation}

In this section, we report on the evaluation of the presented instance-generation approach, in particular the two prompting strategies introduced in Section~\ref{sec:approach}. Using experiments, we study whether they generate valid instances with respect to well-formedness and conformance, whether the generated instance set is diverse, and whether the generated instances are plausible and consistent as snapshots of real-world domains. 


\subsection{Research Questions}

We instantiate our approach with the two prompting strategies introduced in Section~\ref{sec:approach}, IL and CoT, and analyze our LLM-based approach with the following research questions:
\begin{questions}
    \item \textbf{How effective is LIMoGen at producing well-formed and conforming domain model instances?}
    \begin{questions}
        \item \textbf{Well-formedness:} To what extent do the generated instances comply with the plain object diagram syntax?
        \item \textbf{Conformance:} To what extent do the generated instances conform to the constraints and multiplicities defined in the given domain model?
    \end{questions}

    \item \textbf{How effective is LIMoGen at producing domain model instances that cover the model and are diverse in values and structure?}
    \begin{questions}
        \item \textbf{Domain model coverage:} To what extent do the generated instances cover the elements defined in the domain model?
        \item \textbf{Value diversity:} How much attribute value variation exists within and among the generated instances?
        \item \textbf{Structural diversity:} How much structural variation exists within and among the generated instances?
    \end{questions}

    \item \textbf{How effective is LIMoGen at producing realistic domain model instances, and how can realism be assessed and turned into model improvements?}
    \begin{questions}
        \item \textbf{Automated evaluation:} To what extent do automated domain validation checks and an LLM-as-a-judge classify the generated instances as realistic?
        \item \textbf{Human-based analysis:} What types of realism issues can be identified, and how reliably do the LLM-as-a-judge assessments align with human assessments?
        \item \textbf{Model refinement:} To what extent can the identified realism issues be formalized as additional OCL constraints that can be used to strengthen the domain models?
    \end{questions}
\end{questions}

\subsection{Experimental Setup}
\label{sec:experimental-setup}

In the following, we describe the experimental setup of our evaluation, in particular the domain models used, the instance generation protocol, and the evaluation methodology.

\subsubsection{Domain Models}

\input{tables/domains}

We have built a dataset of ten domain models (see repository~\cite{repo} \texttt{docs/models}). Some of them were created by the authors for their lectures, while the others were taken from ModelSet~\cite{LopezIC22-ModelSet} and adapted for this study, i.e., translated into USE and modified. We selected them taking into account $(i)$ different sizes, $(ii)$ variety of application domains, $(iii)$ the coverage of different modeling constructs such as constraints, $(iv)$ abstraction level (language level vs.\ concrete modeling level), and $(v)$ temporal or domain-specific reasoning, e.g., models which represent time or entail temporal constraints such as predecessor and successor relationships.

Table~\ref{tab:characteristics} summarizes the main characteristics of the dataset. The number of model elements ranges from three classes in the Bank model to 21 in the Restaurant model, from two attributes in the State Machine metamodel to 37 in the Football model, and from three relationships in the Bank, Expenses, and Videoclub models to 18 in the Football model. Five models use inheritance relationships between classes, and seven contain enumerations. 

The State Machine model is the only metamodel; the remaining nine models describe dedicated application domains. All models include OCL constraints, and six require some form of temporal reasoning.

Moreover, all models (except the State Machine model) come with one or several domain-specific realism checks on the values of their attributes (e.g., valid IBANs, phone numbers, or chronologically ordered dates), which are not expressed in the domain models directly but are realized in external validation services which are used to assess the realism of the instances (see the last column of Table~\ref{tab:characteristics}). 


\subsubsection{Generation Protocol} \label{sec:generation-protocol}

\input{tables/llm-config}

\input{tables/instantiation}


For each of the ten domains, we generated 30 instances using Instruction Learning (IL) and 30 using Chain-of-Thought (CoT), for a total of 600 instances. IL generates the 30 instances sequentially within a single conversation. CoT performs six rounds of generation, each producing one instance for each of the five categories: baseline, complex, boundary, edge, and over-constraint. Thus, both strategies contribute the same number of instances per domain. Table~\ref{tab:llm-config} lists the LLM used, sampling parameters, and run dates.

Every tentative SOIL-based instance model was parsed and checked with USE, and a failed check was returned to the generator for up to two feedback-based repair rounds. Since over-constraint instances violate the model on purpose, they only receive syntax repair (\secref{sec:approach:CoT-learn}). Table~\ref{tab:sizeGenInstances} reports the total number of objects, attribute assignments, and association links across the 30 instances generated for each domain and strategy. All 600 generated instances are available in our repository~\cite{repo} in \texttt{docs/instances}.

\subsubsection{Evaluation Methodology}
\label{sec:eval-metrics}

This subsection defines the methodology, including the metrics we use to answer each research question. Unless stated otherwise, all metrics are computed on the 30 instances generated for each domain and strategy (as described in Subsection~\ref{sec:generation-protocol}). Whenever we aggregate a metric over several instances, we report the mean $\pm$ standard deviation.





\paragraph{RQ1 - Validity}
First, we study whether a generated SOIL script produces a valid instance of its associated domain model. In particular, we look at two complementary aspects. \emph{Well-formedness} asks whether USE can parse and execute the SOIL commands to construct the instance, i.e., the object diagram. \emph{Conformance} asks whether the object diagram that results satisfies the multiplicities and OCL invariants of the class diagram.

For \textbf{well-formedness} (RQ1.1), we count SOIL commands that USE cannot parse or execute successfully and the repair attempts until an instance is fully well-formed or the repair budget is consumed. After the repair loop, we report the sums of failing SOIL commands for each domain model. For each repair round, we report the number of correction calls and the number of instances accepted at that attempt, and we classify the corresponding issues in a taxonomy.



For \textbf{conformance} (RQ1.2), we count multiplicity and OCL invariant violations reported by the USE built-in checker and the repair attempts until an instance fully conforms or the repair budget is consumed. After the repair loop, we report the sums of failing multiplicity constraints and OCL invariants for each domain model, and separately for the CoT over-constraint instances, whose violations are intended. For each repair round, we report the number of correction calls and the number of instances accepted in that attempt, and we classify the corresponding issues in a taxonomy.


Let $E^{\mathrm{wf}}_i$, $E^{\mathrm{mult}}_i$, and $E^{\mathrm{inv}}_i$ be the well-formedness errors, multiplicity violations, and failed invariants USE reports for instance $i$, respectively. After the repair loop, we report the sums as depicted in Formula \ref{formula:rq1-errors}.

\begin{equation}
  \begin{gathered}
  E^{x}=\sum_{i=1}^{n} E^{x}_i,\quad x\in\{\mathrm{wf},\mathrm{mult},\mathrm{inv}\}\\
  n=\text{number of evaluated instances}
  \end{gathered}
\label{formula:rq1-errors}
\end{equation}


\paragraph{RQ2 - Coverage and diversity}
For RQ2, we examine whether repeated generation explores the instance space or repeatedly selects the same instance. To this end, we measure coverage and the two facets of diversity defined in \secref{sec:motivation}. 

For \textbf{coverage} (RQ2.1), we compute, for every generated instance, the proportion of concrete classes, attributes, and associations of its domain model that it instantiates, and aggregate the values by domain and strategy.

Let $M_{\mathrm{cl}}$ be the set of instantiable classes of a domain model (abstract classes excluded), $M_{\mathrm{att}}$ the set of (class, attribute) pairs on those classes, with inherited attributes counted on each subclass, and $M_{\mathrm{rel}}$ the set of associations, compositions, and aggregations. For a generated instance $i$ and element type $t\in\{\mathrm{cl},\mathrm{att}, \mathrm{rel}\}$, be the elements that $i$ uses at least once: a $\texttt{!new}$ operator call for instantiating a given class, an assignment operator call $\texttt{:=}$ for assigning a value to a given attribute, or an $\texttt{!insert}$ operator call for instantiating a link for a given relationship. 
The per-instance coverage, i.e., how much a single instance is covering the domain model, is given in the following formula.
\begin{equation}
  C_{i,t}=100\\\times\frac{|I_{i,t}|}{|M_t|}
  \label{formula:coverage}
\end{equation}

For answering RQ2.1, we report two aggregated values per domain and strategy.
First, \emph{within-instance coverage} is the mean and standard deviation of $C_{i,t}$ over the 30 instances for a given domain model.
Second, \emph{across-instance coverage} checks whether the 30 instances together, i.e., the union of covered elements, reach every element as expressed in the following formula.
\begin{equation}
  C_t^{\mathrm{across}}=100\times\frac{\bigl|\bigcup_{i=1}^{30} I_{i,t}\bigr|}{|M_t|}
  \label{formula:coverage-across}
\end{equation}

For \textbf{value diversity} (RQ2.2), different attribute types call for different treatment. String, Integer, and Real attributes have large value spaces; thus, we measure pairwise distance between values. The custom data types in our domain models, (e.g. Date), are built in SOIL from a string through a constructor call, e.g., \texttt{Date('2026-02-10')}, so we extract that string and treat it as a String value. In contrast, enumeration and Boolean attributes take their values from a small, fixed set of literals, in which pairwise distance only indicates whether two values differ. For these, we use the Shannon entropy of the value distribution instead, considering every enumeration and Boolean attribute of the domain models.

Let $V=\{v_1,\ldots,v_n\}$ be a bag of same-type attribute values (numeric or string) that we compare pairwise. For within-instance diversity, $V$ collects the values within a single instance; for across-instance diversity, it collects the values across all 30 instances of a domain and strategy. 
We define diversity as the average distance over all unordered pairs:
\begin{equation}
  D_T(V)=
  \frac{1}{\binom{n}{2}}
  \sum_{1 \leq i < j \leq n} d_T(v_i,v_j),
  \qquad n \geq 2
  \label{formula:diversity}
\end{equation}
For numeric values and exact string comparison,
$d_T(v_i,v_j)=0$ if $v_i=v_j$, and $1$ otherwise.
For strings, we additionally use the normalized Levenshtein distance:
\[
d_{\mathrm{lev}}(v_i,v_j)=
\frac{\mathrm{lev}(v_i,v_j)}
     {\max(|v_i|,|v_j|)}
\]
Consequently, 0 denotes identical values, and values closer to 1 indicate greater diversity. For an exact comparison, a score of 0.95 means that, on average, 95\% of all value pairs are different. For the Levenshtein distance, a score of 0.85 means that on average 85\% of the characters differ, i.e., the values are different words rather than variants of the same word.

We further assess the application distribution of enumeration and Boolean literals using normalized Shannon evenness~\cite{pielou1966}, which divides the Shannon entropy~\cite{shannon1948}
by its maximum possible value.
Let $\mathcal{L}=\{\ell_1,\ldots,\ell_m\}$ be the $m$
literals defined by an enumeration, or $\mathcal{L}=\{\mathit{true},\mathit{false}\}$ for a Boolean. For a given enumeration- or Boolean-valued attribute in
a given generated instance, let $c_j$ be the number of occurrences of literal
$\ell_j$, let $N=\sum_{j=1}^{m}c_j$, and let
$p_j=c_j/N$ when $N>0$. The set of active literals is
$\mathcal{A}=\{\ell_j\in\mathcal{L}\mid c_j>0\}$, with
$k=|\mathcal{A}|$. Its Shannon entropy is
\begin{equation}
  H=-\sum_{\ell_j\in\mathcal{A}}p_j\log_2 p_j
  \label{formula:shannon-entropy}
\end{equation}
If $N=0$, we define the entropy and both following evenness scores as zero.
Entropy is zero when only one literal occurs and reaches its maximum when the occurrences are uniformly distributed. We normalize the entropy using two different maxima:
\begin{equation}
  \begin{aligned}
    E_{\mathrm{active}} &=
    \begin{cases}
      H/\log_2 k, & k\geq 2\\
      0,          & k<2
    \end{cases}\\
    E_{\mathrm{all}} &=
    \begin{cases}
      H/\log_2 m, & m\geq 2\\
      0,          & m<2
    \end{cases}
  \end{aligned}
  \label{formula:shannon-evenness}
\end{equation}

Both scores are in $[0,1]$. The active-literal score measures how evenly the instance distributes occurrences among the literals it uses, whereas the all-literal score additionally penalizes literals that it does not use. 
For example, if an enumeration defines four literals but an instance uses two of them equally often, then $H=1$, $E_{\mathrm{active}}=1$, and $E_{\mathrm{all}}=1/2$. Thus, the selected literals are perfectly balanced, but the instance covers only part of the enumeration. 
Moreover, an instance with $N$ values can use at most $\min(N,m)$ literals, so $E_{\mathrm{all}}$ cannot exceed $E_{\mathrm{all}}^{\max}=\log_2\min(N,m)/\log_2 m$, which is zero if $\min(N,m)<2$. Likewise, $E_{\mathrm{active}}$ cannot exceed $E_{\mathrm{active}}^{\max}$, which is one if $\min(N,m)\geq 2$ and zero otherwise. 
We report an upper bound to distinguish low scores due to small instances from an actual concentration on few literals.
A score of 1 means that all literals occur equally often and 0 that only one literal occurs. For instance, 0.5 for an enumeration with four literals corresponds to two literals used equally often.

We calculate both evenness scores and their upper bounds for every enumeration and Boolean attribute in each of the 30 generated instances per strategy. Unlike for coverage, values of inherited attributes are pooled under the class that declares the attribute (e.g., \texttt{vehicleTypeCode} combines the values of both \texttt{Vehicle} and \texttt{Truck} objects), since evenness concerns the literals of the attribute and not the class that uses it. When a domain has several such attributes, we first average their scores per instance, separately for enumerations and Booleans. Tables~\ref{tab:enum-distribution} and~\ref{tab:bool-distribution} report the mean and standard deviation of both scores with their upper bounds for enumerations and Booleans, respectively. For a Boolean, both scores always coincide, since an attribute with two literals uses either both or only one of them.

For \textbf{structural diversity} (RQ2.3), we represent every instance as an undirected labeled multigraph: objects are nodes labeled by their class, and links are edges labeled by their association\footnote{In USE the links can only be instantiated once between the same two objects}. We use the unit-cost Graph Edit Distance (GED)~\cite{SanfeliuFu83,RiesenBunke09}: each node or edge insertion, deletion, or label substitution costs~$1$. 
$\mathrm{GED}(G_i,G_j)$ is the minimum total cost of transforming $G_i$ into $G_j$ under these operations, and $\mathrm{GED}(G,\emptyset)=|V_G|+|E_G|$.
We normalize the raw distance as 
\begin{equation}
  D_{\mathrm{GED}}(G_i,G_j)=
  \frac{\mathrm{GED}(G_i,G_j)}
  {\mathrm{GED}(G_i,\emptyset)+\mathrm{GED}(G_j,\emptyset)}
  \label{formula:ged-diversity}
\end{equation}
and set $D_{\mathrm{GED}}(\emptyset,\emptyset)=0$.
The score lies in $[0,1]$: $0$ denotes identical labeled graphs, and larger values denote greater structural diversity.
The denominator is the cost of turning one instance into the other from scratch, i.e., deleting all its objects and links and inserting all the objects and links of the other. A score of 0.25, therefore, means that the cheapest transformation requires a quarter of that effort. Attribute values are not considered, as they are covered by value diversity.

Since computing exact GED is NP-hard~\cite{BlumenthalBGBB20}, we compute for every pair a certified interval $[L,U]$ with $L\le\mathrm{GED}\le U$ using established heuristics (cf.\ Appendix~\ref{app:ged}).
For each domain and strategy we summarize $D_{\mathrm{GED}}$ by the mean~$\pm$~standard deviation over the $\binom{30}{2}=435$ unordered pairs among the 30 instances, substituting $U$ for $\mathrm{GED}$ in \eqref{formula:ged-diversity}.

\paragraph{RQ3 - Realism}
Since the realism of values and structures cannot be captured by analyzing pure model information alone, we propose dedicated checks that use additional sources, an LLM-as-a-judge, and an expert audit. Finally, we also study the potential of the feedback gained to improve the domain models in a subsequent model-refinement step.

For \textbf{automated evaluation} (RQ3.1), we use two complementary methods as introduced in the previous paragraph. First, we used domain-specific checks that validate selected values with regular expressions, libraries, external APIs, and cross-attribute comparisons. Examples include regular expressions for phone numbers, websites, email addresses, X (Twitter) usernames, and license plates; validators for IBANs and BICs\footnote{Bank-Account Validator, \url{https://gitlab.com/schegge-projects/bank-account-validator}}; API geocoding\footnote{Geoapify API, \url{https://www.geoapify.com/geocoding-api}} for addresses and coordinates, where an address counts as found if the match confidence is above 0.7; date comparisons; and the OMDb API\footnote{OMDb API, \url{https://www.omdbapi.com/}} for movies and series.


The domain-specific realism checks provide a general understanding of the realism of isolated values, but not of the realism of the instance as a whole, its structure, or how the values relate to each other. A complementary method should therefore assess each instance as a whole, run without human intervention, and scale to large sets of instances, thereby making it reusable in agentic pipelines. Hence, our second method submits each domain model, together with each generated instance, to Gemini 3.1 Pro as an LLM-as-a-judge (see the configuration in Table~\ref{tab:llm-config}). We chose a different vendor than for the generator to ensure that the judge does not share OpenAI's training and decoding stack. Furthermore, each instance is judged independently, without access to the generation conversation. The system prompt frames the LLM as a software modeler in review mode
performing \textit{``semantic reality checking''}: it must read the class diagram, map objects to classes, apply common-sense knowledge checks to values and relationships, and return one of \emph{realistic}, \emph{unrealistic}, or \emph{doubtful} labels together with a short justification that refers to the objects, values, or relationships that motivate the verdict. The prompt contains one worked example per label. The complete prompt and all judge responses are available in~\cite{repo} in \texttt{docs/ai-prompts}.

For \textbf{human-based analysis} (RQ3.2), we drew a stratified random sample of 100 judge outputs. The sample contained ten instances per domain, 50 per generation strategy, and 50 per judge verdict (realistic/unrealistic); within each verdict, 25 instances came from IL and 25 from CoT. CoT category scenarios were not an additional stratification factor, as we prioritized balance across domain, generation strategy, and judge verdict: the 50 CoT cases comprised 15 baseline, 10 complex, 14 boundary, and 11 edge instances. Over-constraint instances were excluded because they violate the model by construction and are usually classified as unrealistic by the judge.

The four authors reviewed the 100 judges' outputs (verdicts and justifications) together in a synchronous meeting. Each justification was split into individual statements, and every statement received a consensus decision. For the 50 unrealistic verdicts, each statement reports a problem, which we coded for two properties. The first is \emph{traceability}, i.e., whether every object, value, or relationship cited by the judge actually exists in the instance or was hallucinated by the judge. The second is \emph{confirmation}, i.e., whether the authors consider the reported problem a genuine realism issue of the instance, coded as \emph{yes} (the problem holds as stated), \emph{partial} (only part of the problem holds, e.g., it is overstated or applies to only some of the cited elements), or \emph{no} (the authors do not consider it a problem). For the 50 realistic verdicts, statements were classified as \emph{specific} or \emph{generic}. Specific statements name an object, value, or relationship of the instance and were checked for traceability. Generic statements assess the instance as a whole without naming any element, so they were checked only for validity, i.e., whether the authors agree with the assessment.

Finally, for \textbf{model refinement} (RQ3.3), we use the human-confirmed unrealistic instances from RQ3.2 to strengthen the domain models. The identified unrealistic instances are well-formed and conforming models; thus, the question arises whether missing knowledge can be added to the domain models, i.e., the class diagrams, to better guide the LLM-based generation process. We group recurring problems and encode those expressible in OCL as additional invariants for the domain models. Although the invariants should target the confirmed unrealistic cases, we run them on all 100 audited instances and compare the invariant verdict with the judge verdict and with the expert decision. This shows whether invariants can detect instances the experts consider unrealistic, flag instances the audit may have overlooked as unrealistic, and whether certain invariants are too strict.

\input{tables/feedback}

\andrei{run ocls againts all intances: should we add this?}
To see how the invariants behave beyond the sample they were derived from, we also run them on the remaining 440 instances that were not audited (over-constraint instances excluded, i.e., 600-100-60), compare their verdict with the judge verdict, and inspect the judge-realistic instances that fail an invariant.

%% file: tables/domains.tex
\begin{table*}[ht]
\caption{Characteristics of the domain models used for the evaluation}
\label{tab:characteristics}
\footnotesize
\setlength{\tabcolsep}{1.5pt}
\renewcommand{\arraystretch}{1.12}
\begin{tabularx}{\textwidth}{@{}>{\raggedright\arraybackslash}p{2.35cm}*{7}{P{0.72cm}}P{1.35cm}P{1.02cm}P{0.78cm}>{\raggedright\arraybackslash}X@{}}
\toprule
\multirow{2}{*}{Domain}
  & \multicolumn{7}{c}{Size (Number of elements)}
  & \multicolumn{4}{c}{Domain characteristics} \\
\cmidrule(lr){2-8}\cmidrule(lr){9-12}
& \multicolumn{1}{c}{Cl.} & \multicolumn{1}{c}{Abs.} & \multicolumn{1}{c}{Attr.}
& \multicolumn{1}{c}{Rel.} & \multicolumn{1}{c}{Inh.} & \multicolumn{1}{c}{Enum}
& \multicolumn{1}{c}{Dt}
& \begin{tabular}[c]{@{}c@{}}Abstr.\\ Level\end{tabular}
& \multicolumn{1}{c}{Constr.}
& \begin{tabular}[c]{@{}c@{}}Temp.\\ Reas.\end{tabular}
& \multicolumn{1}{Y}{\begin{tabular}[c]{@{}c@{}}Realism\\checks\textsuperscript{a}\end{tabular}} \\
\midrule
Address Book     & 7  & 1 & 14 & 5  & 2  & 2 & 1 & model     & 1 & No  & Phone, website, email, address \\
\midrule
Bank             & 3  & 0 & 8  & 3  & 0  & 0 & 0 & model     & 2 & No  & IBAN, BIC code, country \\
\midrule
Hotel Management & 7  & 0 & 22 & 6  & 0  & 0 & 0 & model     & 3 & Yes & Dates (checkin\textgreater{}=start, checkout\textless{}=end) \\
\midrule
Expenses         & 4  & 0 & 11 & 3  & 0  & 2 & 1 & model     & 2 & Yes & Dates (end\textgreater{}=start) \\
\midrule
Pickup Net       & 6  & 0 & 10 & 8  & 0  & 1 & 0 & model     & 4 & No  & Address, latitude, longitude, X username \\
\midrule
State Machine    & 5  & 1 & 2  & 6  & 3  & 0 & 0 & metamodel & 2 & Yes & - \\
\midrule
Vehicle Rental   & 7  & 1 & 27 & 5  & 3  & 2 & 1 & model     & 5 & Yes & Address, license plate, valid home phone \\
\midrule
Videoclub        & 6  & 0 & 8  & 3  & 2  & 1 & 1 & model     & 3 & Yes & Production title, genre, type (movie or series), actors, rental year \textgreater{}= release year \\
\midrule
Football         & 16 & 0 & 37 & 18 & 0  & 3 & 0 & model     & 4 & Yes & Dates (matches, trainings, notes, objectives) \\
\midrule
Restaurant       & 21 & 2 & 34 & 13 & 11 & 8 & 2 & model     & 4 & No  & Dates (reservation, birth), phone number \\
\bottomrule
\end{tabularx}

\medskip
\footnotesize
(a) Only attributes for which dedicated checks are defined are listed.

Cl. Classes, Abs. Abstract Classes, Attr. Attributes, Rel. Relationships, Inh. Inheritance, Enum. Enumerations, Dt. Datatypes, Abstr. Level. Abstraction Level, Constr. OCL Constraints, Temp. Reas. Temporal Reasoning.
\vspace{-.2cm}
\end{table*}

%% file: tables/llm-config.tex
\begin{table*}[ht]
\caption{LLM configurations}
\label{tab:llm-config}
\scriptsize
\setlength{\tabcolsep}{5pt}
\begin{tabularx}{\textwidth}{@{}>{\raggedright\arraybackslash}p{2.6cm} *{2}{>{\raggedright\arraybackslash}X}@{}}
\toprule
 & Generator (IL, CoT) & LLM as a Judge \\
\midrule
Model
  & GPT 5.2
  & Gemini 3.1 Pro \\
Identifier
  & \texttt{openai/gpt-5.2}
  & \texttt{google/gemini-3.1-pro-preview} \\
Snapshot
  & \texttt{gpt-5.2-2025-12-11}
  & \texttt{preview} \\
Access
  & OpenRouter
  & OpenRouter \\
Temperature\textsuperscript{\dag}
  & Not applicable with reasoning
  & Google default (\texttt{1.0}) \\
TopP\textsuperscript{\dag}
  & Not applicable with reasoning
  & Google default (\texttt{0.95}) \\
Reasoning\textsuperscript{\dag}
  & OpenRouter default (\texttt{medium})
  & OpenRouter default (\texttt{medium}) \\
Dates
  & 16--23 Feb 2026
  & 10--17 Mar 2026 \\
Memory
  & Sliding window of 128k tokens\par
    {\leftskip=1em IL: one session per domain\par
    CoT: one session for the analysis, one per category, and one shared by all instantiations\par}
  & No chat memory\par
    {\leftskip=1em one session per request\par} \\
\bottomrule
\end{tabularx}

\medskip
\footnotesize
\textsuperscript{\dag} We did not set these sampling hyperparameters in the API requests; thus, the listed values are the provider defaults that we used for the experiments.

\end{table*}

%% file: tables/instantiation.tex
\begin{table*}[t]
\caption{Number of generated model elements for Instruction Learning (IL) and Chain-of-Thought (CoT)}
\label{tab:sizeGenInstances}
\centering
\scriptsize
\setlength{\tabcolsep}{2pt}
\renewcommand{\arraystretch}{1.12}
\begin{tabular*}{\textwidth}{@{\extracolsep{\fill}}l*{6}{N{0000}}@{}}
\toprule
Domain
  & \multicolumn{3}{c}{Instruction Learning (IL)}
  & \multicolumn{3}{c}{Chain-of-Thought (CoT)} \\
\cmidrule(lr){2-4}\cmidrule(l){5-7}
& \thd{\#objects} & \thd{\#attributes} & \thd{\#associations}
& \thd{\#objects} & \thd{\#attributes} & \thd{\#associations} \\
\midrule
Address Book     & 792  & 2275 & 1029 & 962  & 2619 & 1292 \\
Bank             & 394  & 1041 & 613  & 434  & 1159 & 643  \\
Hotel Management & 315  & 1014 & 283  & 487  & 1522 & 455  \\
Expenses         & 341  & 1247 & 425  & 449  & 1408 & 506  \\
Pickup Net       & 678  & 1138 & 867  & 749  & 1287 & 1028 \\
State Machine    & 1450 & 1450 & 2444 & 733  & 733  & 1227 \\
Vehicle Rental   & 449  & 2566 & 703  & 415  & 2439 & 604  \\
Videoclub        & 508  & 895  & 556  & 573  & 999  & 723  \\
Football         & 2587 & 6352 & 2880 & 3096 & 7486 & 3522 \\
Restaurant       & 1362 & 4094 & 1488 & 1535 & 4497 & 1700 \\
\midrule
Total            & 8876 & 22072 & 11288 & 9433 & 24149 & 11700 \\
\bottomrule
\end{tabular*}
\vspace{-.2cm}
\end{table*}

%% file: tables/feedback.tex
\begin{table*}[ht]
\caption{USE Feedback: correction calls and instances accepted at each attempt}
\label{tab:llm-corrections}
\centering
\scriptsize
\setlength{\tabcolsep}{2pt}
\renewcommand{\arraystretch}{1.12}
\begin{tabular*}{\textwidth}{@{\extracolsep{\fill}}l*{2}{N{0}N{0}N{0}N{00}N{0}N{0}}@{}}
\toprule
Domain &
\multicolumn{6}{c}{Instruction Learning (IL)} &
\multicolumn{6}{c}{Chain of Thought (CoT)} \\
\cmidrule(lr){2-7}\cmidrule(l){8-13}
& \multicolumn{3}{c}{Correction calls} & \multicolumn{3}{c}{Accepted att.} &
  \multicolumn{3}{c}{Correction calls} & \multicolumn{3}{c}{Accepted att.} \\
\cmidrule(lr){2-4}\cmidrule(lr){5-7}\cmidrule(lr){8-10}\cmidrule(l){11-13}
& \thd{Wf.} & \thd{Conf.} & \thd{Tot.} & \thd{1} & \thd{2} & \thd{3}
& \thd{Wf.} & \thd{Conf.} & \thd{Tot.} & \thd{1} & \thd{2} & \thd{3} \\
\midrule
Address Book     & 2 & 0 & 2 & 29 & 0 & 1 & 4 & 0 & 4 & 26 & 4 & 0 \\
Bank             & 0 & 0 & 0 & 30 & 0 & 0 & 1 & 0 & 1 & 29 & 1 & 0 \\
Hotel Management & 0 & 4 & 4 & 26 & 4 & 0 & 2 & 0 & 2 & 29 & 0 & 1 \\
Expenses         & 1 & 0 & 1 & 29 & 1 & 0 & 5 & 0 & 5 & 25 & 5 & 0 \\
Pickup Net       & 1 & 0 & 1 & 29 & 1 & 0 & 6 & 0 & 6 & 25 & 4 & 1 \\
State Machine    & 2 & 3 & 5 & 25 & 5 & 0 & 2 & 0 & 2 & 28 & 2 & 0 \\
Vehicle Rental   & 0 & 0 & 0 & 30 & 0 & 0 & 2 & 0 & 2 & 28 & 2 & 0 \\
Videoclub        & 0 & 0 & 0 & 30 & 0 & 0 & 3 & 0 & 3 & 27 & 3 & 0 \\
Football         & 1 & 3 & 4 & 26 & 4 & 0 & 2 & 0 & 2 & 28 & 2 & 0 \\
Restaurant       & 2 & 2 & 4 & 26 & 4 & 0 & 2 & 3 & 5 & 26 & 3 & 1 \\
\midrule
Total   & 9 & 12 & 21 & 280 & 19 & 1
                 & 29 & 3 & 32 & 271 & 26 & 3 \\
\bottomrule
\end{tabular*}

\medskip
\begin{minipage}{\textwidth}
\raggedright
\footnotesize
Wf. Well-formedness, Conf. Conformance, Tot. Total, Accepted att. Accepted at attempt (1 means no USE feedback needed).
\end{minipage}

\end{table*}

%% file: sections/Results.tex
\subsection{Results}
\label{sec:evaluation-results}

In this subsection, we present the results from executing the outlined experimental setup to address the aforementioned RQs. 

\subsubsection{RQ1 - Validity}

\paragraph{RQ1.1 - Well-formedness}

\input{tables/taxonomy-wf}

Both prompting strategies produce well-formed object diagrams for every given domain model once the proposed repair loop has run (cf. Table~\ref{tab:llm-corrections}). At command level, USE processed all 42,236 IL SOIL commands and all 45,282 CoT SOIL commands of the final instances, and the generation processes required only a small number of repair requests to achieve these results.

Table~\ref{tab:llm-corrections} quantifies the repair effort in more detail. Across all domain models, USE well-formedness diagnostics triggered 9 correction calls for IL and 29 for CoT. Table~\ref{tab:wf-taxonomy} classifies the issues behind these calls into four categories. 

\emph{Concrete syntax} is the main category (5 IL, 12 CoT), and it is dominated by string quoting: realistic values such as \texttt{O'Connell} or \texttt{d'Italie} contain apostrophes, which the LLM escapes as \texttt{''}, a form that SOIL does not foresee. The few remaining syntax failures are truncated commands, an illegal identifier, and an assignment without \texttt{:=}. \emph{Type mismatch} failures (2 IL, 7 CoT) mostly reference objects that were never created, and in two cases a rewrite during the repair loop dropped the corresponding \texttt{!new} commands. \emph{Rejected link insertions} (2 IL, 8 CoT) are either duplicate links or a composition part inserted into two wholes.\dominik{what do you mean? the composite/parent?} Finally, two CoT instances set an integer attribute to $-2^{31}$, which makes USE abort with an internal error, as this value exceeds the limits of the integer data type.

Over-constraint instances are also fully well-formed after running the repair process. However, deliberately violating the domain model also seems to challenge basic syntactic concerns in the generation process, as 45\% (13 of 29) of the CoT well-formedness repairs come from this category.

\paragraph{RQ1.2 - Conformance}

\input{tables/taxonomy-cf}

The instance generation approach using the two prompting strategies, including the repair loop, was also effective in achieving conformance. The 300 IL instances produced no errors in 11,288 multiplicity checks and 900 invariant validations using only a small number of repairs. After excluding the 60 over-constraint instances, the remaining 240 CoT instances produced no errors in 10,370 multiplicity checks and 720 invariant validations, also using only a small number of repairs.

Taking a closer look at the taxonomy of the conformance errors, we see that all 15 errors (12 IL + 3 CoT) are multiplicity failures (cf.\ Table~\ref{tab:conf-taxonomy}), which can be classified into two main groups: $(i)$ missing required link, i.e., the object is not connected to another one where the lower bound is at least 1, and $(ii)$ extra link, i.e., the object is connected to at least two other objects where the upper bound is 1.

Analyzing the 60 over-constraint instances that are intended to fail, 53 violate at least one multiplicity or invariant. In total, 61 multiplicity violations and 26 invariant violations were recognized. The other 7 describe scenarios the LLM treated as over-constrained, although the model in fact permits them. We further analyze some over-constrained instances in RQ3.1.

\manuel{todo: manuel : review this paragraph}
Finally, comparing the correction calls of both strategies (Tables~\ref{tab:wf-taxonomy} and~\ref{tab:conf-taxonomy}), IL needs fewer well-formedness repairs (9 vs.\ 29) and CoT fewer conformance repairs (3 vs.\ 12). In IL, a single prompt asks the LLM both to decide the content of an instance and to write it in SOIL. On the one hand, this lets the LLM stay with values and structures it knows how to express in SOIL, which explains the few well-formedness repairs. On the other hand, the LLM has to respect multiplicities and invariants while writing the instance, without having reasoned about them beforehand, and this is where its 12 conformance repairs occur. CoT splits these two tasks. First, the LLM analyzes the domain model in NL (Listing~\ref{lst:prompt-analysis}), describing for each association how many objects can be linked at both ends and which invariants must hold. Then, based on this analysis, it writes the scenarios in NL, and only afterward translates them into SOIL. Since the constraints were made explicit before any scenario was written, most scenarios already respect them, and only 3 conformance repairs are needed. However, the scenarios are written without any knowledge of SOIL and contain richer values, such as names with apostrophes, which are a likely source of the quoting errors discussed above.

\begin{mdframed}[style=rqanswer,nobreak=true]
{\textbf{Answering RQ1:}} The presented approach is effective at producing well-formed and conforming instances. After the repair loop, all instances are well-formed and conform to their domain models, and more than 90\% already did so on the first attempt. In the over-constraint category, 88\% of the 60 instances contain the intended conformance violations.
\end{mdframed}

\subsubsection{RQ2 - Coverage and diversity}

\paragraph{RQ2.1 - Coverage}

\input{tables/coverage}

Table~\ref{tab:coverage} shows consistently high coverage, but no strategy dominates in every domain. IL prompting shows class, attribute, and association coverage ranges from 87.1--100\%, 88.3--100\%, and 83.3--100\%, respectively. The corresponding CoT prompting coverage results ranges are 81.7--100\%, 80.6--100\%, and 85.3--100\% for classes, attributes, and associations, respectively. Both strategies cover every modeled class, attribute, and association in the Bank domain. IL also achieves full coverage across all three element types for the State Machine and Football domain models, while CoT is particularly strong in the Hotel Management domain. The lowest class and attribute coverage occurs for CoT on the Videoclub model; the lowest association coverage occurs for IL on the Hotel Management model and for CoT on the Address Book model. 
Across instances, the union of the 30 instances per strategy reaches every element in nine domains. The remaining gaps can help us understand and improve the domain models. In the Restaurant model, the concrete superclass \texttt{Person} and its attribute \texttt{name} are never instantiated directly in any of the 600 instances, because the LLM always instantiates \texttt{Cook}, \texttt{Waiter}, and \texttt{Manager} amongst the other subclasses. A human would probably do the same unless it is not clear which role the person is playing. Thus, \texttt{Person} is a candidate to be declared abstract rather than left concrete. A similar case is the superclass \texttt{Cassette} in the Videoclub model. It is concrete, yet missing from 16 of the 30 IL instances and from 29 of the 30 CoT instances, which instead instantiate \texttt{Movie} and \texttt{Series}. Declaring it abstract would match how the domain is actually used.\dominik{the last sentence is weird} The other per-instance gaps are optional elements where their usage depends on the scenario. \texttt{RoomExtra} is absent from 12 of the 30 IL Hotel Management instances, and \texttt{BusDriver} from 8 of the 30 IL Restaurant instances. Both appear in the union of those 30 instances; thus, they are instantiable by the LLM, but they are not classes that every scenario has to include, as the domain model also states.

\paragraph{RQ2.2 - Value diversity.}


\input{tables/diversity}

Generated values rarely repeat, neither within an instance nor across instances, as can be seen in Table~\ref{tab:semantic-similarity}. IL and CoT differ by at most 0.07 for numeric diversity and normalized Levenshtein distance. Therefore, we report their combined values directly. Exact string diversity is at least 0.95 within instances and rounds to 1.00 across instances in every domain, i.e., at most 5\% of the string pairs within an instance and fewer than 0.2\% of the pairs across the 30 instances of a domain model are identical. Normalized Levenshtein distance ranges from 0.78 to 0.91 within and from 0.80 to 0.92 across instances, so even non-identical strings differ in most of their characters and are not small variations of a common template such as \texttt{Name1} and \texttt{Name2}, whose distance is only 0.2. Values with a fixed format, such as dates, are the exception, since two different dates differ in only a few characters, which lowers the scores of the domains where they are frequent. Numeric values repeat slightly more often, with within-instance scores between 0.91 and 0.99 (Address Book has no numeric attributes), although part of this repetition might also be prescribed by the models. For example, the Expenses model, the domain with the lowest score, has an invariant that requires the amount of an expense to equal the sum of its bills. Thus, an expense with a single bill necessarily has to repeat the same amount.

We also want to emphasize that realistic data naturally contains some repetition, such as small quantities in some domains, frequent first names, or values that the model forces to be equal. Therefore, we consider high diversity scores, but not maximal diversity scores, as the required outcome.

\begin{figure*}[ht]
  \centering
  \includegraphics[width=0.8\linewidth]{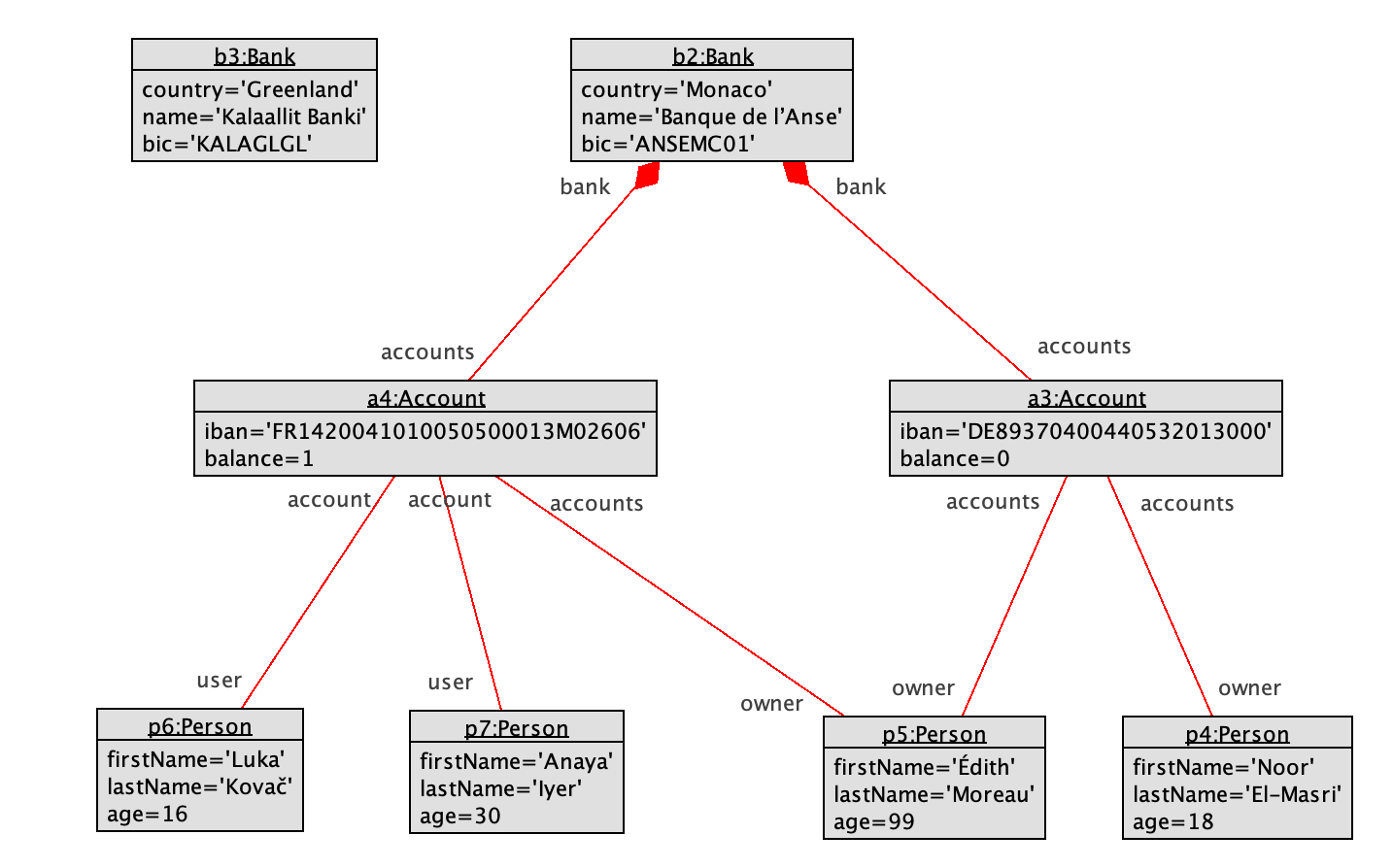}
  \caption{Edge instance \cotinst{bank}{1}{edge}{\texttt{Bank edge1}} generated with CoT. Its values are valid but unusual. Both IBANs pass every automated check, yet a Monegasque bank holding a German and a French IBAN is incoherent.}
  \label{fig:edge-bank}
\end{figure*}

For example, Figure~\ref{fig:edge-bank} shows an edge instance generated with CoT for Bank that illustrates value diversity. It combines unusual but plausible values, namely an owner aged exactly 18, an owner aged 99, balances of 0 and 1, and a 16-year-old account user, which the model allows because only owners must be adults. 

\input{tables/shannon}
\input{tables/shannon-bool}

Enumeration literals are used almost uniformly across instances, whereas within single instances evenness varies strongly between domains (see Table~\ref{tab:enum-distribution}). Across instances, all-literal evenness ranges from 0.91 to 1.00 for both prompting strategies. Within instances, it ranges from 0.42 (IL for the Restaurant model and CoT for the Vehicle Rental model) to 0.90 (IL for the Address Book and Football models). Low within-instance values do not necessarily indicate repetition, since an instance with few values of an enumeration-typed attribute cannot use many literals. Taking this limit into account, the within-instance score stays within 0.12 of its upper bound $E_{\mathrm{all}}^{\max}$ in every domain. In Vehicle Rental, for example, the enumeration attribute \texttt{Individual.driverLicenseState} has eight literals, but an instance contains only 1.6 individual customers on average, and they almost always hold licenses from different states. With two customers, $E_{\mathrm{all}}$ cannot exceed $\log_2 2/\log_2 8=1/3$, and with a single customer it is zero, so the low scores of this attribute, 0.20 (IL) and 0.17 (CoT), are close to their upper bounds of 0.21 and 0.18 and reflect the small number of customers rather than a concentration on particular states. The 30 instances together use all eight states and cover the full set of literals almost evenly. Neither prompting strategy consistently results in more even literal usage. IL is higher in five of the seven domains with enumerations and CoT in the Videoclub and Restaurant domains, with the largest differences in the Address Book (0.90 vs.\ 0.70 in favor of IL) and Videoclub (0.71 vs.\ 0.56 in favor of CoT) models. CoT's explicit steering therefore does not imply a more even distribution of literals within individual instances.

Boolean values are distributed less evenly than enumeration literals (see Table~\ref{tab:bool-distribution}). Across instances, their evenness is 0.76 (IL) and 0.82 (CoT), compared to 0.94 and 0.95 for enumerations. Within instances, it is 0.52 and 0.48, with upper bounds of 0.73 and 0.72. Most of this gap comes from a few attributes for which the LLM almost always chooses the same value. In the IL instances of the Restaurant model, \texttt{FoodItem.purchaseFlag} is \texttt{true} for 206 of 209 food items, and in those of the Hotel Management model, \texttt{Room.occupied} is \texttt{false} for 58 of 64 rooms, which results in within-instance scores of only 0.03 and 0.02 for these two domains. This may reflect realistic defaults (e.g., most bookings are not canceled), but it also shows that the LLM rarely chooses the less common value unless it is explicitly asked to. CoT is considerably more even in exactly these two domains (0.28 and 0.25 within, 0.77 and 0.81 across), probably because its edge, boundary, and invalid scenarios ask for less typical situations, whereas IL is more even in the Vehicle Rental, Videoclub, and Football domains, where CoT instances are smaller and scenarios that focus on other aspects, such as most invalid instances, keep the Booleans at a single value (e.g., no suspended client in four of the six invalid Videoclub instances).



\paragraph{RQ2.3 - Structural diversity}

\input{tables/ged}

Unlike coverage or value diversity, structural diversity differs between the prompting strategies; CoT produces structurally more diverse instances than IL (see Table~\ref{tab:structural-diversity}). Averaged over the ten domains, $D_{\mathrm{GED}}$ is 0.26 for IL and 0.41 for CoT, an increase of approximately 60\%. CoT is more diverse in nine of the ten domains, with the largest differences for Address Book (0.21 vs.\ 0.56), Vehicle Rental (0.26 vs.\ 0.47), and State Machine (0.24 vs.\ 0.45). The only exception is the Hotel Management domain, where IL is slightly more diverse (0.35 vs.\ 0.30). Transforming one IL instance into another thus costs about 26\% as much as rebuilding it from scratch, and for CoT, about 41\%. Both strategies request diversity: the IL follow-up prompt asks for an instance that is structurally and semantically different from the previous ones, while CoT additionally prescribes a scenario category. The difference, therefore, measures the effect of explicit category steering (which seems more effective) on top of a generic diversity request. In particular, boundary, edge, and complex scenarios encourage changes in model size and topology that a generic diversity request does not specify.

\manuel{todo: manuel : review this paragraph}
Figure~\ref{fig:edge-bank} shows an edge instance generated with CoT for the Bank domain that illustrates this effect. At the structural level, which \(D_{\mathrm{GED}}\) compares, it is small (8 objects and 7 links, against an average of 13 objects and 20 links for the IL Bank instances) and includes an unusual case, a bank without any accounts. As a result, its mean \(D_{\mathrm{GED}}\) to the 30 IL Bank instances is 0.40, compared to 0.25 among the IL Bank instances themselves. Structurally, it thus differs more from the IL instances than these differ from each other.


\begin{mdframed}[style=rqanswer,nobreak=true]
{\textbf{Answering RQ2:}} \manuel{todo: manuel : review this paragraph} Our approach is effective at producing instances that cover the domain model and are diverse in values and structure. Combined, the instances of each domain reach every model element in nine of the ten domains, and the remaining gaps point to classes that could be declared abstract. Values rarely repeat, except for some Boolean attributes, whose values may reflect realistic defaults, since real data also contains repetition. Structurally, transforming one instance into another costs, on average, 26\% (IL) and 41\% (CoT) as much as rebuilding it from scratch, so CoT's scenario categories lead to clearly more varied structures.
\end{mdframed}

\subsubsection{RQ3 - Realism} 
\label{sec:rq3-realism}

\paragraph{RQ3.1 - Automated evaluation}

\input{tables/errors-checks}
\input{tables/errors-judge}

The number of individual values that pass a dedicated deterministic external realism check is shown in Table~\ref{tab:realism-checks}. The LLM-as-a-judge realism rates for a whole instance are reported in Table~\ref{tab:realism-judge}. 

\manuel{todo: manuel : review this paragraph : added incomplete APIs to the threats section}
Format checks for the individual values usually pass. IL produces almost no invalidly formatted phone numbers, emails, IBANs, BICs, countries, and X usernames. However, stricter checks that require sophisticated knowledge of the domain show that some generated values are invalid. The IBAN(API) row reports the results of checking the country prefix, the length prescribed for that country, the check digits, and the country-specific account structure, as defined by the IBAN standard ISO 13616. Only 35 of 141 IL and 37 of 143 CoT IBANs pass the full check. The issues are incorrect digit checks on otherwise well-formed IBANs, incorrect country lengths, and IBANs invented for countries that do not use them, such as the US or Japan.

The external APIs for addresses and productions show a similar gap (although their results are approximations because the underlying data sources may be incomplete; see \secref{sec:threats}). The Address (API) row indicates whether the geocoding service finds the address. For IL, most addresses resolve in the Address Book model (133 of 166), the Pickup Net model (149 of 172), and the Vehicle Rental model (157 of 187). For CoT, fewer addresses resolve (123 of 197 in the Address Book model, 66 of 199 in the Pickup Net model, and 126 of 174 in the Vehicle Rental model), with the lowest rates in edge scenarios. For the Videoclub model, IL finds 23 of the 124 titles and 1 of the 39 actors in OMDb, and CoT finds 6 of the 168 titles and none of the 10 actors.

\begin{figure*}[ht!]
    \centering
    \includegraphics[width=0.8\linewidth]{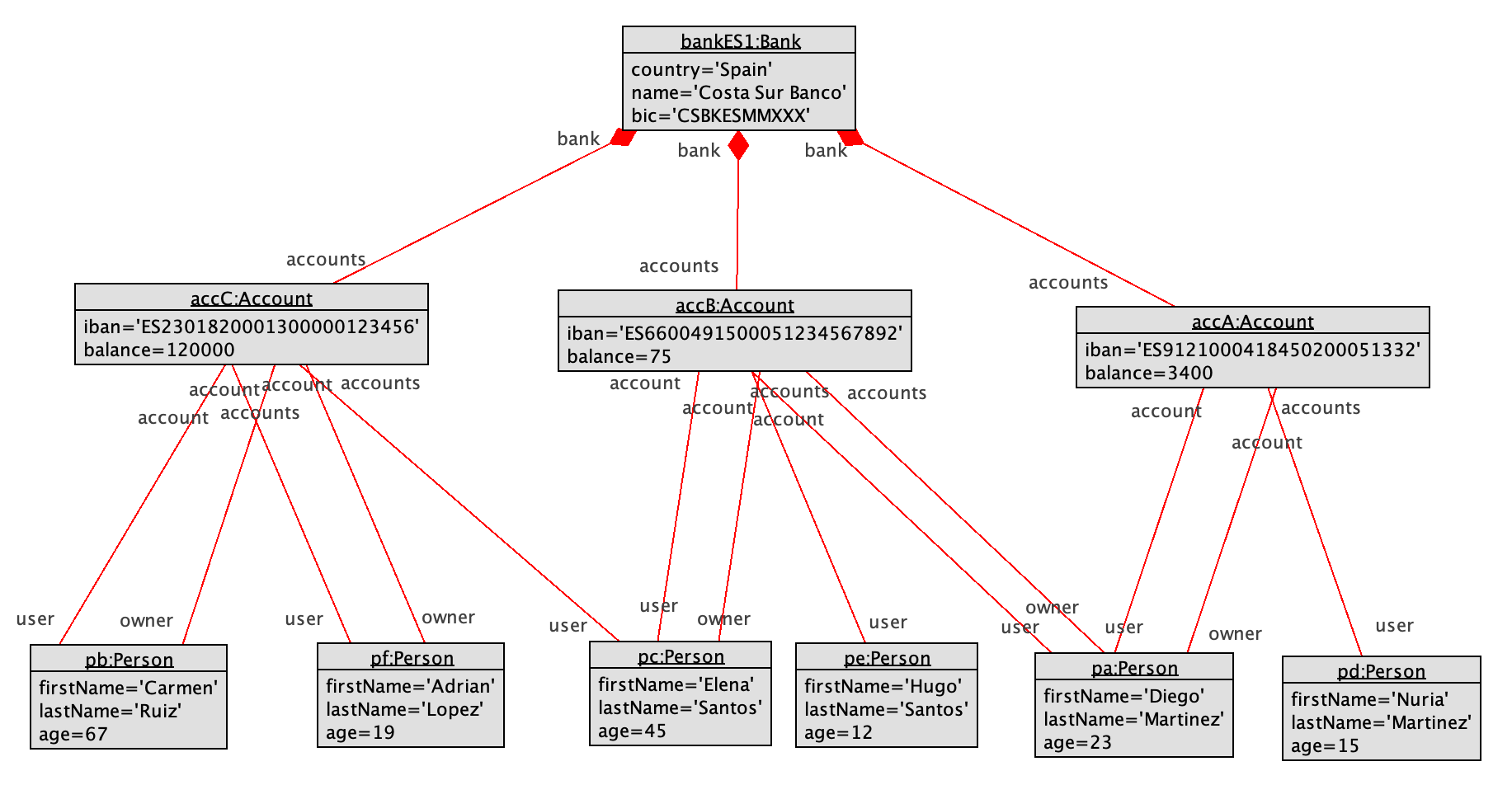}
    \caption{Well-formed instance \ilinst{bank}{2}{\texttt{Bank gen2}} generated with IL. All three accounts of the Spanish bank \texttt{Costa Sur Banco} have IBANs with the \texttt{ES} prefix and the correct length of 24, and the BIC \texttt{CSBKESMMXXX} is well formed. The bank, IBANs and its BIC are fictitious although they conform our checks.}
    \label{fig:wf-iban-not-real}
\end{figure*}

We want to emphasize that these outcomes do not necessarily make a value implausible. Fictitious but well-formed addresses, accounts\manuel{general and domain-specific rules should be considered for accounts I would say - it is a bit different to an address which is not existing but could be a valid address value}\andrei{should we remove 'accounts'?}, and productions are often preferable for privacy-preserving synthetic data, and remain useful for teaching, testing, and training (see Figure~\ref{fig:wf-iban-not-real}). Even instances that are not fully realistic can be useful. The edge instance in Figure~\ref{fig:edge-bank} combines an incoherent bank and IBAN country with unusual but valid values, which makes it a good test case precisely because it lies outside typical scenarios and might signal the need for constraining the domain model.


Comparisons between values within the same instance also pass frequently. All date comparisons in Table~\ref{tab:realism-checks} succeed for the IL instances. For CoT, the failures are confined to edge scenarios. In the Hotel Management model, 7 of 14 edge reservations fail the check-in, and 10 of 14 fail the check-out comparison because their dates are empty, are not dates at all (e.g., \texttt{`N/A'} or \texttt{`next Friday after Maghrib'}), or fall outside the booking period. Three (out of 13) edge expenses also end before they start. The coordinates of the Pickup Net addresses that the geocoding service finds also match, except for one IL address out of 149. The Football and Restaurant date checks only test whether a value can be parsed as a valid date, and all pass.

Analyzing the instance realism as a whole, the LLM-as-a-judge labels 215 out of 300 IL instances as realistic (71.7\%) and 131 out of 300 CoT instances (43.7\%). It never uses the \textit{``doubtful''} label. Comparing by category, baseline and complex CoT perform similarly to IL, scoring 46/60 (76.7\%) and 45/60 (75.0\%), respectively. However, the boundary, edge, and over-constraint categories are lower, at 23/60 (38.3\%), 4/60 (6.7\%), and 13/60 (21.7\%), respectively. These three categories are designed to be atypical, and the judge tends to score atypical instances as unrealistic. Thirteen over-constraint instances were labeled realistic; of these, 8 actually violated constraints, while the remaining 5 actually conformed to the domain model. 

\begin{figure}[ht]
    \centering
    \includegraphics[width=\linewidth]{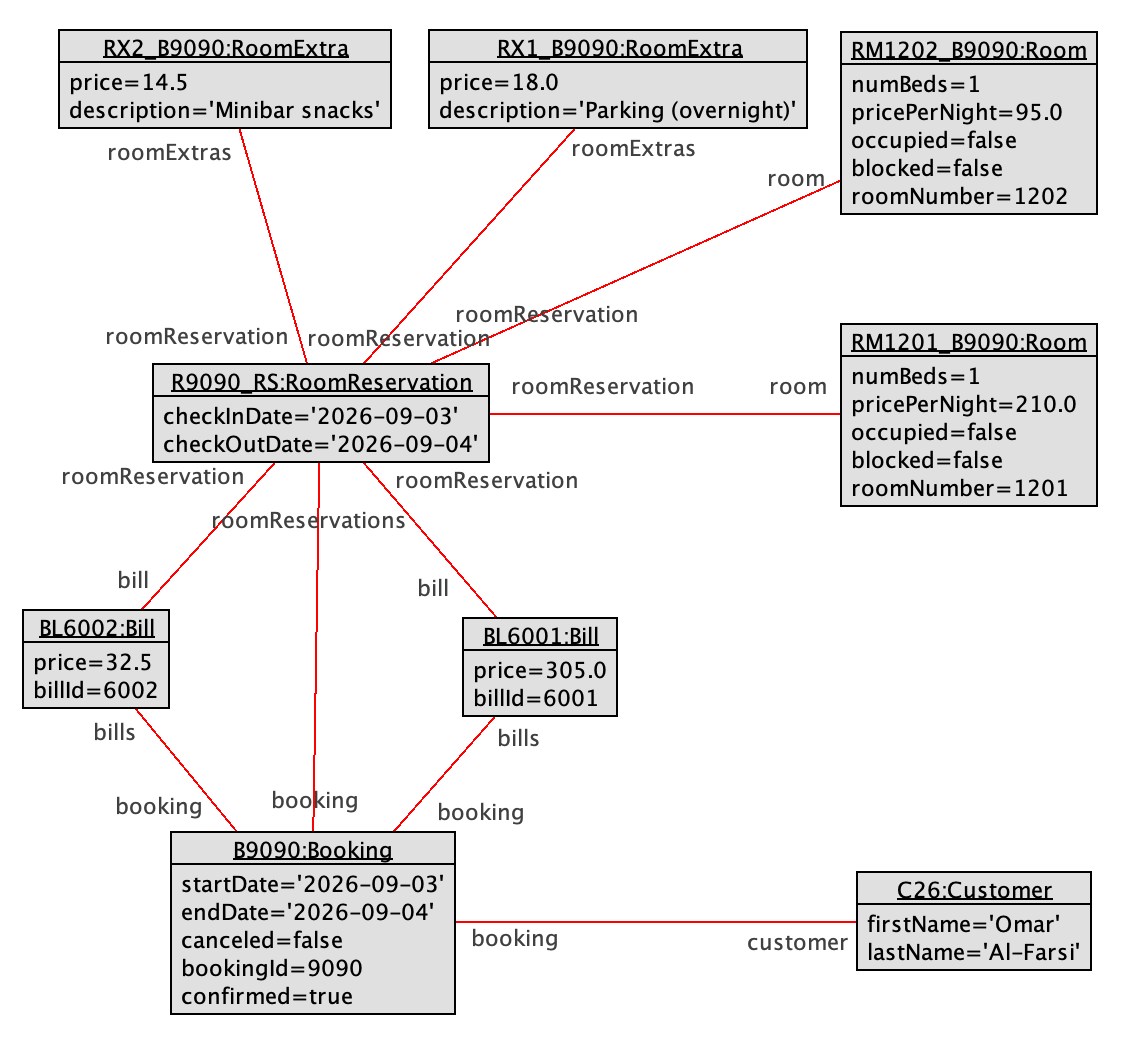}
    \caption{Over-constraint instance \cotinst{hotelmanagement}{5}{invalid}{\texttt{Hotel Management over-constraint5}}, labeled realistic by the judge, with two bills for one room reservation.}
    \label{fig:ovc-example-hotel}
\end{figure}

When analyzing these eight instances manually, we can take \cotinst{hotelmanagement}{5}{invalid}{\texttt{Hotel Management over-constraint5}} as a typical case (see Figure~\ref{fig:ovc-example-hotel}). Omar Al-Farsi stays one night on a business trip. His company pays for the rooms, and he pays for the parking and the minibar himself, so the single room reservation has two bills. Hotels commonly split bills this way. However, \texttt{BillRoomReservation} requires exactly one bill per room reservation. The same model already allows a booking to have \texttt{0..*} bills (\texttt{BookingBill}). It only forbids two of those bills from referring to the same reservation. The scenario is ordinary, and the judge is right to mark it as realistic, and the domain model is overconstrained.

The other instances follow a similar pattern. Some violations show upper bounds that are too low, i.e., the model allows fewer links than occur in reality. For example, in \cotinst{football}{1}{invalid}{\texttt{Football over-constraint1}}, a team hosts two of its three cup matches, although \texttt{LocalMatch} allows only one. In \cotinst{football}{3}{invalid}{\texttt{Football over-constraint3}}, a 17-year-old is registered with both the first team and the reserves. In \cotinst{hotelmanagement}{1}{invalid}{\texttt{Hotel Management over-constraint1}}, a guest comes back for a second booking, but the model does not permit it. Other violations show lower bounds that are too high, i.e., the model makes associations mandatory that are optional in reality. For example, \cotinst{football}{5}{invalid}{\texttt{Football over-constraint5}} has a match abandoned after 29 minutes because of lightning, which therefore has no report. \cotinst{hotelmanagement}{1}{invalid}{\texttt{Hotel Management over-constraint1}} has an empty room, and \cotinst{restaurant}{4}{invalid}{\texttt{Restaurant over-constraint4}} has a table with no reservation. \cotinst{myexpenses}{4}{invalid}{\texttt{Expenses over-constraint4}} has a draft receipt and a comment not yet attached to an expense, although \texttt{DRAFT} is one of the model's own bill statuses. In \cotinst{pickupnet}{5}{invalid}{\texttt{Pickup Net over-constraint5}}, a new shipment and a new customer have no station, and a rural delivery address described by landmarks has no geolocation. The \texttt{NEW} status and the optional driver suggest that a station could also be optional. As with the coverage gaps discussed for RQ2, the over-constraint category provides useful feedback to the modeler, helping identify constraints that are too strict and which parts of the model are easy to misread.

\input{tables/judge}

\manuel{todo: manuel : review this paragraph}
Finally, we compare the realism checks with the LLM-as-a-Judge verdicts for each instance. We only consider the format/rule checks of Table~\ref{tab:realism-checks} and exclude the API checks, since the judge is instructed, in line with our definition of realism, to assess whether values are plausible, and it has no access to the external data sources needed to check whether they exist. This comparison covers eight domains and 480 instances (30 per strategy and domain). When at least one format/rule check fails, the judge also labels the instance unrealistic in 39 of 53 cases (74\%; IL 10/14, CoT 29/39). Conversely, the judge labels 176 of the 427 instances (41\%) that pass all these checks as unrealistic, and both methods agree that the remaining 251 instances are realistic. This shows that the two methods are complementary, since the format/rule checks only validate single values or fixed comparisons (e.g., dates), while the judge also reports inconsistencies across the whole instance, such as implausible links between objects. However, the 14 of 53 instances that fail a check but are labeled realistic (e.g., six CoT Pickup Net instances with X usernames longer than the 15-character limit, or four IL Vehicle Rental instances with license plates such as \texttt{CA-TRK-93001}) show that the judge also misses value-level errors.

\paragraph{RQ3.2 - Human-based analysis}

From the 100 instances analyzed by the experts, the explanations of the 50 instances labeled unrealistic contain 82 individual problems. Every object, value, or relationship named in a problem exists in the corresponding instance, i.e., the judge does not hallucinate objects, relationships, or attributes the instance does not have. The experts fully confirm 60 problems and partially confirm seven. We count both as confirmed, so 67 of the 82 problems (81.7\%) are confirmed and 15 (18.3\%) are rejected and not considered as actual issues. At instance level, 44 of the 50 instances labeled unrealistic contain at least one confirmed problem, whereas for the remaining six instances all reported problems are rejected. 

The explanations of the 50 instances labeled realistic contain 198 statements, of which 63 are specific and 135 are generic. All 63 specific statements refer to elements present in the instance, thus again there are no hallucinations. Of the 135 generic statements, experts accept 133. The two rejected statements occur in two different instances. Thus, 48 of the 50 realistic verdicts contain no statement experts reject.
In summary, there is expert agreement with the judge in 92 of the 100 audited instances. 

The categorization in Table~\ref{tab:expert-judge} shows that 48 of the 67 confirmed problems (71.6\%) are \textit{inconsistent values}, i.e., logical contradictions between values such as temporal contradictions and cross-attribute violations. \textit{Implausible values}, which are valid but very unlikely, account for a further 11 problems (16.4\%). The remaining problems are \textit{incorrect values} that violate the domain data type (4, 6.0\%), \textit{duplicated instances}, i.e., distinct objects that actually represent the same entity (3, 4.5\%), and one \textit{incomplete model} (1.5\%) which does not provide enough information to reason about it.

\input{tables/judge-2}

\manuel{todo: manuel : review this paragraph : added further explanation about the rejected problems}
The agreement between experts and judge on the reported problems is domain-dependent (Table~\ref{tab:expert-judge-by-domain}): every reported problem is confirmed for the Address Book, Expenses, and Restaurant domains, whereas only 37.5\% are confirmed for the Bank domain. The 15 rejected problems fall into four categories (Table~\ref{tab:expert-judge-disagreement}). Most of them (10, 66.7\%) are unsupported claims, where the judge bases its verdict on facts that the model does not contain. For example, it considers the account users in \ilinst{bank}{16}{\texttt{Bank gen16}} unrelated to the owner because their names suggest another nationality, although nationality cannot be inferred from a name. Several of these claims reveal some biases of the judge, such as inferring nationality from names (\ilinst{bank}{16}{\texttt{Bank gen16}}, \ilinst{bank}{20}{\texttt{Bank gen20}}) or assuming that a women's team and a men's team never play against each other (\ilinst{football}{5}{\texttt{Football gen5}}). In three cases (20.0\%), the judge turns a common convention into a universal rule, e.g., it rejects an 18-year-old in \ilinst{bank}{27}{\texttt{Bank gen27}} who uses four accounts owned by other adults. The remaining two problems occur in State Machine instances. In \cotinst{statemachine}{4}{boundary}{\texttt{State Machine boundary4}}, the judge applies the semantics of a technical process to a social process. It rejects a carnival gathering state with five outgoing transitions on the same timer as non-deterministic, although a crowd can split up towards several destinations at once. In \cotinst{statemachine}{3}{boundary}{\texttt{State Machine boundary3}}, it flags a timer value of 2147483647 as lasting decades, although the model never states that time is measured in seconds or any other unit.

\paragraph{RQ3.3 - Model refinement}
\input{figs/venn-ocl-experts}
The judge explanations and human adjudication provide actionable feedback about missing documented domain knowledge. We formalized recurrent, expressible domain rules as 48 additional OCL invariants across all ten domain models (all are available in~\cite{repo} in \texttt{docs/experts-judge}). Following the audit, we consider 46 instances unrealistic, i.e., the 44 judge-unrealistic instances with at least one confirmed problem, plus the two judge-realistic instances whose explanation we rejected. \manuel{we should explain a bit more which invariants we have defined. Some are going beyond what is normally done with OCL.} \andrei{please check docs/experts-judge/experts-judge-eval.xlsx second sheet is it enough with that?}

Agreement with the judge (83/100) is partly circular\manuel{what agreement is circular - the agreement about the failing constraints?}, since the invariants were written from the same audit for which the judge had already flagged the problems.\manuel{why, could be even a double-check if all problems have been found} We therefore compare the invariants\manuel{the invariants or the evaluation results of the invariants on the instances} with the expert labels. When applied to the 100 audited instances, the enriched domain models report at least one violation in 45 instances\manuel{all constraints or only the new ones?}, and their verdict agrees with the experts on 85 of the 100 instances. 

Figure~\ref{fig:venn-ocl-experts} splits this comparison by the judge verdict. Of the 50 instances the judge labels realistic, the experts agree with the judge on 48 and the added OCL on 44, and both agree on 43. Of the 50 instances the judge labels unrealistic, the experts agree on 44 and the added OCL on 39, and both agree on 37. The instances where only one of them agrees with the judge are the 15 disagreements listed in Table~\ref{tab:ocl-disagreement}. In the remaining five instances, both disagree with the judge, i.e., the experts and the added OCL agree with each other but not with the judge. Of the 46 instances we consider unrealistic, 38 fail at least one invariant. The eight that pass the invariants are expected, as we only formalized recurring problems that OCL can express.\manuel{are these the instances which are produced for domain models without additional OCL constraints?} Some\manuel{what do we refer here - the realism problems?} depend on knowledge outside the model. For instance, a drone that starts charging right after an emergency landing is implausible only because of what the state names mean.\manuel{example may be difficult to understand if the model is not known... it is also interesting why an LLM is not detecting such a problem if they are so intelligent tools} Others would require an arbitrary threshold, such as a maximum number of guests per waiter to flag one waiter who has to serve for 120 guests. In others, the problem is visible only in names and free text.\manuel{what is free text - string-based values?} A refund stored as a positive amount passes all invariants, because the bills still add up, and only the object name and the free-text category reveal that it is a refund.\manuel{again, the example is difficult to understand..} Finally, two problems could be encoded but were left out. An empty address in a Pickup Net instance can be legitimate in test-oriented instances\manuel{do we mean as a negative example which we do not like to see... here realism is now differently interpreted - is a negative test instance a realistic model in test contexts?}, and the bank in Fig.~\ref{fig:edge-bank} would require a country-name-to-code mapping that is fragile as an invariant.\manuel{why is this fragile as an invariant - is it hard to define? I think we could realize it}

Conversely, 7 of the 54 instances we\manuel{maybe we should describe here that the 54 instances was the result of the expert review?} consider realistic fail for at least one invariant. Six of them contain real defects that the audit missed, since for realistic verdicts we reviewed primarily the judge's statements rather than every detail of the whole instance. For example, four Bank instances use IBAN prefixes of countries not included in the SWIFT IBAN registry (e.g., US, AU, MA, CL). Only one failure reveals an invariant that is too strict. \texttt{SumOfExpenses} requires every bill to charge all rooms of a reservation for all nights plus the extras but this does not take into account canceled bookings, discounts or royalty points. The invariants therefore not only encode the confirmed feedback but also expose problems that both the judge and the experts overlooked. The refinement converts a large portion of qualitative feedback into executable domain knowledge and exposes underspecification in the original class diagrams. 

\andrei{run ocls againts all intances: should we add this?}
Since the invariants were derived from the audited sample, we also ran them on the 440 non-audited instances, again excluding the over-constraint ones, to see how they behave on unseen instances. There are no expert labels for these instances, so we compare the invariants with the judge only. The enriched models report at least one violation in 180 of them (40.9\%), close to the 45\% of the audited sample. They flag 119 of the 157 instances the judge considers unrealistic (75.8\%, against 78\% in the audited sample), so the invariants detect the same kind of problems beyond the instances they were written from. However, they also flag 61 of the 283 judge-realistic instances (21.6\%, against 12\%), which lowers the agreement with the judge from 83\% to 77.5\%. In 31 of them, the invariants expose defects that the judge missed. For example, 15 Bank instances use IBAN prefixes of non-IBAN countries (e.g., JP, IN, KE) or a BIC containing a space, nine Address Book instances use companies in coworker or boss relationships, and one Expenses instance adds a bill in US dollars to bills in Brazilian reais without conversion. Twenty-one cases are debatable. In eight Football instances the goals of the players add up to less than the final score, which might be because the instance records only some players of each match, or maybe own goals (i.e., when a player accidentally puts the ball into their own net, it counts toward the other team's score, but it doesn't count as a goal for any player). In eight Hotel instances the bill differs from the formula of \texttt{SumOfExpenses}, e.g., only one of two reserved rooms is charged, or a confirmed booking has a zero bill, which might be either an error or a case the formula does not cover, such as a discount that the instance does not record. Four State Machine instances contain a state without transitions, three of them deliberately in boundary and edge scenarios and one as the only state of a separate machine. In one Address Book instance, a coworker has no recorded employer and seems to be external to the company. The remaining nine cases reveal invariants that are too strict. \texttt{SumOfExpenses} (3 cases) does not allow zero bills for canceled or unconfirmed bookings or for nights paid with loyalty points, even when the instance records the reason, \texttt{RealisticPricing} (4) requires a positive daily rate although some rentals might be charged per mile, and \texttt{CoworkerSameCompany} (2) requires coworkers to share an employer recorded in the instance and ignores subdivisions. The classification of each instance is available in~\cite{repo} in \texttt{docs/experts-judge/ocl-flagged-judge.xlsx}. Hence, the invariants generalize to unseen instances and keep exposing defects that the judge overlooks, while also showing which of them should be relaxed.

\begin{mdframed}[style=rqanswer,nobreak=true]
{\textbf{Answering RQ3:}} Realism depends strongly on both the domain and the requested scenario. By combining automated, humnan, and LLM-as-a-judge checks we detected relational and semantic inconsistencies in the instances. 
The judge's findings are grounded, as the experts fully or partially confirm 81.7\% of the reported problems, and they are useful for refining the domain model, since the newly added OCL invariants derived from them agree with the experts on 85\% of the audited instances and even expose six defects that both the judge and the experts overlooked. On the 440 non-audited instances, they still flag 75.8\% of the judge-unrealistic ones and expose 31 further defects missed by the judge, although three invariants turn out to be too strict.
\end{mdframed}

\input{tables/ocl-disagreement}

%% file: tables/taxonomy-wf.tex
\begin{table*}[ht]
\caption{Taxonomy of the well-formedness correction calls failures}
\label{tab:wf-taxonomy}
\centering
\scriptsize
\setlength{\tabcolsep}{4pt}
\renewcommand{\arraystretch}{1.05}
\begin{tabular}{llrr>{\raggedright\arraybackslash}p{0.20\textwidth}>{\raggedright\arraybackslash}p{0.32\textwidth}}
\toprule
\textbf{Family} & Failure & IL & CoT & Example & Instances \\
\midrule
\multicolumn{2}{l}{\textbf{Concrete syntax}}
  & \textbf{5} & \textbf{12} & & \\
& String quoting\textsuperscript{1}
  & 2 & 11 & \texttt{O''Connell}, \texttt{d''Italie}, \texttt{''USD''}, \texttt{83''}
  & \ilinst{addressbook}{6}{\texttt{Address Book gen6}}, \cotinst{addressbook}{2}{complex}{\texttt{Address Book complex2}}; \ilinst{restaurant}{3}{\texttt{Restaurant gen3}}, \cotinst{restaurant}{3}{edge}{\texttt{Restaurant edge3}}; \cotinst{football}{1}{invalid}{\texttt{Football over-constraint1}}, \cotinst{football}{1}{complex}{\texttt{Football complex1}}; \cotinst{myexpenses}{2}{edge}{\texttt{Expenses edge2}}, \cotinst{myexpenses}{2}{complex}{\texttt{Expenses complex2}}, \cotinst{myexpenses}{3}{edge}{\texttt{Expenses edge3}}, \cotinst{myexpenses}{6}{complex}{\texttt{Expenses complex6}}; \cotinst{pickupnet}{6}{invalid}{\texttt{Pickup Net over-constraint6}}; \cotinst{videoclub}{2}{baseline}{\texttt{Videoclub baseline2}}, \cotinst{videoclub}{6}{edge}{\texttt{Videoclub edge6}} \\
& Truncated command
  & 2 & 0 & \texttt{!new GeoLocation('geo}
  & \ilinst{addressbook}{6}{\texttt{Address Book gen6}}, \ilinst{pickupnet}{1}{\texttt{Pickup Net gen1}} \\
& Illegal identifier
  & 0 & 1 & \texttt{AB-Emptyish-1}
  & \cotinst{addressbook}{1}{boundary}{\texttt{Address Book boundary1}} \\
& Assignment as expression
  & 1 & 0 & \texttt{!i15\_teamSBY\_R.name} with no \texttt{:=}
  & \ilinst{football}{24}{\texttt{Football gen24}} \\
\midrule
\multicolumn{2}{l}{\textbf{Type mismatch}}
  & \textbf{2} & \textbf{7} & & \\
& Undefined object\textsuperscript{2}
  & 2 & 4 & \texttt{curUSD}, \texttt{w47}, \texttt{S1}
  & \ilinst{myexpenses}{3}{\texttt{Expenses gen3}}, \ilinst{restaurant}{18}{\texttt{Restaurant gen18}}, \cotinst{hotelmanagement}{6}{boundary}{\texttt{Hotel Management boundary6}}, \cotinst{pickupnet}{1}{baseline}{\texttt{Pickup Net baseline1}}, \cotinst{pickupnet}{1}{edge}{\texttt{Pickup Net edge1}}, \cotinst{pickupnet}{6}{invalid}{\texttt{Pickup Net over-constraint6}} \\
& Undefined enum literal
  & 0 & 1 & \texttt{\#NY}
  & \cotinst{vehiclerental}{2}{invalid}{\texttt{Vehicle Rental over-constraint2}} \\
& Wrong class at association
  & 0 & 1 & a \texttt{Cook} inserted at the \texttt{Chef} end of \texttt{ChefCook}
  & \cotinst{restaurant}{1}{complex}{\texttt{Restaurant complex1}} \\
& Attribute type mismatch
  & 0 & 1 & \texttt{Set\{\#Action, \#Comedy\}} assigned to \texttt{Genre}
  & \cotinst{videoclub}{6}{invalid}{\texttt{Videoclub over-constraint6}} \\
\midrule
\multicolumn{2}{l}{\textbf{Links insert rejected}}
  & \textbf{2} & \textbf{8} & & \\
& Duplicate link
  & 2 & 0 & the same \texttt{TransitionState} link inserted twice
  & \ilinst{statemachine}{20}{\texttt{State Machine gen20}}, \ilinst{statemachine}{24}{\texttt{State Machine gen24}} \\
& Shared composition part\textsuperscript{3}
  & 0 & 8 & one part inserted into two wholes
  & \cotinst{addressbook}{1}{invalid}{\texttt{Address Book over-constraint1}}, \cotinst{addressbook}{2}{invalid}{\texttt{Address Book over-constraint2}}; \cotinst{bank}{4}{invalid}{\texttt{Bank over-constraint4}}; \cotinst{myexpenses}{3}{invalid}{\texttt{Expenses over-constraint3}}; \cotinst{pickupnet}{1}{invalid}{\texttt{Pickup Net over-constraint1}}, \cotinst{pickupnet}{2}{invalid}{\texttt{Pickup Net over-constraint2}}; \cotinst{statemachine}{2}{invalid}{\texttt{State Machine over-constraint2}}, \cotinst{statemachine}{5}{invalid}{\texttt{State Machine over-constraint5}} \\
\midrule
\multicolumn{2}{l}{\textbf{Boundary values}}
  & \textbf{0} & \textbf{2} & & \\
& Integer minimum\textsuperscript{4}
  & 0 & 2 & \texttt{roomNumber} and \texttt{idNumber} set to \texttt{-2147483648}
  & \cotinst{hotelmanagement}{6}{boundary}{\texttt{Hotel Management boundary6}}, \cotinst{vehiclerental}{4}{edge}{\texttt{Vehicle Rental edge4}} \\
\midrule
\multicolumn{2}{l}{\textbf{Total}}
  & \textbf{9} & \textbf{29} & & \\
\bottomrule
\end{tabular}

\medskip
\begin{minipage}{0.98\textwidth}
\footnotesize
\textsuperscript{1} SOIL strings are single-quoted, and \texttt{''} does not escape a quote.
\textsuperscript{2} \cotinst{hotelmanagement}{6}{boundary}{\texttt{Hotel Management boundary6}} and \cotinst{pickupnet}{6}{invalid}{\texttt{Pickup Net over-constraint6}} are second passes where the rewrite dropped the \texttt{!new} lines.
\textsuperscript{3} All eight calls are over-constraint instances.
\textsuperscript{4} The value fits a Java \texttt{int}, but USE parses the unsigned literal \texttt{2147483648} before applying the minus sign, so the literal overflows and USE aborts with an internal error. This is a defect of USE, not of the generated instance.
\end{minipage}
\end{table*}

%% file: tables/taxonomy-cf.tex
\begin{table*}[ht]
\caption{Taxonomy of the conformance correction calls failures}
\label{tab:conf-taxonomy}
\centering
\scriptsize
\setlength{\tabcolsep}{4pt}
\renewcommand{\arraystretch}{1.05}
\begin{tabular}{llrr>{\raggedright\arraybackslash}p{0.24\textwidth}>{\raggedright\arraybackslash}p{0.22\textwidth}}
\toprule
Family & Failure & IL & CoT & Example & Instances \\
\midrule
\multicolumn{2}{l}{\textbf{Missing required link, lower bound}}
  & \textbf{8} & \textbf{0} & & \\
& Player with no training objective
  & 2 & 0 & \texttt{TrainingObjectivePlayer}, \texttt{1..*} & \ilinst{football}{11}{\texttt{Football gen11}}, \ilinst{football}{29}{\texttt{Football gen29}} \\
& Room with no reservation
  & 1 & 0 & \texttt{RoomReservationRoom}, \texttt{1} & \ilinst{hotelmanagement}{8}{\texttt{Hotel Management gen8}} \\
& Reservation with no bill
  & 2 & 0 & \texttt{BillRoomReservation}, \texttt{1} & \ilinst{hotelmanagement}{24}{\texttt{Hotel Management gen24}}, \ilinst{hotelmanagement}{26}{\texttt{Hotel Management gen26}} \\
& Bus driver with no banquet
  & 1 & 0 & \texttt{BanquetBusDriver}, \texttt{1} & \ilinst{restaurant}{5}{\texttt{Restaurant gen5}} \\
& State machine with no start state
  & 2 & 0 & \texttt{StateMachineStartsStates}, \texttt{1} & \ilinst{statemachine}{1}{\texttt{State Machine gen1}}, \ilinst{statemachine}{22}{\texttt{State Machine gen22}} \\
\midrule
\multicolumn{2}{l}{\textbf{Extra link, upper bound exceeded}}
  & \textbf{4} & \textbf{3} & & \\
& Cook supervised by two chefs
  & 1 & 3 & \texttt{ChefCook}, \texttt{1} & \ilinst{restaurant}{9}{\texttt{Restaurant gen9}}, \cotinst{restaurant}{1}{baseline}{\texttt{Restaurant baseline1}}, \cotinst{restaurant}{2}{boundary}{\texttt{Restaurant boundary2}}, \cotinst{restaurant}{3}{edge}{\texttt{Restaurant edge3}} \\
& Reservation with two bills
  & 1 & 0 & \texttt{BillRoomReservation}, \texttt{1} & \ilinst{hotelmanagement}{2}{\texttt{Hotel Management gen2}} \\
& Team visiting two matches
  & 1 & 0 & \texttt{VisitorMatch}, \texttt{0..1} & \ilinst{football}{27}{\texttt{Football gen27}} \\
& Transition with two source states
  & 1 & 0 & \texttt{StateTransition}, \texttt{1} & \ilinst{statemachine}{28}{\texttt{State Machine gen28}} \\
\midrule
\multicolumn{2}{l}{\textbf{Total}\textsuperscript{1}}
  & \textbf{12} & \textbf{3} & & \\
\bottomrule
\end{tabular}

\medskip
\begin{minipage}{0.98\textwidth}
\footnotesize
\textsuperscript{1} No call contains an invariant failure. Every diagnostic is a multiplicity violation.
\end{minipage}
\end{table*}

%% file: tables/coverage.tex
\begin{table*}[t]
\caption{Within- and across-instance coverage of domain-model elements under IL and CoT (\%)}
\label{tab:coverage}
\scriptsize
\setlength{\tabcolsep}{4pt}
\renewcommand{\arraystretch}{1.12}
\begin{tabularx}{\textwidth}{@{}>{\raggedright\arraybackslash}p{2.3cm}>{\centering\arraybackslash}p{1.15cm}*{6}{>{\centering\arraybackslash}X}@{}}
\toprule
\multirow{2}{*}{Domain} & \multirow{2}{*}{Scope} & \multicolumn{3}{c}{IL} & \multicolumn{3}{c}{CoT} \\
\cmidrule(lr){3-5}\cmidrule(lr){6-8}
 & & \#cl & \#att & \#rel & \#cl & \#att & \#rel \\
\midrule
\multirow{2}{*}{Address Book}
 & Within & 97.2 $\pm$ 7.7 & 96.1 $\pm$ 10.1 & 98.7 $\pm$ 7.3 & 87.8 $\pm$ 25.1 & 85.9 $\pm$ 29.3 & 85.3 $\pm$ 30.1 \\
 & Across & 100.0 & 100.0 & 100.0 & 100.0 & 100.0 & 100.0 \\
\midrule
\multirow{2}{*}{Bank}
 & Within & 100.0 $\pm$ 0.0 & 100.0 $\pm$ 0.0 & 100.0 $\pm$ 0.0 & 100.0 $\pm$ 0.0 & 100.0 $\pm$ 0.0 & 100.0 $\pm$ 0.0 \\
 & Across & 100.0 & 100.0 & 100.0 & 100.0 & 100.0 & 100.0 \\
\midrule
\multirow{2}{*}{\begin{tabular}[c]{@{}l@{}}Hotel\\ Management\end{tabular}}
 & Within & 87.1 $\pm$ 19.3 & 88.8 $\pm$ 17.3 & 83.3 $\pm$ 28.0 & 98.6 $\pm$ 4.4 & 99.1 $\pm$ 2.8 & 98.3 $\pm$ 5.1 \\
 & Across & 100.0 & 100.0 & 100.0 & 100.0 & 100.0 & 100.0 \\
\midrule
\multirow{2}{*}{Expenses}
 & Within & 90.0 $\pm$ 14.1 & 94.8 $\pm$ 11.9 & 86.7 $\pm$ 18.8 & 95.0 $\pm$ 16.6 & 95.2 $\pm$ 17.3 & 93.3 $\pm$ 22.1 \\
 & Across & 100.0 & 100.0 & 100.0 & 100.0 & 100.0 & 100.0 \\
\midrule
\multirow{2}{*}{Pickup Net}
 & Within & 100.0 $\pm$ 0.0 & 100.0 $\pm$ 0.0 & 98.3 $\pm$ 4.3 & 97.2 $\pm$ 9.9 & 97.0 $\pm$ 10.2 & 95.4 $\pm$ 14.9 \\
 & Across & 100.0 & 100.0 & 100.0 & 100.0 & 100.0 & 100.0 \\
\midrule
\multirow{2}{*}{State Machine}
 & Within & 100.0 $\pm$ 0.0 & 100.0 $\pm$ 0.0 & 100.0 $\pm$ 0.0 & 96.7 $\pm$ 12.7 & 96.7 $\pm$ 12.7 & 95.0 $\pm$ 17.0 \\
 & Across & 100.0 & 100.0 & 100.0 & 100.0 & 100.0 & 100.0 \\
\midrule
\multirow{2}{*}{Vehicle Rental}
 & Within & 97.8 $\pm$ 5.8 & 96.6 $\pm$ 9.0 & 100.0 $\pm$ 0.0 & 93.3 $\pm$ 9.4 & 93.7 $\pm$ 8.8 & 93.3 $\pm$ 20.6 \\
 & Across & 100.0 & 100.0 & 100.0 & 100.0 & 100.0 & 100.0 \\
\midrule
\multirow{2}{*}{Videoclub}
 & Within & 88.9 $\pm$ 10.1 & 88.3 $\pm$ 11.7 & 97.8 $\pm$ 12.2 & 81.7 $\pm$ 6.7 & 80.6 $\pm$ 9.4 & 100.0 $\pm$ 0.0 \\
 & Across & 100.0 & 100.0 & 100.0 & 100.0 & 100.0 & 100.0 \\
\midrule
\multirow{2}{*}{Football}
 & Within & 100.0 $\pm$ 0.0 & 100.0 $\pm$ 0.0 & 100.0 $\pm$ 0.0 & 98.8 $\pm$ 3.4 & 98.9 $\pm$ 2.9 & 98.7 $\pm$ 3.8 \\
 & Across & 100.0 & 100.0 & 100.0 & 100.0 & 100.0 & 100.0 \\
\midrule
\multirow{2}{*}{Restaurant}
 & Within & 88.8 $\pm$ 4.5 & 91.7 $\pm$ 5.9 & 95.6 $\pm$ 6.3 & 90.7 $\pm$ 6.6 & 94.1 $\pm$ 6.9 & 96.7 $\pm$ 9.0 \\
 & Across & 94.7 & 98.3 & 100.0 & 94.7 & 98.3 & 100.0 \\
\midrule
\multirow{2}{*}{All}
 & Within & 95.0 $\pm$ 10.2 & 95.6 $\pm$ 9.8 & 96.0 $\pm$ 13.0 & 94.0 $\pm$ 12.8 & 94.1 $\pm$ 14.0 & 95.6 $\pm$ 16.0 \\
 & Across & 98.7 & 99.5 & 100.0 & 98.7 & 99.5 & 100.0 \\
\bottomrule
\end{tabularx}

\medskip
\footnotesize
\#cl. Classes, \#att. Attributes, \#rel. Relationships. All-Within averages the coverage of each of the 300 instances of a strategy; All-Across divides the number of distinct elements instantiated in the 300 instances by the number of elements in the ten domain models.
\end{table*}

%% file: tables/diversity.tex
\begin{table*}[ht]
\caption{Cumulative diversity results for the 600 generated model instances through IL and CoT}
\label{tab:semantic-similarity}
\centering
\scriptsize
\setlength{\tabcolsep}{2pt}
\renewcommand{\arraystretch}{1.12}
\begin{tabular*}{\textwidth}{@{\extracolsep{\fill}}ll*{3}{M{0.0000}{0.0000}}@{}}
\toprule
Domain & Scope & \thd{Numeric} & \multicolumn{2}{c}{String} \\
\cmidrule(l){4-5}
 & & & \thd{Exact match} & \thd{Norm. Lev. dist.} \\
\midrule
Address Book & Within & -- & 0.9813 $\pm$ 0.0810 & 0.8940 $\pm$ 0.0684 \\
 & Across & -- & 0.9988 & 0.9147 \\
\midrule
Bank & Within & 0.9920 $\pm$ 0.0217 & 0.9945 $\pm$ 0.0084 & 0.9100 $\pm$ 0.0164 \\
 & Across & 0.9839 & 0.9994 & 0.9180 \\
\midrule
Hotel Management & Within & 0.9463 $\pm$ 0.0480 & 0.9504 $\pm$ 0.0357 & 0.7812 $\pm$ 0.0646 \\
 & Across & 0.9691 & 0.9986 & 0.8207 \\
\midrule
Expenses & Within & 0.9112 $\pm$ 0.2294 & 0.9559 $\pm$ 0.1305 & 0.7952 $\pm$ 0.1151 \\
 & Across & 0.9905 & 0.9988 & 0.8326 \\
\midrule
Pickup Net & Within & 0.9936 $\pm$ 0.0323 & 0.9992 $\pm$ 0.0036 & 0.8406 $\pm$ 0.0214 \\
 & Across & 0.9996 & 1.0000 & 0.8849 \\
\midrule
State Machine & Within & 0.9168 $\pm$ 0.1398 & 0.9994 $\pm$ 0.0022 & 0.8316 $\pm$ 0.0378 \\
 & Across & 0.9118 & 0.9995 & 0.8580 \\
\midrule
Vehicle Rental & Within & 0.9564 $\pm$ 0.1220 & 0.9966 $\pm$ 0.0090 & 0.8048 $\pm$ 0.0322 \\
 & Across & 0.9806 & 0.9984 & 0.7999 \\
\midrule
Videoclub & Within & 0.9325 $\pm$ 0.0796 & 0.9879 $\pm$ 0.0181 & 0.8747 $\pm$ 0.0299 \\
 & Across & 0.9238 & 0.9997 & 0.8850 \\
\midrule
Football & Within & 0.9377 $\pm$ 0.0650 & 0.9866 $\pm$ 0.0088 & 0.8157 $\pm$ 0.0101 \\
 & Across & 0.9459 & 0.9992 & 0.8327 \\
\midrule
Restaurant & Within & 0.9729 $\pm$ 0.0308 & 0.9984 $\pm$ 0.0023 & 0.8791 $\pm$ 0.0092 \\
 & Across & 0.9802 & 0.9998 & 0.8902 \\
\midrule
All & Within & 0.9511 $\pm$ 0.1096 & 0.9850 $\pm$ 0.0528 & 0.8426 $\pm$ 0.0661 \\
 & Across & 0.9778 & 0.9998 & 0.8992 \\
\bottomrule
\end{tabular*}

\medskip
\begin{minipage}{\textwidth}
\raggedright
\footnotesize
Values are in the range [0..1], the closer to 1, the higher the diversity is. All-Within is the mean $\pm$ std over the 600 instances; All-Across pools the values of each of the 600 instances.
\end{minipage}
\vspace{-.2cm}
\end{table*}

%% file: tables/shannon.tex
\begin{table*}[t]
\caption{Within- and across-instance distribution of enumeration values (Shannon evenness) of the generated model instances}
\label{tab:enum-distribution}
\scriptsize
\setlength{\tabcolsep}{4pt}
\renewcommand{\arraystretch}{1.1}
\begin{tabular*}{\textwidth}{@{\extracolsep{\fill}}l*{6}{c}@{}}
\toprule
\multirow{3}{*}{Domain} & \multicolumn{4}{c}{Within \textcolor{black!50}{(max.)}} & \multicolumn{2}{c}{Across \textcolor{black!50}{(max.)}} \\
\cmidrule(lr){2-5}\cmidrule(lr){6-7}
 & \multicolumn{2}{c}{All literals $E_{\mathrm{all}}$} & \multicolumn{2}{c}{Active literals $E_{\mathrm{active}}$} & \multirow{2}{*}{IL} & \multirow{2}{*}{CoT} \\
\cmidrule(lr){2-3}\cmidrule(lr){4-5}
 & IL & CoT & IL & CoT & & \\
\midrule
Address Book & \begin{tabular}[t]{@{}c@{}}0.90 $\pm$ 0.11\\[-1pt]\textcolor{black!50}{(0.98)}\end{tabular} & \begin{tabular}[t]{@{}c@{}}0.70 $\pm$ 0.33\\[-1pt]\textcolor{black!50}{(0.78)}\end{tabular} & \begin{tabular}[t]{@{}c@{}}0.92 $\pm$ 0.09\\[-1pt]\textcolor{black!50}{(0.98)}\end{tabular} & \begin{tabular}[t]{@{}c@{}}0.73 $\pm$ 0.33\\[-1pt]\textcolor{black!50}{(0.80)}\end{tabular} & \begin{tabular}[t]{@{}c@{}}0.96\\[-1pt]\textcolor{black!50}{(1.00)}\end{tabular} & \begin{tabular}[t]{@{}c@{}}0.91\\[-1pt]\textcolor{black!50}{(1.00)}\end{tabular} \\
\addlinespace[3pt]
Expenses & \begin{tabular}[t]{@{}c@{}}0.79 $\pm$ 0.22\\[-1pt]\textcolor{black!50}{(0.91)}\end{tabular} & \begin{tabular}[t]{@{}c@{}}0.74 $\pm$ 0.29\\[-1pt]\textcolor{black!50}{(0.81)}\end{tabular} & \begin{tabular}[t]{@{}c@{}}0.88 $\pm$ 0.24\\[-1pt]\textcolor{black!50}{(0.93)}\end{tabular} & \begin{tabular}[t]{@{}c@{}}0.86 $\pm$ 0.29\\[-1pt]\textcolor{black!50}{(0.90)}\end{tabular} & \begin{tabular}[t]{@{}c@{}}0.93\\[-1pt]\textcolor{black!50}{(1.00)}\end{tabular} & \begin{tabular}[t]{@{}c@{}}0.94\\[-1pt]\textcolor{black!50}{(1.00)}\end{tabular} \\
\addlinespace[3pt]
Pickup Net & \begin{tabular}[t]{@{}c@{}}0.84 $\pm$ 0.26\\[-1pt]\textcolor{black!50}{(0.91)}\end{tabular} & \begin{tabular}[t]{@{}c@{}}0.77 $\pm$ 0.29\\[-1pt]\textcolor{black!50}{(0.79)}\end{tabular} & \begin{tabular}[t]{@{}c@{}}0.92 $\pm$ 0.25\\[-1pt]\textcolor{black!50}{(1.00)}\end{tabular} & \begin{tabular}[t]{@{}c@{}}0.92 $\pm$ 0.25\\[-1pt]\textcolor{black!50}{(0.93)}\end{tabular} & \begin{tabular}[t]{@{}c@{}}0.98\\[-1pt]\textcolor{black!50}{(1.00)}\end{tabular} & \begin{tabular}[t]{@{}c@{}}1.00\\[-1pt]\textcolor{black!50}{(1.00)}\end{tabular} \\
\addlinespace[3pt]
Vehicle Rental & \begin{tabular}[t]{@{}c@{}}0.46 $\pm$ 0.10\\[-1pt]\textcolor{black!50}{(0.55)}\end{tabular} & \begin{tabular}[t]{@{}c@{}}0.42 $\pm$ 0.26\\[-1pt]\textcolor{black!50}{(0.47)}\end{tabular} & \begin{tabular}[t]{@{}c@{}}0.83 $\pm$ 0.19\\[-1pt]\textcolor{black!50}{(0.88)}\end{tabular} & \begin{tabular}[t]{@{}c@{}}0.67 $\pm$ 0.33\\[-1pt]\textcolor{black!50}{(0.71)}\end{tabular} & \begin{tabular}[t]{@{}c@{}}0.94\\[-1pt]\textcolor{black!50}{(1.00)}\end{tabular} & \begin{tabular}[t]{@{}c@{}}0.97\\[-1pt]\textcolor{black!50}{(1.00)}\end{tabular} \\
\addlinespace[3pt]
Videoclub & \begin{tabular}[t]{@{}c@{}}0.56 $\pm$ 0.50\\[-1pt]\textcolor{black!50}{(0.57)}\end{tabular} & \begin{tabular}[t]{@{}c@{}}0.71 $\pm$ 0.44\\[-1pt]\textcolor{black!50}{(0.77)}\end{tabular} & \begin{tabular}[t]{@{}c@{}}0.56 $\pm$ 0.50\\[-1pt]\textcolor{black!50}{(0.57)}\end{tabular} & \begin{tabular}[t]{@{}c@{}}0.71 $\pm$ 0.44\\[-1pt]\textcolor{black!50}{(0.77)}\end{tabular} & \begin{tabular}[t]{@{}c@{}}0.97\\[-1pt]\textcolor{black!50}{(1.00)}\end{tabular} & \begin{tabular}[t]{@{}c@{}}1.00\\[-1pt]\textcolor{black!50}{(1.00)}\end{tabular} \\
\addlinespace[3pt]
Football & \begin{tabular}[t]{@{}c@{}}0.90 $\pm$ 0.07\\[-1pt]\textcolor{black!50}{(0.98)}\end{tabular} & \begin{tabular}[t]{@{}c@{}}0.88 $\pm$ 0.11\\[-1pt]\textcolor{black!50}{(0.95)}\end{tabular} & \begin{tabular}[t]{@{}c@{}}0.94 $\pm$ 0.03\\[-1pt]\textcolor{black!50}{(1.00)}\end{tabular} & \begin{tabular}[t]{@{}c@{}}0.92 $\pm$ 0.07\\[-1pt]\textcolor{black!50}{(0.97)}\end{tabular} & \begin{tabular}[t]{@{}c@{}}0.94\\[-1pt]\textcolor{black!50}{(1.00)}\end{tabular} & \begin{tabular}[t]{@{}c@{}}0.95\\[-1pt]\textcolor{black!50}{(1.00)}\end{tabular} \\
\addlinespace[3pt]
Restaurant & \begin{tabular}[t]{@{}c@{}}0.42 $\pm$ 0.14\\[-1pt]\textcolor{black!50}{(0.49)}\end{tabular} & \begin{tabular}[t]{@{}c@{}}0.46 $\pm$ 0.20\\[-1pt]\textcolor{black!50}{(0.52)}\end{tabular} & \begin{tabular}[t]{@{}c@{}}0.52 $\pm$ 0.14\\[-1pt]\textcolor{black!50}{(0.57)}\end{tabular} & \begin{tabular}[t]{@{}c@{}}0.56 $\pm$ 0.23\\[-1pt]\textcolor{black!50}{(0.60)}\end{tabular} & \begin{tabular}[t]{@{}c@{}}0.93\\[-1pt]\textcolor{black!50}{(1.00)}\end{tabular} & \begin{tabular}[t]{@{}c@{}}0.95\\[-1pt]\textcolor{black!50}{(1.00)}\end{tabular} \\
\midrule
\textbf{All} & \begin{tabular}[t]{@{}c@{}}0.70 $\pm$ 0.31\\[-1pt]\textcolor{black!50}{(0.77)}\end{tabular} & \begin{tabular}[t]{@{}c@{}}0.67 $\pm$ 0.32\\[-1pt]\textcolor{black!50}{(0.73)}\end{tabular} & \begin{tabular}[t]{@{}c@{}}0.80 $\pm$ 0.30\\[-1pt]\textcolor{black!50}{(0.85)}\end{tabular} & \begin{tabular}[t]{@{}c@{}}0.77 $\pm$ 0.32\\[-1pt]\textcolor{black!50}{(0.81)}\end{tabular} & \begin{tabular}[t]{@{}c@{}}0.94\\[-1pt]\textcolor{black!50}{(1.00)}\end{tabular} & \begin{tabular}[t]{@{}c@{}}0.95\\[-1pt]\textcolor{black!50}{(1.00)}\end{tabular} \\
\bottomrule
\end{tabular*}

\medskip
\footnotesize
Values are in $[0,1]$; closer to 1 means a more even distribution. Within is the mean $\pm$ standard deviation over the 30 instances of each strategy and Across pools the 30 instances. The upper bound on the evenness given the number of values is shown below each score in gray parentheses. Across instances, every literal occurs at least once, so $E_{\mathrm{active}}$ equals $E_{\mathrm{all}}$ therefore a single value is shown.
\vspace{-.2cm}
\end{table*}

%% file: tables/shannon-bool.tex
\begin{table}[t]
\caption{Within- and across-instance distribution of Boolean values (Shannon evenness) of the generated model instances}
\label{tab:bool-distribution}
\scriptsize
\setlength{\tabcolsep}{1.5pt}
\renewcommand{\arraystretch}{1.1}
\begin{tabular*}{\columnwidth}{@{\extracolsep{\fill}}l*{4}{c}@{}}
\toprule
\multirow{2}{*}{Domain} & \multicolumn{2}{c}{Within \textcolor{black!50}{(max.)}} & \multicolumn{2}{c}{Across \textcolor{black!50}{(max.)}} \\
\cmidrule(lr){2-3}\cmidrule(lr){4-5}
 & IL & CoT & IL & CoT \\
\midrule
\begin{tabular}[t]{@{}l@{}}Hotel\\[-1pt]Management\end{tabular} & \begin{tabular}[t]{@{}c@{}}0.02 $\pm$ 0.09\\[-1pt]\textcolor{black!50}{(0.32)}\end{tabular} & \begin{tabular}[t]{@{}c@{}}0.25 $\pm$ 0.22\\[-1pt]\textcolor{black!50}{(0.53)}\end{tabular} & \begin{tabular}[t]{@{}c@{}}0.67\\[-1pt]\textcolor{black!50}{(1.00)}\end{tabular} & \begin{tabular}[t]{@{}c@{}}0.81\\[-1pt]\textcolor{black!50}{(1.00)}\end{tabular} \\
\addlinespace[3pt]
\begin{tabular}[t]{@{}l@{}}Vehicle\\[-1pt]Rental\end{tabular} & \begin{tabular}[t]{@{}c@{}}0.78 $\pm$ 0.27\\[-1pt]\textcolor{black!50}{(0.87)}\end{tabular} & \begin{tabular}[t]{@{}c@{}}0.59 $\pm$ 0.37\\[-1pt]\textcolor{black!50}{(0.72)}\end{tabular} & \begin{tabular}[t]{@{}c@{}}0.95\\[-1pt]\textcolor{black!50}{(1.00)}\end{tabular} & \begin{tabular}[t]{@{}c@{}}0.91\\[-1pt]\textcolor{black!50}{(1.00)}\end{tabular} \\
\addlinespace[3pt]
Videoclub & \begin{tabular}[t]{@{}c@{}}0.92 $\pm$ 0.06\\[-1pt]\textcolor{black!50}{(1.00)}\end{tabular} & \begin{tabular}[t]{@{}c@{}}0.66 $\pm$ 0.41\\[-1pt]\textcolor{black!50}{(0.93)}\end{tabular} & \begin{tabular}[t]{@{}c@{}}0.93\\[-1pt]\textcolor{black!50}{(1.00)}\end{tabular} & \begin{tabular}[t]{@{}c@{}}0.87\\[-1pt]\textcolor{black!50}{(1.00)}\end{tabular} \\
\addlinespace[3pt]
Football & \begin{tabular}[t]{@{}c@{}}0.84 $\pm$ 0.15\\[-1pt]\textcolor{black!50}{(0.94)}\end{tabular} & \begin{tabular}[t]{@{}c@{}}0.61 $\pm$ 0.17\\[-1pt]\textcolor{black!50}{(0.83)}\end{tabular} & \begin{tabular}[t]{@{}c@{}}0.92\\[-1pt]\textcolor{black!50}{(1.00)}\end{tabular} & \begin{tabular}[t]{@{}c@{}}0.80\\[-1pt]\textcolor{black!50}{(1.00)}\end{tabular} \\
\addlinespace[3pt]
Restaurant & \begin{tabular}[t]{@{}c@{}}0.03 $\pm$ 0.08\\[-1pt]\textcolor{black!50}{(0.50)}\end{tabular} & \begin{tabular}[t]{@{}c@{}}0.28 $\pm$ 0.24\\[-1pt]\textcolor{black!50}{(0.58)}\end{tabular} & \begin{tabular}[t]{@{}c@{}}0.42\\[-1pt]\textcolor{black!50}{(1.00)}\end{tabular} & \begin{tabular}[t]{@{}c@{}}0.77\\[-1pt]\textcolor{black!50}{(1.00)}\end{tabular} \\
\midrule
\textbf{All} & \begin{tabular}[t]{@{}c@{}}0.52 $\pm$ 0.43\\[-1pt]\textcolor{black!50}{(0.73)}\end{tabular} & \begin{tabular}[t]{@{}c@{}}0.48 $\pm$ 0.34\\[-1pt]\textcolor{black!50}{(0.72)}\end{tabular} & \begin{tabular}[t]{@{}c@{}}0.76\\[-1pt]\textcolor{black!50}{(1.00)}\end{tabular} & \begin{tabular}[t]{@{}c@{}}0.82\\[-1pt]\textcolor{black!50}{(1.00)}\end{tabular} \\
\bottomrule
\end{tabular*}

\medskip
\footnotesize
Values are in $[0,1]$; closer to 1 means a more even distribution. Within is the mean $\pm$ standard deviation over the 30 instances of each strategy and Across pools the 30 instances. The upper bound on the evenness given the number of values is shown below each score in gray parentheses. With two literals, $E_{\mathrm{active}}$ equals $E_{\mathrm{all}}$, so a single value is shown.
\vspace{-.2cm}
\end{table}

%% file: tables/ged.tex
\begin{table}[t]
\caption{Structural diversity (normalized GED) of the generated model instances}
\label{tab:structural-diversity}
\centering
\scriptsize
\setlength{\tabcolsep}{2pt}
\renewcommand{\arraystretch}{1.12}
\begin{tabular*}{\linewidth}{@{\extracolsep{\fill}}l*{2}{M{0.00}{0.00}}@{}}
\toprule
Domain & \thd{IL} & \thd{CoT} \\
\midrule
Address Book     & 0.21 $\pm$ 0.07 & 0.56 $\pm$ 0.26 \\
Bank             & 0.25 $\pm$ 0.06 & 0.39 $\pm$ 0.14 \\
Hotel Management & 0.35 $\pm$ 0.20 & 0.30 $\pm$ 0.13 \\
Expenses         & 0.32 $\pm$ 0.15 & 0.44 $\pm$ 0.21 \\
Pickup Net       & 0.23 $\pm$ 0.08 & 0.38 $\pm$ 0.18 \\
State Machine    & 0.24 $\pm$ 0.10 & 0.45 $\pm$ 0.22 \\
Vehicle Rental   & 0.26 $\pm$ 0.06 & 0.47 $\pm$ 0.16 \\
Videoclub        & 0.28 $\pm$ 0.07 & 0.41 $\pm$ 0.15 \\
Football         & 0.22 $\pm$ 0.12 & 0.37 $\pm$ 0.16 \\
Restaurant       & 0.21 $\pm$ 0.06 & 0.33 $\pm$ 0.12 \\
\midrule
All              & 0.26 $\pm$ 0.12 & 0.41 $\pm$ 0.19 \\
\bottomrule
\end{tabular*}

\medskip
\footnotesize
Values are in $[0,1]$; larger values indicate greater structural diversity. Each cell is the mean $\pm$ std over instance pairs; All pools the pairs of all ten domains.
\vspace{-.2cm}
\end{table}

%% file: tables/errors-checks.tex
\begin{table*}[t]
\caption{Successful realism checks for model instances generated through IL and CoT}
\label{tab:realism-checks}
\centering
\scriptsize
\setlength{\tabcolsep}{2pt}
\renewcommand{\arraystretch}{1.12}
\begin{tabular*}{\textwidth}{@{\extracolsep{\fill}}llF{0000}*{5}{F{000}}N{00.0\%}@{}}
\toprule
Domain & Check & \thd{IL} & \multicolumn{6}{c}{CoT category} \\
\cmidrule(l){4-9}
& & & \thd{Baseline} & \thd{Complex} & \thd{Boundary} & \thd{Edge} & \thd{Over-constraint} & \thd{CoT} \\
\midrule
Address Book & Phone\textsuperscript{*} & 201/201 & 42/42 & 74/74 & 15/16 & 15/18 & 27/28 & 97.2\% \\
 & Website & 116/116 & 42/42 & 74/74 & 9/9 & 12/17 & 27/27 & 97.0\% \\
 & Emails & 201/201 & 42/42 & 74/74 & 12/12 & 13/18 & 27/27 & 97.1\% \\
 & Address(API)\textsuperscript{*} & 133/166 & 33/44 & 70/97 & 3/6 & 2/28 & 15/22 & 62.4\% \\
\midrule
Bank & IBAN(API)\textsuperscript{*} & 35/141 & 9/21 & 20/58 & 3/26 & 4/27 & 1/11 & 25.9\% \\
 & IBAN\textsuperscript{*} & 141/141 & 21/21 & 58/58 & 20/26 & 27/27 & 11/11 & 95.8\% \\
 & BIC\textsuperscript{*} & 62/63 & 12/12 & 23/23 & 18/18 & 15/15 & 10/11 & 98.7\% \\
 & Country\textsuperscript{*} & 63/63 & 12/12 & 23/23 & 18/18 & 15/15 & 11/11 & 100\% \\
\midrule
Hotel & Check in $\geq$ start date\textsuperscript{*} & 48/48 & 13/13 & 17/17 & 16/16 & 7/14 & 9/9 & 89.9\% \\
Management & Check out $\leq$ end date\textsuperscript{*} & 48/48 & 13/13 & 17/17 & 16/16 & 4/14 & 9/9 & 85.5\% \\
\midrule
Expenses & End date $\geq$ start date\textsuperscript{*} & 50/50 & 7/7 & 27/27 & 8/8 & 10/13 & 7/7 & 95.2\% \\
\midrule
Pickup Net & Address(API)\textsuperscript{*} & 149/172 & 21/28 & 26/77 & 4/25 & 2/43 & 13/26 & 33.2\% \\
 & Lat/Lo(API)\textsuperscript{\dag} & 148/149 & 21/21 & 26/26 & 4/4 & 2/2 & 13/13 & 100\% \\
 & X username & 80/80 & 11/15 & 24/26 & 6/11 & 17/17 & 10/10 & 86.1\% \\
\midrule
Vehicle Rental & Address(API)\textsuperscript{*} & 157/187 & 29/29 & 47/53 & 19/35 & 11/34 & 20/23 & 72.4\% \\
 & License plate\textsuperscript{*} & 65/115 & 19/20 & 48/48 & 12/21 & 15/22 & 10/10 & 86.0\% \\
 & Home phone & 48/48 & 8/8 & 17/17 & 9/9 & 9/9 & 4/4 & 100\% \\
\midrule
Videoclub & Title(API)\textsuperscript{*} & 23/124 & 1/29 & 2/54 & 1/28 & 2/36 & 0/21 & 3.6\% \\
 & Type(API)\textsuperscript{\dag} & 13/23 & 1/1 & 1/2 & 1/1 & 1/2 & -- & 66.7\% \\
 & Genre(API)\textsuperscript{\dag} & 4/6 & 0/1 & 0/1 & 0/1 & -- & -- & 0\% \\
 & Actors(API)\textsuperscript{\dag} & 1/39 & 0/1 & 0/4 & 0/3 & 0/2 & -- & 0\% \\
 & Release(API) $>$ Rental\textsuperscript{\dag} & 19/23 & -- & 2/2 & 1/1 & 1/1 & -- & 100\% \\
\midrule
Football & Dates\textsuperscript{*} & 1063/1063 & 270/270 & 418/418 & 229/229 & 204/204 & 163/163 & 100\% \\
\midrule
Restaurant & Dates\textsuperscript{*} & 395/395 & 83/83 & 142/142 & 85/85 & 97/97 & 56/56 & 100\% \\
 & Phone number\textsuperscript{*} & 395/395 & 83/83 & 142/142 & 85/85 & 96/97 & 56/56 & 99.8\% \\
\midrule
All & Format/rule checks & 98.3\% & 99.3\% & 99.8\% & 96.4\% & 93.1\% & 99.5\% & 98.0\% \\
 & API checks & 66.2\% & 65.7\% & 51.9\% & 27.7\% & 14.3\% & 53.4\% & 44.5\% \\
\bottomrule
\end{tabular*}

\medskip
\begin{minipage}{\textwidth}
\raggedright
\footnotesize
State Machine has no value-level realism checks. ``--'' means that no value was available to check.
\textsuperscript{\dag} Checked only for the productions or addresses that the API finds.
\textsuperscript{*} Attributes that are mandatory in reality, for which empty values count as failures. For the remaining checks, empty values are skipped.
\end{minipage}
\vspace{-.2cm}
\end{table*}

%% file: tables/errors-judge.tex
\begin{table*}[ht]
\caption{Realism rate by LLM-as-a-judge for model instances generated through IL and CoT}
\label{tab:realism-judge}
\centering
\scriptsize
\setlength{\tabcolsep}{2pt}
\renewcommand{\arraystretch}{1.12}
\begin{tabular*}{\textwidth}{@{\extracolsep{\fill}}l*{7}{N{00.0\%}}@{}}
\toprule
Domain & \thd{IL} & \multicolumn{6}{c}{CoT category} \\
\cmidrule(l){3-8}
& & \thd{Baseline} & \thd{Complex} & \thd{Boundary} & \thd{Edge} & \thd{Over-constraint} & \thd{CoT} \\
\midrule
Address Book & 20.0\% & 16.7\% & 100\% & 50.0\% & 16.7\% & 16.7\% & 40.0\% \\
Bank & 90.0\% & 100\% & 83.3\% & 16.7\% & 0\% & 16.7\% & 43.3\% \\
Hotel Management & 90.0\% & 100\% & 83.3\% & 0\% & 0\% & 33.3\% & 43.3\% \\
Expenses & 33.3\% & 66.7\% & 100\% & 50.0\% & 0\% & 33.3\% & 50.0\% \\
Pickup Net & 83.3\% & 100\% & 100\% & 16.7\% & 0\% & 16.7\% & 46.7\% \\
State Machine & 73.3\% & 66.7\% & 66.7\% & 50.0\% & 50.0\% & 16.7\% & 50.0\% \\
Vehicle Rental & 36.7\% & 66.7\% & 16.7\% & 0\% & 0\% & 0\% & 16.7\% \\
Videoclub & 96.7\% & 100\% & 83.3\% & 50.0\% & 0\% & 16.7\% & 50.0\% \\
Football & 96.7\% & 100\% & 100\% & 100\% & 0\% & 50.0\% & 70.0\% \\
Restaurant & 96.7\% & 50.0\% & 16.7\% & 50.0\% & 0\% & 16.7\% & 26.7\% \\
\midrule
All & 71.7\% & 76.7\% & 75.0\% & 38.3\% & 6.7\% & 21.7\% & 43.7\% \\
\bottomrule
\end{tabular*}

\medskip
\begin{minipage}{\textwidth}
\raggedright
\footnotesize
Percentage of instances judged realistic by the LLM-as-a-judge. IL has 30 instances per domain; each CoT category has six instances per domain (30 per domain in total).
\end{minipage}
\vspace{-.2cm}
\end{table*}

%% file: tables/judge.tex
\begin{table*}[ht]
\caption{Categorization of confirmed problems ($n=67$)}
\label{tab:expert-judge}
\centering
\scriptsize
\setlength{\tabcolsep}{4pt}
\renewcommand{\arraystretch}{1.12}
\begin{tabularx}{\textwidth}{@{}>{\raggedright\arraybackslash}p{3.3cm}>{\raggedright\arraybackslash\hsize=0.8\hsize}X>{\raggedright\arraybackslash\hsize=1.2\hsize}X>{\raggedleft\arraybackslash}p{1.8cm}@{}}
\toprule
Problem & Description & Example & \multicolumn{1}{c}{Confirmed} \\
\midrule
Inconsistent values  & Logical contradiction (cross-attribute violations) & \mbox{\cotinst{bank}{1}{edge}{\texttt{Bank edge1}}} (Fig.~\ref{fig:edge-bank})\newline Bank in Monaco whose accounts have German and French IBANs & 48 (71.6\%) \\
Implausible values   & Valid but implausible values & \mbox{\cotinst{hotelmanagement}{2}{boundary}{\texttt{Hotel Management boundary2}}}\newline 250 free presidential suites with 12 beds each in one hotel & 11 (16.4\%) \\
Incorrect values     & Invalid values (domain data type violated) & \mbox{\cotinst{football}{3}{edge}{\texttt{Football edge3}}}\newline Match event at time $-1$; football shirt number 0 & 4 (6.0\%) \\
Duplicated instances & Distinct instances in one model represent the same entity & \mbox{\cotinst{restaurant}{5}{boundary}{\texttt{Restaurant boundary5}}}\newline Two cooks with the same name, date of birth, and phone number & 3 (4.5\%) \\
Incomplete model     & Information provided by one model is not enough for meaningful reasoning & \mbox{\ilinst{statemachine}{12}{\texttt{State Machine gen12}}}\newline State machines that contain no states or transitions & 1 (1.5\%) \\
\midrule
Total                & & & 67 (100\%) \\
\bottomrule
\end{tabularx}

\medskip
\begin{minipage}{\textwidth}
\raggedright
\footnotesize
Four authors jointly reviewed a stratified random sample and reached each decision by consensus. For unrealistic instances, each judge-reported problem was validated individually. Percentages are over confirmed problems (Yes or Partial). Each example names its domain and instance, i.e., the IL generation number (\texttt{gen}), or the CoT category and scenario number.
\end{minipage}
\vspace{-.2cm}
\end{table*}

%% file: tables/judge-2.tex
\begin{table}[t]
\caption{Expert confirmation rate of judge-reported problems per domain (unrealistic sample)}
\label{tab:expert-judge-by-domain}
\centering
\scriptsize
\setlength{\tabcolsep}{2pt}
\renewcommand{\arraystretch}{1.12}
\begin{tabular*}{\linewidth}{@{\extracolsep{\fill}}lN{00}N{00}N{00.0\%}@{}}
\toprule
Domain & \thd{Reported} & \thd{Confirmed} & \thd{Rate} \\
\midrule
Address Book     & 10 & 10 & 100\% \\
Bank             & 8  & 3  & 37.5\% \\
Hotel Management & 11 & 10 & 90.9\% \\
Expenses         & 10 & 10 & 100\% \\
Pickup Net       & 8  & 7  & 87.5\% \\
State Machine    & 8  & 6  & 75.0\% \\
Vehicle Rental   & 9  & 7  & 77.8\% \\
Videoclub        & 2  & 1  & 50.0\% \\
Football         & 11  & 8 & 72.7\% \\
Restaurant       & 5  & 5  & 100\% \\
\midrule
All              & 82 & 67 & 81.7\% \\
\bottomrule
\end{tabular*}

\medskip
\footnotesize
Confirmed problems (Yes or Partial) out of the problems reported by the judge. Higher rates indicate domains where the judge aligns better with expert assessment.
\vspace{-.2cm}
\end{table}


\begin{table*}[t]
\caption{Reasons for expert disagreement with the LLM-as-a-judge (unrealistic sample)}
\label{tab:expert-judge-disagreement}
\centering
\scriptsize
\setlength{\tabcolsep}{4pt}
\renewcommand{\arraystretch}{1.12}
\begin{tabularx}{\textwidth}{@{}>{\raggedright\arraybackslash}p{4.1cm}>{\raggedright\arraybackslash\hsize=0.8\hsize}X>{\raggedright\arraybackslash\hsize=1.2\hsize}X>{\raggedleft\arraybackslash}p{1.8cm}@{}}
\toprule
Disagreement reason & Description & Example & \multicolumn{1}{c}{Count} \\
\midrule
Unsupported claim        & The judge bases its verdict on facts that the model does not contain & \mbox{\ilinst{bank}{16}{\texttt{Bank gen16}}}\newline Account users judged unrelated to the owner because their names suggest another nationality & 10 (66.7\%) \\
Overly generic           & The judge turns a common convention into a universal rule & \mbox{\ilinst{bank}{27}{\texttt{Bank gen27}}}\newline An 18-year-old who can use four accounts owned by other adults rejected as implausible & 3 (20.0\%) \\
Domain paradigm mismatch & The judge applies the semantics of a different kind of system & \mbox{\cotinst{statemachine}{4}{boundary}{\texttt{State Machine boundary4}}}\newline Carnival gathering state with five transitions on the same timer rejected as non-deterministic, although a crowd can split up at once & 1 (6.7\%) \\
Assumed time unit        & The judge assumes a unit that the model leaves unspecified & \mbox{\cotinst{statemachine}{3}{boundary}{\texttt{State Machine boundary3}}}\newline Extreme timer value flagged without evidence that time is measured in seconds & 1 (6.7\%) \\
\midrule
Total                    & & & 15 (100\%) \\
\bottomrule
\end{tabularx}

\medskip
\begin{minipage}{\textwidth}
\raggedright
\footnotesize
Based on 15 judge-reported problems rejected by experts (Problem exists = No). Counts sum to the rejection set; some problems carry multiple disagreement cues. Percentages are over the rejected problems. Each example names its domain and instance, i.e., the IL generation number (\texttt{gen}), or the CoT category and scenario number.
\end{minipage}
\vspace{-.2cm}
\end{table*}

%% file: figs/venn-ocl-experts.tex
\begin{figure}[t]
\centering
\colorlet{venOCL}{blue!65!black}
\colorlet{venEXP}{orange!85!black}
\tikzset{
  vencount/.style={font=\fontsize{9}{10}\selectfont\bfseries, inner sep=1pt},
  venregion/.style={font=\fontsize{5}{5.8}\selectfont, text=black!60, inner sep=1pt, align=center},
  venset/.style={font=\fontsize{6.8}{7.8}\selectfont\bfseries, align=center, inner sep=1pt},
}
\newcommand{\vennpanel}[7]{%
  \begin{tikzpicture}
    \draw[black!60, fill=black!3, rounded corners=6pt] (-3.75,-1.75) rectangle (3.75,1.45);
    \node[font=\fontsize{6.8}{7.8}\selectfont, anchor=north west, align=left] at (-3.65,1.38)
      {\textbf{Judge} \textbf{#1}\\50 instances};
    \fill[venOCL!12] (-0.75,-0.3) circle (1.2cm);
    \fill[venEXP!15] (0.75,-0.3) circle (1.2cm);
    \begin{scope}
      \clip (-0.75,-0.3) circle (1.2cm);
      \fill[violet!20] (0.75,-0.3) circle (1.2cm);
    \end{scope}
    \draw[venOCL, line width=0.7pt] (-0.75,-0.3) circle (1.2cm);
    \draw[venEXP, line width=0.7pt] (0.75,-0.3) circle (1.2cm);
    \node[venset, text=venOCL, anchor=east] at (-2.0,-0.2) {OCL\\agree\\#1\\(#6)};
    \node[venset, text=venEXP, anchor=west] at (2.0,-0.2) {Experts\\agree\\#1\\(#7)};
    \node[vencount] at (-1.25,0.15) {#2};
    \node[venregion, anchor=north] at (-1.25,-0.1) {only\\OCL\\agree};
    \node[vencount] at (0,0.15) {#3};
    \node[venregion, anchor=north] at (0,-0.1) {both\\agree};
    \node[vencount] at (1.25,0.15) {#4};
    \node[venregion, anchor=north] at (1.25,-0.1) {only\\experts\\agree};
    \node[vencount, anchor=south east] at (3.62,-1.6) {#5};
    \node[venregion, anchor=south east] at (3.32,-1.56) {both disagree};
  \end{tikzpicture}%
}
\begin{subfigure}{\columnwidth}
\centering
\vennpanel{realistic}{1}{43}{5}{1}{44}{48}
\caption{Instances accepted as realistic}
\label{fig:venn-ocl-experts-r}
\end{subfigure}

\medskip
\begin{subfigure}{\columnwidth}
\centering
\vennpanel{unrealistic}{2}{37}{7}{4}{39}{44}
\caption{Instances rejected as unrealistic}
\label{fig:venn-ocl-experts-u}
\end{subfigure}
\caption{Added OCL invariants and experts on the 100 audited instances, split by the judge verdict. OCL considers an instance realistic if it satisfies all 48 added invariants and unrealistic if it violates at least one.}
\label{fig:venn-ocl-experts}
\end{figure}

%% file: tables/ocl-disagreement.tex
\begin{table*}[t]
\caption{Disagreements between the added OCL invariants and the expert labels}
\label{tab:ocl-disagreement}
\centering
\scriptsize
\setlength{\tabcolsep}{4pt}
\renewcommand{\arraystretch}{1.05}
\begin{tabularx}{\textwidth}{llr>{\raggedright\arraybackslash}X>{\raggedright\arraybackslash}p{0.19\textwidth}}
\toprule
Family & Reason & \# & Example & Instances \\
\midrule
\multicolumn{2}{l}{\textbf{Unrealistic, passes all invariants}} & \textbf{8} & & \\
& Knowledge outside the model & 2 & A drone charges right after an emergency landing; timers without a time unit & \cotinst{statemachine}{6}{baseline}{\texttt{State Machine baseline6}}, \cotinst{statemachine}{5}{complex}{\texttt{State Machine complex5}} \\
& Arbitrary threshold & 2 & One waiter for 120 guests; a credit-card payment in 1900 & \ilinst{restaurant}{12}{\texttt{Restaurant gen12}}, \cotinst{myexpenses}{6}{boundary}{\texttt{Expenses boundary6}} \\
& Visible only in names and free text & 2 & A refund stored as a positive amount; room service on a canceled booking & \ilinst{myexpenses}{25}{\texttt{Expenses gen25}}, \ilinst{hotelmanagement}{14}{\texttt{Hotel Management gen14}} \\
& Expressible but not encoded & 2 & An empty address; a bank in Monaco with German and French IBANs & \cotinst{pickupnet}{5}{boundary}{\texttt{Pickup Net boundary5}}, \cotinst{bank}{1}{edge}{\texttt{Bank edge1}} \\
\midrule
\multicolumn{2}{l}{\textbf{Realistic, fails an invariant}} & \textbf{7} & & \\
& Defect missed by the audit & 6 & IBAN prefixes of non-IBAN countries (\texttt{ValidIbanCountry}); an assigned shipment without a driver (\texttt{AssignedShouldHaveDriver}); two companies linked as coworkers (\texttt{CoworkerOnlyBetweenPersons}) & \ilinst{bank}{13}{\texttt{Bank gen13}}, \ilinst{bank}{16}{\texttt{Bank gen16}}, \ilinst{bank}{20}{\texttt{Bank gen20}}, \ilinst{bank}{21}{\texttt{Bank gen21}}, \ilinst{pickupnet}{18}{\texttt{Pickup Net gen18}}, \cotinst{addressbook}{1}{complex}{\texttt{Address Book complex1}} \\
& Invariant too strict & 1 & \texttt{SumOfExpenses} charges every room of a reservation for every night & \cotinst{hotelmanagement}{2}{baseline}{\texttt{Hotel Management baseline2}} \\
\midrule
\textbf{Total} & & \textbf{15} & & \\
\bottomrule
\end{tabularx}

\medskip
\begin{minipage}{\textwidth}
\raggedright
\footnotesize
Unrealistic and realistic refer to the expert labels (46 unrealistic, 54 realistic instances).
\end{minipage}
\vspace{-.2cm}
\end{table*}

%% file: sections/Threats.tex
\subsection{Threats to Validity}
\label{sec:threats}

Next, we discuss the four threats to validity of our study~\cite{WohlinRHCB12}. 

\subsubsection{Conclusion Validity}
One threat to conclusion validity is that LLMs show non-deterministic behavior~\cite{song2024good}. To derive relevant results, we generated 30 instances for each domain and strategy and report means with standard deviations, along with per-domain results rather than a single aggregate. In addition, each instance was judged only once, but the expert audit (RQ3.2) compensates for the variability of single verdicts.
Another conclusion validity threat concerns the sample of instances judged by experts; however, it is balanced by design across domains, strategies, and verdicts. Furthermore, the sample size was determined using a 95\% confidence interval and a 10\% error rate, as the study's main focus was not on maximizing the prediction of the judge's quality but on obtaining a general picture of the realism of the generated instances based on qualitative feedback. 

Moreover, to enhance the reproducibility of our results, we used the LLM guidelines checklist reported in~\cite{Baltes26-LLMGuidelines}. Thus, we explicate the exact model identifiers, snapshots, sampling parameters, and run dates (Table~\ref{tab:llm-config}), and we publish all prompt templates, the complete interaction traces of every generation, repair, and judge request, and our tools in our repository~\cite{repo}.

\subsubsection{Internal Validity}
Our work also addresses threats to internal validity. First, the prompts, the one-shot syntax example, the repair policy, and the choice of GPT-5.2 were defined by the authors and may be influenced by our previous experiences. To limit this threat, we used the same generic prompts, syntax example, and repair policy across all 10 domains and both prompting strategies. Thus, no prompt was tuned to a particular domain model. Additionally, the USE feedback loop is part of our approach. Thus, the presented RQ1 results describe both generation and repair. To separate both effects, we report the number of correction calls (Table~\ref{tab:llm-corrections}).

Second, the generator could reproduce memorized instances if the domain models or instances of them were part of its training data, i.e., data leakage. We consider this probability low, as part of the domain models were created by the authors for their lectures, and the models taken from ModelSet~\cite{LopezIC22-ModelSet} were modified for this study, i.e., translated into USE and adapted.
Moreover, to the best of our knowledge, no SOIL instances for these models were publicly available prior to the generation runs.

Third, defects in our tools could distort the results, e.g., by mis-parsing an instance or miscomputing a metric. The well-formedness and conformance are checked by a mature tool (USE~\cite{GogollaBR07}) rather than by our own code. In addition, we built \emph{llm-viewer}\footnote{\url{https://a-coman.github.io/llm-viewer/gpt_5_2/}}, a website that shows, for every instance, the class diagram, the SOIL script and its object diagram, the prompts and responses of each attempt, the computed metrics, and the judge verdict side by side. We used it to manually inspect the instances and validate the reported metrics, and it allows others to do the same.

Fourth, the LLM-as-a-judge may hallucinate or apply its own conventions. Therefore, we validated it with human experts on a stratified sample of 100 outputs (RQ3.2), and the judge comes from a different vendor than the generator and sees each instance in isolation (\secref{sec:eval-metrics}). However, the experts are the authors; they reviewed the judge explanations instead of rating the instances blindly, and they reached decisions by consensus, which may anchor the assessment towards the judge. Likewise, the 48 OCL invariants (RQ3.3) were written by the authors and may not encode the intended problems. To validate them, we applied the enriched models to the audited instances and checked whether they detected the same problems that the judge reported and the experts confirmed. The invariants were derived from the same sample and re-evaluated on the same sample. 
We therefore also ran them on the 440 non-audited instances, where they flag a similar share of judge-unrealistic instances (75.8\% versus 78\%), but these instances have no expert labels for comparison. Additionally, we did not regenerate instances with the enriched models; given the full conformance achieved in RQ1, we expect regenerated instances to satisfy the new invariants, but whether this improves their realism remains to be evaluated.

Moreover, IL and CoT differ beyond the prompting technique, as CoT also prescribes scenario categories. The differences between the strategies are therefore attributable to the strategy as a whole, which is why we report realism per each CoT category.

Finally, we did not compare our approach with a non-LLM generator. The closest candidate is the USE model validator~\cite{KuhlmannG12}, which translates UML/OCL models into the relational logic of Kodkod~\cite{TorlakJ07} and was used by Burgue\~{n}o et al.~\cite{BurguenoCCG19} to generate diverse instances. However, a fair comparison was not feasible. The validator requires an older USE version without data types, its default configuration is unsatisfiable for nine of our ten models, so bounds and value ranges would have to be tuned per model, and enumerating 30 distinct instances is prohibitively slow, as each new solution must be non-isomorphic to all previous ones. Moreover, its attribute values come from generic pools (e.g., \texttt{'string1'}), so realism can only be assessed after creating realistic value pools for each attribute, which is the manual effort our approach aims to avoid. Hence, we cannot quantify how much of the observed validity and structural diversity is specific to the LLM.

\subsubsection{Construct Validity}

In terms of construct validity, each of our metrics captures only part of the property it measures. High coverage does not imply diversity, and high value or structural diversity does not imply meaningful variation. To mitigate this threat, we do not rely on a single measure but combine complementary ones, i.e., coverage within and across instances, exact and Levenshtein distances for values, Shannon evenness for enumerations and Booleans, and GED for structure, and we illustrate them with concrete instances.

Another threat concerns realism, a contextual notion without a ground truth. Our domain-specific checks cover only selected attributes and combine plausibility checks (formats, cross-attribute comparisons) with existence checks (Geoapify, OMDb). Moreover, the checks and their thresholds, such as regular expressions or minimum geocoding confidence, were defined by the authors and may be too lenient or too strict for certain values. In addition, Geoapify (based on OpenStreetMap) and OMDb (community-driven) are incomplete, so existence checks may miss real values or accept invented ones that resemble them. The results of these checks are therefore approximations. However, this affects our assessment of realism only partially, since, following our distinction between realistic and real values (\secref{sec:introduction}), we do not consider a well-formed but non-existent value unrealistic. To further mitigate these threats, we report both kinds of checks separately and combine four evaluation methods: domain checks assess individual values or the relationship between two values; the judge assesses whole instances; the experts assess the judge's feedback; and the OCL invariants recheck the confirmed problems on the instances. Finally, where these methods disagree, we analyzed the cases manually.

\subsubsection{External Validity}

The fact that we have used a dataset covering ten domains, with models of different sizes, language features, and abstraction levels, gives us some confidence that our results may be generalized to other cases. However, the models are relatively small and mostly stem from educational contexts, and we cannot generalize our results to large domain models and/or instances, as LLM performance is known to degrade and token limits apply when their size grows. We also have to be careful about generalization in specialized niche domains with scarce, openly available data, which may not be well supported by existing LLMs.

Furthermore, we only used one generator (GPT-5.2) and one judge (Gemini 3.1 Pro), and did not include an open-weight model. Thus, we cannot generalize to other LLMs. To mitigate this threat in the future, another LLM can be evaluated by changing only the model identifier and re-running the published pipeline (\secref{sec:approach:tools}), and our preliminary experiments with other LLMs indicate that the approach is not tied to GPT-5.2, although we did not evaluate them with the full methodology.

%% file: sections/RelatedWork.tex
\section{Related Work}
\label{sec:relatedwork}

The automated instantiation of models is a long-standing research topic. In the following, we first discuss generators (\secref{sec:related:generators}) that aim to produce valid and at-large instances, then address the realism of generated models (\secref{sec:related:realism}), and finally discuss LLM-based approaches dedicated to modeling and instance generation (\secref{sec:related:llm-generators}), which are the closest to our work.

\subsection{Model Instantiation without LLMs}
\label{sec:related:generators}
Burgue\~{n}o et al.~\cite{BurguenoCCG19} introduce an approach that focuses on the generation of \textit{diverse valid instances} of UML Class Diagrams in the USE notation. Their approach uses model finders~\cite{TorlakJ07}, classifying terms~\cite{HilkenGBV18}, and constraint-strengthening techniques~\cite{ClarisoC18}. Their approach is steered toward generating diverse model instances, i.e., covering all extreme/edge cases of the corresponding model at the higher abstraction level. Compared to our work, this work solely focuses on the structural diversity of valid instances without considering realism. The work introduces the notion of classifying terms as a means of deriving partitions of the search space that classify diverse instantiations (analogous to the equivalence classes of a schema). The USE model validator~\cite{KuhlmannG12}, which underlies this line of work, translates UML/OCL models into relational logic and uses the Kodkod model finder~\cite{TorlakJ07} to compute valid object diagrams. Its attribute values, however, come from generic, configurable value pools (e.g., \texttt{'string1'}) rather than from the domain. Graph solvers such as the VIATRA Solver~\cite{SemerathNV18,Oszkar2019} generate large, consistent, domain-specific models, but they likewise focus on the models' graph structure.

Logic-based analyzers have also been used to explore the possible instances of a model. In~\cite{BenevidesGBA10}, OntoUML models are transformed into logical Alloy~\cite{JacksonAlloy.2006} specifications that an analyzer can use to extensively explore the solution space defined by this specification. By providing a visual simulation of possible instances, the tool enables the identification of patterns and anti-patterns in OntoUML modeling~\cite{SalesG15}. A similar work demonstrates Alloy's analysis capabilities in the context of UML modeling~\cite{AnastasakisBGR07}. In relation to our work, these analyses focus on exploring the solution space of types (in particular, well-formedness) in the model, rather than on the instance-level realism of model instance property values, or the realism of the model as a whole.

Other generators focus more on efficiently producing large models rather than satisfying complex constraints with respect to diversity and realism. An approach to automated metamodel instantiation based on graph grammars has been presented in~\cite{EhrigKT09}. The approach is tested using UML State Diagrams and demonstrates its feasibility for creating syntactically valid metamodel instances. Compared to our work, no attention is given to the diverse and realistic instantiation of attribute values. 
A rule-based, configurable approach to automating model generation, with an emphasis on the diversity of generated models, is proposed in~\cite{NassarKKT20}. The approach aims to achieve scalability by using a prescriptive, rule-based configuration during model instance generation and to ensure diversity in the created instances themselves. A similar motivation underlies the approach presented in~\cite{MougenotDBS09}. The authors use the Boltzmann random sampling theory~\cite{DuchonFLS04} to generate instances from a metamodel that has been transformed into a tree structure.

The EMF random instantiator~\cite{emf-instantiator}, as the name indicates, generates random instances for EMF (Ecore) metamodels using a deterministic, seed-specific random generator. The default configuration uses uniform probability distributions for meta-classes and structural features. A generation configuration holds information such as the meta-classes involved in generation and the probability distributions for generating instances. Compared to our approach, property values are defined according to the statistical distributions in the configuration, aiming for structural comprehensiveness rather than diverse and realistic instance values. Similar approaches have been proposed by Scheidgen~\cite{Scheidgen15} and Popoola et al.~\cite{PopoolaKR16}, who propose a DSL for defining the generator configuration, which is then used to instantiate Ecore metamodel instances.

\subsection{Realism}
\label{sec:related:realism}
None of the generators discussed above aims at realistic models. Sz\'{a}rnyas et al.~\cite{SzarnyasKSV16} propose a collection of metrics to assess the characteristics of model instances. They experimented with 83 models from six different domains and conclude, among others, that \textit{``Relying only upon metamodel-level information is clearly insufficient, real instance models of human engineers are required to characterize the domain''} and that future work is required to develop \textit{``domain-specific model generators that are capable of synthesizing realistic models.''} Varr\'{o} et al.~\cite{VarroSSH18} later characterized model generators along four dimensions: consistency, diversity, scalability, and realism. In this line of work, realism is usually structural, i.e., the graph shape of the generated models resembles that of human-made models, while attribute values are disregarded~\cite{PoloGambinRH26}.

Following this observation, an approach to model generation that uses other models as valid samples is proposed in~\cite{LopezC23}. The approach decomposes a model into atomic edit operations, similar to, e.g.,~\cite{EhrigKT09,PietschYK11,NassarKKT20}, which are then used to train a Graph Neural Network and a Recurrent Neural Network. The generator then samples the edit operations to refine partial models, thereby generating new structurally valid instances that mimic the set of realistic models. The approach requires a corpus of human-made models and is not focused on generating diverse, realistic instance models with diverse and realistic property values.

\subsection{LLM-based Model Instantiation}
\label{sec:related:llm-generators}
Regarding LLMs, related approaches focus on generating valid domain models from natural language descriptions~\cite{AliRB24,CamaraTBV23,SilvaMCKP24}.
For example, Chen et al.~\cite{ChenYCLMV23} propose an LLM-based approach to domain modeling. They experiment with zero-shot, few-shot, and chain-of-thought prompting approaches that use textual domain descriptions as input to GPT-3.5 and GPT-4. The LLMs are then tasked with generating UML Class Diagram domain models represented in a textual EBNF notation. 

At the instance level, KGFiller~\cite{CiattoAMO24} uses an initial schema composed of classes, relationships, and properties, and queries an LLM to generate instances of that schema with instantiated classes, relationships, and properties. Compared to our work, KGFiller uses the LLM as an oracle of domain knowledge, i.e., the generated instances are meant to be facts (e.g., real meals and their ingredients) that enrich the ontology, rather than plausible scenarios that conform to a constrained domain model. \manuel{so the instances are added to the ontology as an ontology contain both? Maybe we should clarify this. Dominik, please have a look here!}

Closest to our work are recent approaches that use LLMs to generate instances of a given metamodel. Pan et al.~\cite{PanPZWK25} observe that LLMs struggle to produce valid XMI directly. Therefore, given an Ecore metamodel and a natural-language specification of a concrete system, the LLM produces a simplified, format-independent conceptual instance model that a compiler then converts to XMI. Their goal is fidelity to the specification (the prompt explicitly forbids inferring missing details), measured as precision and recall against reference instances. OCL constraints, diversity, and realism are not addressed.

Polo-Gamb\'{i}n et al.~\cite{PoloGambinRH26} reformulate instance generation as code synthesis. The metamodel and its well-formedness constraints are encoded as Pydantic classes and validators. The LLM writes Python code to build the instance, and validation errors are fed back to it. By using an iterative approach, the LLM generates consistent models of more than 2.000 elements, but diversity drops significantly as model size grows, and structural realism, measured against human-made models, depends on the domain. Their notion of realism is structural, and they consider realistic attribute values ``an orthogonal challenge''~\cite{PoloGambinRH26}.

Lastly, AMIGO~\cite{HummelRHCKR26} lets an LLM agent build an instance through schema-typed queries, manage, and validate tools that a metamodel-aware server exposes via the Model Context Protocol (MCP). The server applies each operation deterministically, persists conforming XMI, and reports OCL violations, which the agent then repairs. For the Palladio Component Model, 596 of 600 generated instances satisfy all OCL constraints, and their structure is compared with expert-made reference models. Like Pan et al., AMIGO translates a description of an existing system (e.g., architecture and sequence diagrams, deployment configurations) into a model of that system, and its agent is instructed never to invent elements, names, or values.

Compared to our work, Pan et al.\ and AMIGO take the names and values of their instances from the description of an existing system, while Polo-Gamb\'{i}n et al.\ generate only the model structure. LIMoGen generates both the structure and the values, and its only input is the type-level specification, i.e., a class diagram, possibly with OCL constraints. Hence, the LLM has to generate plausible content and vary it across instances. The LLM writes the instances directly in a textual DSL (SOIL), which USE validates, and we evaluate the realism and diversity of the attribute values, their combinations, and each instance as a whole, rather than relying on graph statistics or fidelity to a reference model.

\textbf{Synopsis.}
The overview of the most related works shows that the automated instantiation of models is a long-standing research endeavor in the community. Many approaches based on grammars, constraint solvers, configurations, and machine learning have been proposed, but most focus on the structural properties of the instances, and instance-level properties either follow an a priori defined distribution or are sampled randomly.

Recent LLM-based generators~\cite{PanPZWK25,PoloGambinRH26,HummelRHCKR26} produce consistent instances that are either faithful to a given specification or structurally realistic, but they leave the realism and diversity of attribute values aside. In contrast, LIMoGen focuses on the automated generation of instances whose values, value combinations, and structures are realistic and diverse, and provides a methodology for evaluating this realism with domain checks, an LLM-as-a-judge, and expert review. To the best of our knowledge, this combination has not been addressed so far.

%% file: sections/Conclusion.tex
\section{Conclusion and Future Work}
\label{sec:conclusion}


In this paper, we presented LIMoGen, an approach that combines LLMs with the USE model validation tool to instantiate UML class diagrams as realistic and diverse object diagrams. Our evaluation of two prompting strategies, Instruction Learning (IL) and Chain-of-Thought (CoT), across ten domain models shows that, with USE feedback, all generated instances are well-formed and conformant, with more than 90\% produced on the first attempt. In addition, the generated instances collectively cover the domain models and are diverse in their values and structures. 

Regarding realism, the main challenge is not individual values but their consistency, since most remaining problems stem from inconsistencies among values. An LLM-as-a-judge largely detects them, as the experts confirm 82\% of the problems it reports, although it also shows some biases and limitations, such as inferring nationality from names or turning common conventions into universal rules. Overall, the two strategies complement each other: IL produces mostly realistic instances (71.7\% judged realistic) but structurally similar ones (0.26 Graph Edit Distance diversity), while CoT explores unusual cases through its boundary and edge scenarios, increasing structural diversity (0.41) at the cost of realism (43.7\%).

Beyond instance generation, our results show that evaluating the generated instances can also help improve domain models. Each evaluation dimension contributes to this: coverage points to classes that could be declared abstract; conformance checking of over-constrained scenarios to multiplicities that are too strict; and the realism assessment to implicit domain knowledge, which we encoded as additional OCL invariants that even exposed defects both the judge and the experts overlooked.

In future work, we plan to address larger models and instances, e.g., by combining code synthesis for the structural skeleton~\cite{PoloGambinRH26} with LIMoGen for realistic values. We also plan to bring our approach into an agentic setting in which, following AMIGO~\cite{HummelRHCKR26}, USE, the judge, and the realism checks are exposed as tools via an MCP server, so that agents can verify validity and realism while building an instance. Furthermore, we will evaluate other LLMs, including open-weight models, and study how LIMoGen handles other modeling languages, e.g., SysML. Finally, we will investigate whether the generated instances can be used to fine-tune LLMs for modeling tasks, and how to integrate our approach into AI-assisted modeling editors.

%% file: sections/Appendix.tex
\section{Computing GED bounds}\label{app:ged}

Computing exact GED is NP-hard~\cite{BlumenthalBGBB20}, so for every pair of
instances in RQ2.3 we compute a certified interval $[L,U]$ with $L\le\mathrm{GED}\le U$.
The upper bound $U$ is the cost of a concrete edit path, obtained by the
bipartite assignment (BP) of Riesen and Bunke~\cite{RiesenBunke09}, a single
IPFP descent~\cite{BougleuxBCFGV17} initialized from that assignment, and
swap-based local search (REFINE)~\cite{ZengTWFZ09,BoriaBBG20}.
The lower bound $L$ is BRANCH~\cite{BlumenthalGamper18}.
These heuristics are not symmetric in their two arguments, so we run both
orientations and keep the smaller $U$ and the larger $L$.
A pair is exact when $L=U$.
This combination is among the empirically tightest bound pairs in the survey
of Blumenthal et~al.~\cite{BlumenthalBGBB20}.

%% file: main-sosym.bbl

\begin{thebibliography}{72}
\ifx \bisbn   \undefined \def \bisbn  #1{ISBN #1}\fi
\ifx \binits  \undefined \def \binits#1{#1}\fi
\ifx \bauthor  \undefined \def \bauthor#1{#1}\fi
\ifx \batitle  \undefined \def \batitle#1{#1}\fi
\ifx \bjtitle  \undefined \def \bjtitle#1{#1}\fi
\ifx \bvolume  \undefined \def \bvolume#1{\textbf{#1}}\fi
\ifx \byear  \undefined \def \byear#1{#1}\fi
\ifx \bissue  \undefined \def \bissue#1{#1}\fi
\ifx \bfpage  \undefined \def \bfpage#1{#1}\fi
\ifx \blpage  \undefined \def \blpage #1{#1}\fi
\ifx \burl  \undefined \def \burl#1{\textsf{#1}}\fi
\ifx \doiurl  \undefined \def \doiurl#1{\url{https://doi.org/#1}}\fi
\ifx \betal  \undefined \def \betal{\textit{et al.}}\fi
\ifx \binstitute  \undefined \def \binstitute#1{#1}\fi
\ifx \binstitutionaled  \undefined \def \binstitutionaled#1{#1}\fi
\ifx \bctitle  \undefined \def \bctitle#1{#1}\fi
\ifx \beditor  \undefined \def \beditor#1{#1}\fi
\ifx \bpublisher  \undefined \def \bpublisher#1{#1}\fi
\ifx \bbtitle  \undefined \def \bbtitle#1{#1}\fi
\ifx \bedition  \undefined \def \bedition#1{#1}\fi
\ifx \bseriesno  \undefined \def \bseriesno#1{#1}\fi
\ifx \blocation  \undefined \def \blocation#1{#1}\fi
\ifx \bsertitle  \undefined \def \bsertitle#1{#1}\fi
\ifx \bsnm \undefined \def \bsnm#1{#1}\fi
\ifx \bsuffix \undefined \def \bsuffix#1{#1}\fi
\ifx \bparticle \undefined \def \bparticle#1{#1}\fi
\ifx \barticle \undefined \def \barticle#1{#1}\fi
\bibcommenthead
\ifx \bconfdate \undefined \def \bconfdate #1{#1}\fi
\ifx \botherref \undefined \def \botherref #1{#1}\fi
\ifx \url \undefined \def \url#1{\textsf{#1}}\fi
\ifx \bchapter \undefined \def \bchapter#1{#1}\fi
\ifx \bbook \undefined \def \bbook#1{#1}\fi
\ifx \bcomment \undefined \def \bcomment#1{#1}\fi
\ifx \oauthor \undefined \def \oauthor#1{#1}\fi
\ifx \citeauthoryear \undefined \def \citeauthoryear#1{#1}\fi
\ifx \endbibitem  \undefined \def \endbibitem {}\fi
\ifx \bconflocation  \undefined \def \bconflocation#1{#1}\fi
\ifx \arxivurl  \undefined \def \arxivurl#1{\textsf{#1}}\fi
\csname PreBibitemsHook\endcsname

\bibitem[\protect\citeauthoryear{Belzner et~al.}{2023}]{belzner2023large}
\begin{bchapter}
\bauthor{\bsnm{Belzner}, \binits{L.}},
\bauthor{\bsnm{Gabor}, \binits{T.}},
\bauthor{\bsnm{Wirsing}, \binits{M.}}:
\bctitle{Large language model assisted software engineering: prospects,
  challenges, and a case study}.
In: \bbtitle{International Conference on Bridging the Gap Between AI and
  Reality}.
\bsertitle{Lecture Notes in Computer Science},
pp. \bfpage{355}--\blpage{374}.
\bpublisher{Springer},
\blocation{Cham}
(\byear{2023}).
\doiurl{10.1007/978-3-031-46002-9\_23}
\end{bchapter}
\endbibitem

\bibitem[\protect\citeauthoryear{Hou et~al.}{2024}]{hou2023large}
\begin{barticle}
\bauthor{\bsnm{Hou}, \binits{X.}},
\bauthor{\bsnm{Zhao}, \binits{Y.}},
\bauthor{\bsnm{Liu}, \binits{Y.}},
\bauthor{\bsnm{Yang}, \binits{Z.}},
\bauthor{\bsnm{Wang}, \binits{K.}},
\bauthor{\bsnm{Li}, \binits{L.}},
\bauthor{\bsnm{Luo}, \binits{X.}},
\bauthor{\bsnm{Lo}, \binits{D.}},
\bauthor{\bsnm{Grundy}, \binits{J.}},
\bauthor{\bsnm{Wang}, \binits{H.}}:
\batitle{Large language models for software engineering: A systematic
  literature review}.
\bjtitle{ACM Transactions on Software Engineering and Methodology}
\bvolume{33}(\bissue{8}),
\bfpage{1}--\blpage{79}
(\byear{2024})
\doiurl{10.1145/3695988}
\end{barticle}
\endbibitem

\bibitem[\protect\citeauthoryear{Bork et~al.}{2023}]{CMAI-SMS.23}
\begin{botherref}
\oauthor{\bsnm{Bork}, \binits{D.}},
\oauthor{\bsnm{Ali}, \binits{S.J.}},
\oauthor{\bsnm{Roelens}, \binits{B.}}:
Conceptual modeling and artificial intelligence: {A} systematic mapping study.
CoRR
\textbf{abs/2303.06758}
(2023)
\doiurl{10.48550/ARXIV.2303.06758}
\end{botherref}
\endbibitem

\bibitem[\protect\citeauthoryear{Rädler et~al.}{2025}]{Raedler.MDE-AI.24}
\begin{barticle}
\bauthor{\bsnm{Rädler}, \binits{S.}},
\bauthor{\bsnm{Berardinelli}, \binits{L.}},
\bauthor{\bsnm{Winter}, \binits{K.}},
\bauthor{\bsnm{Rahimi}, \binits{A.}},
\bauthor{\bsnm{Rinderle-Ma}, \binits{S.}}:
\batitle{{Bridging MDE and AI: a systematic review of domain-specific languages
  and model-driven practices in AI software systems engineering}}.
\bjtitle{Software and Systems Modeling}
\bvolume{24}(\bissue{2}),
\bfpage{445}--\blpage{469}
(\byear{2025})
\doiurl{10.1007/s10270-024-01211-y}
\end{barticle}
\endbibitem

\bibitem[\protect\citeauthoryear{Chaaben et~al.}{2023}]{ChaabenBS23}
\begin{bchapter}
\bauthor{\bsnm{Chaaben}, \binits{M.B.}},
\bauthor{\bsnm{Burgue{\~{n}}o}, \binits{L.}},
\bauthor{\bsnm{Sahraoui}, \binits{H.A.}}:
\bctitle{Towards using few-shot prompt learning for automating model
  completion}.
In: \bbtitle{45th {IEEE/ACM} International Conference on Software Engineering:
  New Ideas and Emerging Results, NIER@ICSE, Melbourne, Australia, May 14-20,
  2023},
pp. \bfpage{7}--\blpage{12}.
\bpublisher{{IEEE}},
\blocation{Piscataway, NJ}
(\byear{2023}).
\doiurl{10.1109/ICSE-NIER58687.2023.00008}
\end{bchapter}
\endbibitem

\bibitem[\protect\citeauthoryear{Chen et~al.}{2023}]{ChenYCLMV23}
\begin{bchapter}
\bauthor{\bsnm{Chen}, \binits{K.}},
\bauthor{\bsnm{Yang}, \binits{Y.}},
\bauthor{\bsnm{Chen}, \binits{B.}},
\bauthor{\bsnm{{Hern{\'{a}}ndez L{\'{o}}pez}}, \binits{J.A.}},
\bauthor{\bsnm{Mussbacher}, \binits{G.}},
\bauthor{\bsnm{Varr{\'{o}}}, \binits{D.}}:
\bctitle{Automated domain modeling with large language models: {A} comparative
  study}.
In: \bbtitle{26th {ACM/IEEE} International Conference on Model Driven
  Engineering Languages and Systems, {MODELS} 2023, V{\"{a}}ster{\aa}s, Sweden,
  October 1-6, 2023},
pp. \bfpage{162}--\blpage{172}.
\bpublisher{{IEEE}},
\blocation{Piscataway, NJ}
(\byear{2023}).
\doiurl{10.1109/MODELS58315.2023.00037}
\end{bchapter}
\endbibitem

\bibitem[\protect\citeauthoryear{He et~al.}{2026}]{He.2026}
\begin{bchapter}
\bauthor{\bsnm{He}, \binits{X.}},
\bauthor{\bsnm{Shen}, \binits{J.}},
\bauthor{\bsnm{Chen}, \binits{R.}}:
\bctitle{{LLM}-enhanced stochastic generation of class diagram datasets}.
In: \bbtitle{29th {ACM/IEEE} International Conference on Model Driven
  Engineering Languages and Systems, {MODELS} 2026, M{\'{a}}laga, Spain,
  October 4-9, 2026},
pp. \bfpage{212}--\blpage{223}.
\bpublisher{{ACM}},
\blocation{New York, NY}
(\byear{2026}).
\doiurl{10.1145/3822455.3830323}
\end{bchapter}
\endbibitem

\bibitem[\protect\citeauthoryear{Burgue{\~{n}}o et~al.}{2025}]{BurguenoRSW25}
\begin{barticle}
\bauthor{\bsnm{Burgue{\~{n}}o}, \binits{L.}},
\bauthor{\bsnm{{Di Ruscio}}, \binits{D.}},
\bauthor{\bsnm{Sahraoui}, \binits{H.A.}},
\bauthor{\bsnm{Wimmer}, \binits{M.}}:
\batitle{Automation in model-driven engineering: {A} look back, and ahead}.
\bjtitle{{ACM} Trans. Softw. Eng. Methodol.}
\bvolume{34}(\bissue{5}),
\bfpage{122}
(\byear{2025})
\doiurl{10.1145/3712008}
\end{barticle}
\endbibitem

\bibitem[\protect\citeauthoryear{Burgue{\~{n}}o et~al.}{2019}]{BurguenoCCG19}
\begin{bchapter}
\bauthor{\bsnm{Burgue{\~{n}}o}, \binits{L.}},
\bauthor{\bsnm{Cabot}, \binits{J.}},
\bauthor{\bsnm{Claris{\'{o}}}, \binits{R.}},
\bauthor{\bsnm{Gogolla}, \binits{M.}}:
\bctitle{A systematic approach to generate diverse instantiations for
  conceptual schemas}.
In: \beditor{\bsnm{Laender}, \binits{A.H.F.}},
\beditor{\bsnm{Pernici}, \binits{B.}},
\beditor{\bsnm{Lim}, \binits{E.}},
\beditor{\bsnm{Oliveira}, \binits{J.P.M.}} (eds.)
\bbtitle{Conceptual Modeling - 38th International Conference, {ER} 2019,
  Salvador, Brazil, November 4-7, 2019, Proceedings}.
\bsertitle{Lecture Notes in Computer Science},
vol. \bseriesno{11788},
pp. \bfpage{513}--\blpage{521}.
\bpublisher{Springer},
\blocation{Cham}
(\byear{2019}).
\doiurl{10.1007/978-3-030-33223-5\_42}
\end{bchapter}
\endbibitem

\bibitem[\protect\citeauthoryear{Cabot et~al.}{2007}]{CabotCR07}
\begin{bchapter}
\bauthor{\bsnm{Cabot}, \binits{J.}},
\bauthor{\bsnm{Claris{\'{o}}}, \binits{R.}},
\bauthor{\bsnm{Riera}, \binits{D.}}:
\bctitle{{UMLtoCSP}: a tool for the formal verification of {UML/OCL} models
  using constraint programming}.
In: \bbtitle{22nd {IEEE/ACM} International Conference on Automated Software
  Engineering ({ASE} 2007)},
pp. \bfpage{547}--\blpage{548}.
\bpublisher{{ACM}}, \blocation{???}
(\byear{2007}).
\doiurl{10.1145/1321631.1321737}
\end{bchapter}
\endbibitem

\bibitem[\protect\citeauthoryear{Kuhlmann and Gogolla}{2012}]{KuhlmannG12}
\begin{bchapter}
\bauthor{\bsnm{Kuhlmann}, \binits{M.}},
\bauthor{\bsnm{Gogolla}, \binits{M.}}:
\bctitle{From {UML} and {OCL} to relational logic and back}.
In: \beditor{\bsnm{France}, \binits{R.B.}},
\beditor{\bsnm{Kazmeier}, \binits{J.}},
\beditor{\bsnm{Breu}, \binits{R.}},
\beditor{\bsnm{Atkinson}, \binits{C.}} (eds.)
\bbtitle{Model Driven Engineering Languages and Systems - 15th International
  Conference, {MODELS} 2012, Innsbruck, Austria, September 30-October 5, 2012.
  Proceedings}.
\bsertitle{Lecture Notes in Computer Science},
vol. \bseriesno{7590},
pp. \bfpage{415}--\blpage{431}.
\bpublisher{Springer},
\blocation{Berlin, Heidelberg}
(\byear{2012}).
\doiurl{10.1007/978-3-642-33666-9\_27}
\end{bchapter}
\endbibitem

\bibitem[\protect\citeauthoryear{Semer{\'{a}}th et~al.}{2018}]{SemerathNV18}
\begin{bchapter}
\bauthor{\bsnm{Semer{\'{a}}th}, \binits{O.}},
\bauthor{\bsnm{Nagy}, \binits{A.S.}},
\bauthor{\bsnm{Varr{\'{o}}}, \binits{D.}}:
\bctitle{A graph solver for the automated generation of consistent
  domain-specific models}.
In: \bbtitle{Proceedings of the 40th International Conference on Software
  Engineering, {ICSE} 2018, Gothenburg, Sweden, May 27 - June 03, 2018},
pp. \bfpage{969}--\blpage{980}.
\bpublisher{{ACM}},
\blocation{New York, NY}
(\byear{2018}).
\doiurl{10.1145/3180155.3180186}
\end{bchapter}
\endbibitem

\bibitem[\protect\citeauthoryear{Semeráth et~al.}{2019}]{Oszkar2019}
\begin{bchapter}
\bauthor{\bsnm{Semeráth}, \binits{O.}},
\bauthor{\bsnm{Babikian}, \binits{A.A.}},
\bauthor{\bsnm{Pilarski}, \binits{S.}},
\bauthor{\bsnm{Varró}, \binits{D.}}:
\bctitle{{VIATRA} solver: A framework for the automated generation of
  consistent domain-specific models}.
In: \bbtitle{2019 IEEE/ACM 41st International Conference on Software
  Engineering: Companion Proceedings (ICSE-Companion)},
pp. \bfpage{43}--\blpage{46}
(\byear{2019}).
\doiurl{10.1109/ICSE-Companion.2019.00034}
\end{bchapter}
\endbibitem

\bibitem[\protect\citeauthoryear{Ehrig et~al.}{2009}]{EhrigKT09}
\begin{barticle}
\bauthor{\bsnm{Ehrig}, \binits{K.}},
\bauthor{\bsnm{K{\"{u}}ster}, \binits{J.M.}},
\bauthor{\bsnm{Taentzer}, \binits{G.}}:
\batitle{Generating instance models from meta models}.
\bjtitle{Softw. Syst. Model.}
\bvolume{8}(\bissue{4}),
\bfpage{479}--\blpage{500}
(\byear{2009})
\doiurl{10.1007/S10270-008-0095-Y}
\end{barticle}
\endbibitem

\bibitem[\protect\citeauthoryear{Taentzer}{2012}]{Taentzer12}
\begin{botherref}
\oauthor{\bsnm{Taentzer}, \binits{G.}}:
Instance generation from type graphs with arbitrary multiplicities.
Electron. Commun. Eur. Assoc. Softw. Sci. Technol.
\textbf{47}
(2012)
\doiurl{10.14279/tuj.eceasst.47.727}
\end{botherref}
\endbibitem

\bibitem[\protect\citeauthoryear{Radke et~al.}{2018}]{RadkeABHT18}
\begin{barticle}
\bauthor{\bsnm{Radke}, \binits{H.}},
\bauthor{\bsnm{Arendt}, \binits{T.}},
\bauthor{\bsnm{Becker}, \binits{J.S.}},
\bauthor{\bsnm{Habel}, \binits{A.}},
\bauthor{\bsnm{Taentzer}, \binits{G.}}:
\batitle{Translating essential {OCL} invariants to nested graph constraints for
  generating instances of meta-models}.
\bjtitle{Sci. Comput. Program.}
\bvolume{152},
\bfpage{38}--\blpage{62}
(\byear{2018})
\doiurl{10.1016/j.scico.2017.08.006}
\end{barticle}
\endbibitem

\bibitem[\protect\citeauthoryear{Nassar et~al.}{2020}]{NassarKKT20}
\begin{bchapter}
\bauthor{\bsnm{Nassar}, \binits{N.}},
\bauthor{\bsnm{Kosiol}, \binits{J.}},
\bauthor{\bsnm{Kehrer}, \binits{T.}},
\bauthor{\bsnm{Taentzer}, \binits{G.}}:
\bctitle{Generating large {EMF} models efficiently - {A} rule-based,
  configurable approach}.
In: \bbtitle{Fundamental Approaches to Software Engineering - 23rd
  International Conference, {FASE} 2020}.
\bsertitle{Lecture Notes in Computer Science},
vol. \bseriesno{12076},
pp. \bfpage{224}--\blpage{244}.
\bpublisher{Springer},
\blocation{Cham}
(\byear{2020}).
\doiurl{10.1007/978-3-030-45234-6\_11}
\end{bchapter}
\endbibitem

\bibitem[\protect\citeauthoryear{Ali et~al.}{2013}]{AliIAB13}
\begin{barticle}
\bauthor{\bsnm{Ali}, \binits{S.}},
\bauthor{\bsnm{Iqbal}, \binits{M.Z.}},
\bauthor{\bsnm{Arcuri}, \binits{A.}},
\bauthor{\bsnm{Briand}, \binits{L.C.}}:
\batitle{Generating test data from {OCL} constraints with search techniques}.
\bjtitle{{IEEE} Trans. Software Eng.}
\bvolume{39}(\bissue{10}),
\bfpage{1376}--\blpage{1402}
(\byear{2013})
\doiurl{10.1109/TSE.2013.17}
\end{barticle}
\endbibitem

\bibitem[\protect\citeauthoryear{Soltana et~al.}{2020}]{SoltanaSB20}
\begin{barticle}
\bauthor{\bsnm{Soltana}, \binits{G.}},
\bauthor{\bsnm{Sabetzadeh}, \binits{M.}},
\bauthor{\bsnm{Briand}, \binits{L.C.}}:
\batitle{Practical constraint solving for generating system test data}.
\bjtitle{{ACM} Trans. Softw. Eng. Methodol.}
\bvolume{29}(\bissue{2}),
\bfpage{11}--\blpage{11148}
(\byear{2020})
\doiurl{10.1145/3381032}
\end{barticle}
\endbibitem

\bibitem[\protect\citeauthoryear{Karimi et~al.}{2024}]{KARIMI2024}
\begin{barticle}
\bauthor{\bsnm{Karimi}, \binits{M.}},
\bauthor{\bsnm{Kolahdouz-Rahimi}, \binits{S.}},
\bauthor{\bsnm{Troya}, \binits{J.}}:
\batitle{Yekta: A low-code framework for automated test models generation}.
\bjtitle{SoftwareX}
\bvolume{27},
\bfpage{101850}
(\byear{2024})
\doiurl{10.1016/j.softx.2024.101850}
\end{barticle}
\endbibitem

\bibitem[\protect\citeauthoryear{Semer{\'{a}}th et~al.}{2020}]{SemerathFBV20}
\begin{barticle}
\bauthor{\bsnm{Semer{\'{a}}th}, \binits{O.}},
\bauthor{\bsnm{Farkas}, \binits{R.}},
\bauthor{\bsnm{Bergmann}, \binits{G.}},
\bauthor{\bsnm{Varr{\'{o}}}, \binits{D.}}:
\batitle{Diversity of graph models and graph generators in mutation testing}.
\bjtitle{Int. J. Softw. Tools Technol. Transf.}
\bvolume{22}(\bissue{1}),
\bfpage{57}--\blpage{78}
(\byear{2020})
\doiurl{10.1007/S10009-019-00530-6}
\end{barticle}
\endbibitem

\bibitem[\protect\citeauthoryear{{Hern{\'{a}}ndez L{\'{o}}pez} and
  {S{\'{a}}nchez Cuadrado}}{2023}]{LopezC23}
\begin{barticle}
\bauthor{\bsnm{{Hern{\'{a}}ndez L{\'{o}}pez}}, \binits{J.A.}},
\bauthor{\bsnm{{S{\'{a}}nchez Cuadrado}}, \binits{J.}}:
\batitle{Generating structurally realistic models with deep autoregressive
  networks}.
\bjtitle{{IEEE} Trans. Software Eng.}
\bvolume{49}(\bissue{4}),
\bfpage{2661}--\blpage{2676}
(\byear{2023})
\doiurl{10.1109/TSE.2022.3228630}
\end{barticle}
\endbibitem

\bibitem[\protect\citeauthoryear{Pan et~al.}{2025}]{PanPZWK25}
\begin{bchapter}
\bauthor{\bsnm{Pan}, \binits{F.}},
\bauthor{\bsnm{Petrovic}, \binits{N.}},
\bauthor{\bsnm{Zolfaghari}, \binits{V.}},
\bauthor{\bsnm{Wen}, \binits{L.}},
\bauthor{\bsnm{Knoll}, \binits{A.}}:
\bctitle{{LLM}-enabled instance model generation}.
In: \bbtitle{28th {ACM/IEEE} International Conference on Model Driven
  Engineering Languages and Systems, {MODELS} 2025 - Companion, Grand Rapids,
  MI, USA, October 5-10, 2025},
pp. \bfpage{586}--\blpage{595}.
\bpublisher{{IEEE}},
\blocation{Piscataway, NJ}
(\byear{2025}).
\doiurl{10.1109/MODELS-C68889.2025.00082}
\end{bchapter}
\endbibitem

\bibitem[\protect\citeauthoryear{Polo{-}Gamb{\'{\i}}n
  et~al.}{2026}]{PoloGambinRH26}
\begin{bchapter}
\bauthor{\bsnm{Polo{-}Gamb{\'{\i}}n}, \binits{J.}},
\bauthor{\bsnm{Ruip{\'{e}}rez{-}Valiente}, \binits{J.A.}},
\bauthor{\bsnm{{Hern{\'{a}}ndez L{\'{o}}pez}}, \binits{J.A.}}:
\bctitle{{LLM}-based instance model generation via code synthesis}.
In: \bbtitle{29th {ACM/IEEE} International Conference on Model Driven
  Engineering Languages and Systems, {MODELS} 2026, M{\'{a}}laga, Spain,
  October 4-9, 2026},
pp. \bfpage{176}--\blpage{187}.
\bpublisher{{ACM}},
\blocation{New York, NY}
(\byear{2026}).
\doiurl{10.1145/3822455.3830320}
\end{bchapter}
\endbibitem

\bibitem[\protect\citeauthoryear{Hummel et~al.}{2026}]{HummelRHCKR26}
\begin{bchapter}
\bauthor{\bsnm{Hummel}, \binits{M.}},
\bauthor{\bsnm{Ro{\ss}kothen}, \binits{J.}},
\bauthor{\bsnm{Hagel}, \binits{N.}},
\bauthor{\bsnm{{\c{C}}atalcaya}, \binits{F.}},
\bauthor{\bsnm{Koziolek}, \binits{H.}},
\bauthor{\bsnm{Reussner}, \binits{R.}}:
\bctitle{{AMIGO}: Agentic model instance generation}.
In: \bbtitle{29th {ACM/IEEE} International Conference on Model Driven
  Engineering Languages and Systems, {MODELS} 2026 - Companion, M{\'{a}}laga,
  Spain, October 4-9, 2026},
pp. \bfpage{259}--\blpage{263}.
\bpublisher{{ACM}},
\blocation{New York, NY}
(\byear{2026}).
\doiurl{10.1145/3837062.3838880}
\end{bchapter}
\endbibitem

\bibitem[\protect\citeauthoryear{Wieringa}{2014}]{Wieringa14}
\begin{bbook}
\bauthor{\bsnm{Wieringa}, \binits{R.J.}}:
\bbtitle{Design Science Methodology for Information Systems and Software
  Engineering}.
\bpublisher{Springer},
\blocation{Berlin, Heidelberg}
(\byear{2014}).
\doiurl{10.1007/978-3-662-43839-8}
\end{bbook}
\endbibitem

\bibitem[\protect\citeauthoryear{Gogolla et~al.}{2007}]{GogollaBR07}
\begin{barticle}
\bauthor{\bsnm{Gogolla}, \binits{M.}},
\bauthor{\bsnm{B{\"{u}}ttner}, \binits{F.}},
\bauthor{\bsnm{Richters}, \binits{M.}}:
\batitle{{{USE:} {A} UML-based specification environment for validating {UML}
  and {OCL}}}.
\bjtitle{Sci. Comput. Program.}
\bvolume{69}(\bissue{1-3}),
\bfpage{27}--\blpage{34}
(\byear{2007})
\doiurl{10.1016/j.scico.2007.01.013}
\end{barticle}
\endbibitem

\bibitem[\protect\citeauthoryear{Coman}{2026}]{repo}
\begin{botherref}
\oauthor{\bsnm{Coman}, \binits{A.}}:
Accompanying software repository for the paper: {LLM}-based Generation of
  Diverse and Realistic Domain Model Instances.
\url{https://github.com/a-coman/LIMoGen}.
Accessed: 2026-11-11
(2026)
\end{botherref}
\endbibitem

\bibitem[\protect\citeauthoryear{Budgen et~al.}{2011}]{BudgenBBKP11}
\begin{barticle}
\bauthor{\bsnm{Budgen}, \binits{D.}},
\bauthor{\bsnm{Burn}, \binits{A.J.}},
\bauthor{\bsnm{Brereton}, \binits{O.P.}},
\bauthor{\bsnm{Kitchenham}, \binits{B.A.}},
\bauthor{\bsnm{Pretorius}, \binits{R.}}:
\batitle{Empirical evidence about the {UML:} a systematic literature review}.
\bjtitle{Softw. Pract. Exp.}
\bvolume{41}(\bissue{4}),
\bfpage{363}--\blpage{392}
(\byear{2011})
\doiurl{10.1002/SPE.1009}
\end{barticle}
\endbibitem

\bibitem[\protect\citeauthoryear{Hutchinson et~al.}{2014}]{HUTCHINSON2014144}
\begin{barticle}
\bauthor{\bsnm{Hutchinson}, \binits{J.}},
\bauthor{\bsnm{Whittle}, \binits{J.}},
\bauthor{\bsnm{Rouncefield}, \binits{M.}}:
\batitle{Model-driven engineering practices in industry: Social, organizational
  and managerial factors that lead to success or failure}.
\bjtitle{Science of Computer Programming}
\bvolume{89},
\bfpage{144}--\blpage{161}
(\byear{2014})
\doiurl{10.1016/j.scico.2013.03.017}
\end{barticle}
\endbibitem

\bibitem[\protect\citeauthoryear{Hamid}{2017}]{Hamid17}
\begin{barticle}
\bauthor{\bsnm{Hamid}, \binits{B.}}:
\batitle{A model-driven approach for developing a model repository: Methodology
  and tool support}.
\bjtitle{Future Gener. Comput. Syst.}
\bvolume{68},
\bfpage{473}--\blpage{490}
(\byear{2017})
\doiurl{10.1016/J.FUTURE.2016.04.018}
\end{barticle}
\endbibitem

\bibitem[\protect\citeauthoryear{{Hern{\'{a}}ndez L{\'{o}}pez}
  et~al.}{2022}]{LopezIC22-ModelSet}
\begin{barticle}
\bauthor{\bsnm{{Hern{\'{a}}ndez L{\'{o}}pez}}, \binits{J.A.}},
\bauthor{\bsnm{{C{\'{a}}novas Izquierdo}}, \binits{J.L.}},
\bauthor{\bsnm{{S{\'{a}}nchez Cuadrado}}, \binits{J.}}:
\batitle{{ModelSet}: a dataset for machine learning in model-driven
  engineering}.
\bjtitle{Softw. Syst. Model.}
\bvolume{21}(\bissue{3}),
\bfpage{967}--\blpage{986}
(\byear{2022})
\doiurl{10.1007/S10270-021-00929-3}
\end{barticle}
\endbibitem

\bibitem[\protect\citeauthoryear{Glaser et~al.}{2026a}]{Glaser25-EAModelSet}
\begin{barticle}
\bauthor{\bsnm{Glaser}, \binits{P.-L.}},
\bauthor{\bsnm{Sallinger}, \binits{E.}},
\bauthor{\bsnm{Bork}, \binits{D.}}:
\batitle{{The extended EA ModelSet---a FAIR dataset for researching and
  reasoning enterprise architecture modeling practices}}.
\bjtitle{Software and Systems Modeling}
\bvolume{25}(\bissue{3}),
\bfpage{659}--\blpage{677}
(\byear{2026})
\doiurl{10.1007/s10270-025-01278-1}
\end{barticle}
\endbibitem

\bibitem[\protect\citeauthoryear{Glaser et~al.}{2026b}]{GlaserBB26}
\begin{botherref}
\oauthor{\bsnm{Glaser}, \binits{P.-L.}},
\oauthor{\bsnm{Burgue{\~{n}}o}, \binits{L.}},
\oauthor{\bsnm{Bork}, \binits{D.}}:
A benchmarking framework for model datasets.
CoRR
\textbf{abs/2603.05250}
(2026)
\doiurl{10.48550/ARXIV.2603.05250}
\end{botherref}
\endbibitem

\bibitem[\protect\citeauthoryear{Verbruggen
  et~al.}{2025}]{VerbruggenNGSHCRSB25}
\begin{bchapter}
\bauthor{\bsnm{Verbruggen}, \binits{C.}},
\bauthor{\bsnm{Netz}, \binits{L.}},
\bauthor{\bsnm{Glaser}, \binits{P.-L.}},
\bauthor{\bsnm{Scholz}, \binits{M.}},
\bauthor{\bsnm{Huemer}, \binits{C.}},
\bauthor{\bsnm{Calamo}, \binits{M.}},
\bauthor{\bsnm{Rumpe}, \binits{B.}},
\bauthor{\bsnm{Snoeck}, \binits{M.}},
\bauthor{\bsnm{Bork}, \binits{D.}}:
\bctitle{Toward a community-curated golden dataset of {UML} models}.
In: \bbtitle{28th {ACM/IEEE} International Conference on Model Driven
  Engineering Languages and Systems, {MODELS} 2025 - Companion, Grand Rapids,
  MI, USA, October 5-10, 2025},
pp. \bfpage{43}--\blpage{50}.
\bpublisher{{IEEE}},
\blocation{Piscataway, NJ}
(\byear{2025}).
\doiurl{10.1109/MODELS-C68889.2025.00012}
\end{bchapter}
\endbibitem

\bibitem[\protect\citeauthoryear{Laue and L{\"{a}}uter}{2024}]{Laue.24}
\begin{bchapter}
\bauthor{\bsnm{Laue}, \binits{R.}},
\bauthor{\bsnm{L{\"{a}}uter}, \binits{M.}}:
\bctitle{{Beobachtungen und Einsichten zu Repositorys von BPMN-Modellen}}.
In: \bbtitle{Modellierung 2024, Potsdam, Germany, March 12-15, 2024}.
\bsertitle{{LNI}},
vol. \bseriesno{{P-348}},
pp. \bfpage{157}--\blpage{173}.
\bpublisher{Gesellschaft f{\"{u}}r Informatik e.V.},
\blocation{Bonn}
(\byear{2024})
\end{bchapter}
\endbibitem

\bibitem[\protect\citeauthoryear{Deli{\'{c}} et~al.}{2025}]{DelicAVNB25}
\begin{bchapter}
\bauthor{\bsnm{Deli{\'{c}}}, \binits{A.}},
\bauthor{\bsnm{Ali}, \binits{S.J.}},
\bauthor{\bsnm{Verbruggen}, \binits{C.}},
\bauthor{\bsnm{Neidhardt}, \binits{J.}},
\bauthor{\bsnm{Bork}, \binits{D.}}:
\bctitle{A model cleansing pipeline for model-driven engineering: Mitigating
  the garbage in, garbage out problem for open model repositories}.
In: \bbtitle{28th {ACM/IEEE} International Conference on Model Driven
  Engineering Languages and Systems, {MODELS} 2025, Grand Rapids, MI, USA,
  October 5-10, 2025},
pp. \bfpage{60}--\blpage{71}.
\bpublisher{{IEEE}},
\blocation{Piscataway, NJ}
(\byear{2025}).
\doiurl{10.1109/MODELS67397.2025.00012}
\end{bchapter}
\endbibitem

\bibitem[\protect\citeauthoryear{Joel et~al.}{2026}]{joel_survey_2024}
\begin{barticle}
\bauthor{\bsnm{Joel}, \binits{S.}},
\bauthor{\bsnm{Wu}, \binits{J.J.}},
\bauthor{\bsnm{Fard}, \binits{F.H.}}:
\batitle{A survey on {LLM}-based code generation for low-resource and
  domain-specific programming languages}.
\bjtitle{ACM Transactions on Software Engineering and Methodology}
\bvolume{35}(\bissue{11}),
\bfpage{1}--\blpage{66}
(\byear{2026})
\doiurl{10.1145/3770084}
\end{barticle}
\endbibitem

\bibitem[\protect\citeauthoryear{Wei et~al.}{2022}]{Wei2022CoT}
\begin{bchapter}
\bauthor{\bsnm{Wei}, \binits{J.}},
\bauthor{\bsnm{Wang}, \binits{X.}},
\bauthor{\bsnm{Schuurmans}, \binits{D.}},
\bauthor{\bsnm{Bosma}, \binits{M.}},
\bauthor{\bsnm{Ichter}, \binits{B.}},
\bauthor{\bsnm{Xia}, \binits{F.}},
\bauthor{\bsnm{Chi}, \binits{E.H.}},
\bauthor{\bsnm{Le}, \binits{Q.V.}},
\bauthor{\bsnm{Zhou}, \binits{D.}}:
\bctitle{Chain-of-thought prompting elicits reasoning in large language
  models}.
In: \bbtitle{Proceedings of the 36th International Conference on Neural
  Information Processing Systems}.
\bsertitle{NIPS '22},
pp. \bfpage{24824}--\blpage{24837}.
\bpublisher{Curran Associates Inc.},
\blocation{Red Hook, NY, USA}
(\byear{2022}).
\doiurl{10.52202/068431-1800}
\end{bchapter}
\endbibitem

\bibitem[\protect\citeauthoryear{Ostrand and Balcer}{1988}]{OstrandB88}
\begin{barticle}
\bauthor{\bsnm{Ostrand}, \binits{T.J.}},
\bauthor{\bsnm{Balcer}, \binits{M.J.}}:
\batitle{The category-partition method for specifying and generating functional
  tests}.
\bjtitle{Communications of the ACM}
\bvolume{31}(\bissue{6}),
\bfpage{676}--\blpage{686}
(\byear{1988})
\doiurl{10.1145/62959.62964}
\end{barticle}
\endbibitem

\bibitem[\protect\citeauthoryear{Myers et~al.}{2011}]{Myers2011}
\begin{bbook}
\bauthor{\bsnm{Myers}, \binits{G.J.}},
\bauthor{\bsnm{Sandler}, \binits{C.}},
\bauthor{\bsnm{Badgett}, \binits{T.}}:
\bbtitle{The Art of Software Testing},
\bedition{3rd} edn.
\bpublisher{John Wiley \& Sons},
\blocation{Hoboken, NJ}
(\byear{2011}).
\doiurl{10.1002/9781119202486}
\end{bbook}
\endbibitem

\bibitem[\protect\citeauthoryear{Pielou}{1966}]{pielou1966}
\begin{barticle}
\bauthor{\bsnm{Pielou}, \binits{E.C.}}:
\batitle{The measurement of diversity in different types of biological
  collections}.
\bjtitle{Journal of Theoretical Biology}
\bvolume{13},
\bfpage{131}--\blpage{144}
(\byear{1966})
\doiurl{10.1016/0022-5193(66)90013-0}
\end{barticle}
\endbibitem

\bibitem[\protect\citeauthoryear{Shannon}{1948}]{shannon1948}
\begin{barticle}
\bauthor{\bsnm{Shannon}, \binits{C.E.}}:
\batitle{A mathematical theory of communication}.
\bjtitle{Bell System Technical Journal}
\bvolume{27}(\bissue{3}),
\bfpage{379}--\blpage{423}
(\byear{1948})
\doiurl{10.1002/j.1538-7305.1948.tb01338.x}
\end{barticle}
\endbibitem

\bibitem[\protect\citeauthoryear{Sanfeliu and Fu}{1983}]{SanfeliuFu83}
\begin{barticle}
\bauthor{\bsnm{Sanfeliu}, \binits{A.}},
\bauthor{\bsnm{Fu}, \binits{K.-S.}}:
\batitle{A distance measure between attributed relational graphs for pattern
  recognition}.
\bjtitle{{IEEE} Trans. Syst. Man Cybern.}
\bvolume{{SMC}-13}(\bissue{3}),
\bfpage{353}--\blpage{362}
(\byear{1983})
\doiurl{10.1109/TSMC.1983.6313167}
\end{barticle}
\endbibitem

\bibitem[\protect\citeauthoryear{Riesen and Bunke}{2009}]{RiesenBunke09}
\begin{barticle}
\bauthor{\bsnm{Riesen}, \binits{K.}},
\bauthor{\bsnm{Bunke}, \binits{H.}}:
\batitle{Approximate graph edit distance computation by means of bipartite
  graph matching}.
\bjtitle{Image Vis. Comput.}
\bvolume{27}(\bissue{7}),
\bfpage{950}--\blpage{959}
(\byear{2009})
\doiurl{10.1016/j.imavis.2008.04.004}
\end{barticle}
\endbibitem

\bibitem[\protect\citeauthoryear{Blumenthal et~al.}{2020}]{BlumenthalBGBB20}
\begin{barticle}
\bauthor{\bsnm{Blumenthal}, \binits{D.B.}},
\bauthor{\bsnm{Boria}, \binits{N.}},
\bauthor{\bsnm{Gamper}, \binits{J.}},
\bauthor{\bsnm{Bougleux}, \binits{S.}},
\bauthor{\bsnm{Brun}, \binits{L.}}:
\batitle{Comparing heuristics for graph edit distance computation}.
\bjtitle{{VLDB} J.}
\bvolume{29}(\bissue{1}),
\bfpage{419}--\blpage{458}
(\byear{2020})
\doiurl{10.1007/s00778-019-00544-1}
\end{barticle}
\endbibitem

\bibitem[\protect\citeauthoryear{Wohlin et~al.}{2012}]{WohlinRHCB12}
\begin{bbook}
\bauthor{\bsnm{Wohlin}, \binits{C.}},
\bauthor{\bsnm{Runeson}, \binits{P.}},
\bauthor{\bsnm{H{\"{o}}st}, \binits{M.}},
\bauthor{\bsnm{Ohlsson}, \binits{M.C.}},
\bauthor{\bsnm{Regnell}, \binits{B.}},
\bauthor{\bsnm{Wessl{\'{e}}n}, \binits{A.}}:
\bbtitle{Experimentation in Software Engineering}.
\bpublisher{Springer},
\blocation{Berlin, Heidelberg}
(\byear{2012}).
\doiurl{10.1007/978-3-642-29044-2}
\end{bbook}
\endbibitem

\bibitem[\protect\citeauthoryear{Song et~al.}{2025}]{song2024good}
\begin{bchapter}
\bauthor{\bsnm{Song}, \binits{Y.}},
\bauthor{\bsnm{Wang}, \binits{G.}},
\bauthor{\bsnm{Li}, \binits{S.}},
\bauthor{\bsnm{Lin}, \binits{B.Y.}}:
\bctitle{The good, the bad, and the greedy: Evaluation of {LLMs} should not
  ignore non-determinism}.
In: \bbtitle{Proceedings of the 2025 Conference of the Nations of the Americas
  Chapter of the Association for Computational Linguistics: Human Language
  Technologies (Volume 1: Long Papers)},
pp. \bfpage{4195}--\blpage{4206}.
\bpublisher{Association for Computational Linguistics},
\blocation{Albuquerque, New Mexico}
(\byear{2025}).
\doiurl{10.18653/v1/2025.naacl-long.211}
\end{bchapter}
\endbibitem

\bibitem[\protect\citeauthoryear{Baltes et~al.}{2026}]{Baltes26-LLMGuidelines}
\begin{barticle}
\bauthor{\bsnm{Baltes}, \binits{S.}},
\bauthor{\bsnm{Angermeir}, \binits{F.}},
\bauthor{\bsnm{Arora}, \binits{C.}},
\bauthor{\bsnm{{Mu{\~n}oz Bar{\'o}n}}, \binits{M.}},
\bauthor{\bsnm{Chen}, \binits{C.}},
\bauthor{\bsnm{B{\"o}hme}, \binits{L.}},
\bauthor{\bsnm{Calefato}, \binits{F.}},
\bauthor{\bsnm{Ernst}, \binits{N.}},
\bauthor{\bsnm{Falessi}, \binits{D.}},
\bauthor{\bsnm{Fitzgerald}, \binits{B.}},
\bauthor{\bsnm{Fucci}, \binits{D.}},
\bauthor{\bsnm{He}, \binits{J.}},
\bauthor{\bsnm{Treude}, \binits{C.}},
\bauthor{\bsnm{Kalinowski}, \binits{M.}},
\bauthor{\bsnm{Lambiase}, \binits{S.}},
\bauthor{\bsnm{Russo}, \binits{D.}},
\bauthor{\bsnm{Lungu}, \binits{M.}},
\bauthor{\bsnm{{Martinez Montes}}, \binits{C.}},
\bauthor{\bsnm{Prechelt}, \binits{L.}},
\bauthor{\bsnm{Ralph}, \binits{P.}},
\bauthor{\bsnm{{van Tonder}}, \binits{R.}},
\bauthor{\bsnm{Wagner}, \binits{S.}}:
\batitle{Guidelines for empirical studies in software engineering involving
  large language models}.
\bjtitle{Empirical Software Engineering}
\bvolume{32}(\bissue{2}),
\bfpage{64}
(\byear{2026})
\doiurl{10.1007/s10664-026-10922-3}
\end{barticle}
\endbibitem

\bibitem[\protect\citeauthoryear{Torlak and Jackson}{2007}]{TorlakJ07}
\begin{bchapter}
\bauthor{\bsnm{Torlak}, \binits{E.}},
\bauthor{\bsnm{Jackson}, \binits{D.}}:
\bctitle{Kodkod: {A} relational model finder}.
In: \beditor{\bsnm{Grumberg}, \binits{O.}},
\beditor{\bsnm{Huth}, \binits{M.}} (eds.)
\bbtitle{Tools and Algorithms for the Construction and Analysis of Systems,
  13th International Conference, {TACAS} 2007, Held as Part of the Joint
  European Conferences on Theory and Practice of Software, {ETAPS} 2007 Braga,
  Portugal, March 24 - April 1, 2007, Proceedings}.
\bsertitle{Lecture Notes in Computer Science},
vol. \bseriesno{4424},
pp. \bfpage{632}--\blpage{647}.
\bpublisher{Springer},
\blocation{Berlin, Heidelberg}
(\byear{2007}).
\doiurl{10.1007/978-3-540-71209-1\_49}
\end{bchapter}
\endbibitem

\bibitem[\protect\citeauthoryear{Hilken et~al.}{2018}]{HilkenGBV18}
\begin{barticle}
\bauthor{\bsnm{Hilken}, \binits{F.}},
\bauthor{\bsnm{Gogolla}, \binits{M.}},
\bauthor{\bsnm{Burgue{\~{n}}o}, \binits{L.}},
\bauthor{\bsnm{Vallecillo}, \binits{A.}}:
\batitle{Testing models and model transformations using classifying terms}.
\bjtitle{Softw. Syst. Model.}
\bvolume{17}(\bissue{3}),
\bfpage{885}--\blpage{912}
(\byear{2018})
\doiurl{10.1007/S10270-016-0568-3}
\end{barticle}
\endbibitem

\bibitem[\protect\citeauthoryear{Claris{\'{o}} and Cabot}{2018}]{ClarisoC18}
\begin{bchapter}
\bauthor{\bsnm{Claris{\'{o}}}, \binits{R.}},
\bauthor{\bsnm{Cabot}, \binits{J.}}:
\bctitle{Fixing defects in integrity constraints via constraint mutation}.
In: \beditor{\bsnm{Bertolino}, \binits{A.}},
\beditor{\bsnm{Amaral}, \binits{V.}},
\beditor{\bsnm{Rupino}, \binits{P.}},
\beditor{\bsnm{Vieira}, \binits{M.}} (eds.)
\bbtitle{11th International Conference on the Quality of Information and
  Communications Technology, {QUATIC} 2018, Coimbra, Portugal, September 4-7,
  2018},
pp. \bfpage{74}--\blpage{82}.
\bpublisher{{IEEE}},
\blocation{Piscataway, NJ}
(\byear{2018}).
\doiurl{10.1109/QUATIC.2018.00020}
\end{bchapter}
\endbibitem

\bibitem[\protect\citeauthoryear{Benevides et~al.}{2010}]{BenevidesGBA10}
\begin{barticle}
\bauthor{\bsnm{Benevides}, \binits{A.B.}},
\bauthor{\bsnm{Guizzardi}, \binits{G.}},
\bauthor{\bsnm{Braga}, \binits{B.F.B.}},
\bauthor{\bsnm{Almeida}, \binits{J.P.A.}}:
\batitle{Validating modal aspects of {OntoUML} conceptual models using
  automatically generated visual world structures}.
\bjtitle{J. Univers. Comput. Sci.}
\bvolume{16}(\bissue{20}),
\bfpage{2904}--\blpage{2933}
(\byear{2010})
\doiurl{10.3217/JUCS-016-20-2904}
\end{barticle}
\endbibitem

\bibitem[\protect\citeauthoryear{Jackson}{2006}]{JacksonAlloy.2006}
\begin{bbook}
\bauthor{\bsnm{Jackson}, \binits{D.}}:
\bbtitle{Software Abstractions - Logic, Language, and Analysis}.
\bpublisher{{MIT} Press},
\blocation{Cambridge, MA}
(\byear{2006})
\end{bbook}
\endbibitem

\bibitem[\protect\citeauthoryear{Sales and Guizzardi}{2015}]{SalesG15}
\begin{barticle}
\bauthor{\bsnm{Sales}, \binits{T.P.}},
\bauthor{\bsnm{Guizzardi}, \binits{G.}}:
\batitle{Ontological anti-patterns: empirically uncovered error-prone
  structures in ontology-driven conceptual models}.
\bjtitle{Data Knowl. Eng.}
\bvolume{99},
\bfpage{72}--\blpage{104}
(\byear{2015})
\doiurl{10.1016/J.DATAK.2015.06.004}
\end{barticle}
\endbibitem

\bibitem[\protect\citeauthoryear{Anastasakis et~al.}{2007}]{AnastasakisBGR07}
\begin{bchapter}
\bauthor{\bsnm{Anastasakis}, \binits{K.}},
\bauthor{\bsnm{Bordbar}, \binits{B.}},
\bauthor{\bsnm{Georg}, \binits{G.}},
\bauthor{\bsnm{Ray}, \binits{I.}}:
\bctitle{{UML2Alloy}: {A} challenging model transformation}.
In: \beditor{\bsnm{Engels}, \binits{G.}},
\beditor{\bsnm{Opdyke}, \binits{B.}},
\beditor{\bsnm{Schmidt}, \binits{D.C.}},
\beditor{\bsnm{Weil}, \binits{F.}} (eds.)
\bbtitle{Model Driven Engineering Languages and Systems, 10th International
  Conference, MoDELS 2007, Nashville, USA, September 30 - October 5, 2007,
  Proceedings}.
\bsertitle{Lecture Notes in Computer Science},
vol. \bseriesno{4735},
pp. \bfpage{436}--\blpage{450}.
\bpublisher{Springer},
\blocation{Berlin, Heidelberg}
(\byear{2007}).
\doiurl{10.1007/978-3-540-75209-7\_30}
\end{bchapter}
\endbibitem

\bibitem[\protect\citeauthoryear{Mougenot et~al.}{2009}]{MougenotDBS09}
\begin{bchapter}
\bauthor{\bsnm{Mougenot}, \binits{A.}},
\bauthor{\bsnm{Darrasse}, \binits{A.}},
\bauthor{\bsnm{Blanc}, \binits{X.}},
\bauthor{\bsnm{Soria}, \binits{M.}}:
\bctitle{Uniform random generation of huge metamodel instances}.
In: \beditor{\bsnm{Paige}, \binits{R.F.}},
\beditor{\bsnm{Hartman}, \binits{A.}},
\beditor{\bsnm{Rensink}, \binits{A.}} (eds.)
\bbtitle{Model Driven Architecture - Foundations and Applications, 5th European
  Conference, {ECMDA-FA} 2009, Enschede, The Netherlands, June 23-26, 2009.
  Proceedings}.
\bsertitle{Lecture Notes in Computer Science},
vol. \bseriesno{5562},
pp. \bfpage{130}--\blpage{145}.
\bpublisher{Springer},
\blocation{Berlin, Heidelberg}
(\byear{2009}).
\doiurl{10.1007/978-3-642-02674-4\_10}
\end{bchapter}
\endbibitem

\bibitem[\protect\citeauthoryear{Duchon et~al.}{2004}]{DuchonFLS04}
\begin{barticle}
\bauthor{\bsnm{Duchon}, \binits{P.}},
\bauthor{\bsnm{Flajolet}, \binits{P.}},
\bauthor{\bsnm{Louchard}, \binits{G.}},
\bauthor{\bsnm{Schaeffer}, \binits{G.}}:
\batitle{Boltzmann samplers for the random generation of combinatorial
  structures}.
\bjtitle{Comb. Probab. Comput.}
\bvolume{13}(\bissue{4--5}),
\bfpage{577}--\blpage{625}
(\byear{2004})
\doiurl{10.1017/S0963548304006315}
\end{barticle}
\endbibitem

\bibitem[\protect\citeauthoryear{{AtlanMod Team}}{}]{emf-instantiator}
\begin{botherref}
\oauthor{\bsnm{{AtlanMod Team}}}:
{EMF} random instantiator.
\url{https://github.com/atlanmod/mondo-atlzoo-benchmark/tree/master/fr.inria.atlanmod.instantiator}
\end{botherref}
\endbibitem

\bibitem[\protect\citeauthoryear{Scheidgen}{2015}]{Scheidgen15}
\begin{bchapter}
\bauthor{\bsnm{Scheidgen}, \binits{M.}}:
\bctitle{Generation of large random models for benchmarking}.
In: \beditor{\bsnm{Kolovos}, \binits{D.S.}},
\beditor{\bsnm{{Di Ruscio}}, \binits{D.}},
\beditor{\bsnm{Matragkas}, \binits{N.D.}},
\beditor{\bsnm{{S{\'{a}}nchez Cuadrado}}, \binits{J.}},
\beditor{\bsnm{R{\'{a}}th}, \binits{I.}},
\beditor{\bsnm{Tisi}, \binits{M.}} (eds.)
\bbtitle{Proceedings of the 3rd Workshop on Scalable Model Driven Engineering
  Part of the Software Technologies: Applications and Foundations {(STAF} 2015)
  Federation of Conferences, L'Aquila, Italy, July 23, 2015}.
\bsertitle{{CEUR} Workshop Proceedings},
vol. \bseriesno{1406},
pp. \bfpage{1}--\blpage{10}.
\bpublisher{CEUR-WS.org},
\blocation{Aachen}
(\byear{2015}).
\burl{https://ceur-ws.org/Vol-1406/paper1.pdf}
\end{bchapter}
\endbibitem

\bibitem[\protect\citeauthoryear{Popoola et~al.}{2016}]{PopoolaKR16}
\begin{bchapter}
\bauthor{\bsnm{Popoola}, \binits{S.}},
\bauthor{\bsnm{Kolovos}, \binits{D.S.}},
\bauthor{\bsnm{Rodriguez}, \binits{H.H.}}:
\bctitle{{EMG:} {A} domain-specific transformation language for synthetic model
  generation}.
In: \beditor{\bsnm{{Van Gorp}}, \binits{P.}},
\beditor{\bsnm{Engels}, \binits{G.}} (eds.)
\bbtitle{Theory and Practice of Model Transformations - 9th International
  Conference, ICMT@STAF 2016, Vienna, Austria, July 4-5, 2016, Proceedings}.
\bsertitle{Lecture Notes in Computer Science},
vol. \bseriesno{9765},
pp. \bfpage{36}--\blpage{51}.
\bpublisher{Springer},
\blocation{Cham}
(\byear{2016}).
\doiurl{10.1007/978-3-319-42064-6\_3}
\end{bchapter}
\endbibitem

\bibitem[\protect\citeauthoryear{Sz{\'{a}}rnyas et~al.}{2016}]{SzarnyasKSV16}
\begin{bchapter}
\bauthor{\bsnm{Sz{\'{a}}rnyas}, \binits{G.}},
\bauthor{\bsnm{Kov{\'{a}}ri}, \binits{Z.}},
\bauthor{\bsnm{Sal{\'{a}}nki}, \binits{{\'{A}}.}},
\bauthor{\bsnm{Varr{\'{o}}}, \binits{D.}}:
\bctitle{Towards the characterization of realistic models: evaluation of
  multidisciplinary graph metrics}.
In: \beditor{\bsnm{Baudry}, \binits{B.}},
\beditor{\bsnm{Combemale}, \binits{B.}} (eds.)
\bbtitle{Proceedings of the {ACM/IEEE} 19th International Conference on Model
  Driven Engineering Languages and Systems, Saint-Malo, France, October 2-7,
  2016},
pp. \bfpage{87}--\blpage{94}.
\bpublisher{{ACM}},
\blocation{New York, NY}
(\byear{2016}).
\doiurl{10.1145/2976767.2976786}
\end{bchapter}
\endbibitem

\bibitem[\protect\citeauthoryear{Varr{\'{o}} et~al.}{2018}]{VarroSSH18}
\begin{bchapter}
\bauthor{\bsnm{Varr{\'{o}}}, \binits{D.}},
\bauthor{\bsnm{Semer{\'{a}}th}, \binits{O.}},
\bauthor{\bsnm{Sz{\'{a}}rnyas}, \binits{G.}},
\bauthor{\bsnm{Horv{\'{a}}th}, \binits{{\'{A}}.}}:
\bctitle{Towards the automated generation of consistent, diverse, scalable and
  realistic graph models}.
In: \beditor{\bsnm{Heckel}, \binits{R.}},
\beditor{\bsnm{Taentzer}, \binits{G.}} (eds.)
\bbtitle{Graph Transformation, Specifications, and Nets - In Memory of Hartmut
  Ehrig}.
\bsertitle{Lecture Notes in Computer Science},
vol. \bseriesno{10800},
pp. \bfpage{285}--\blpage{312}.
\bpublisher{Springer},
\blocation{Cham}
(\byear{2018}).
\doiurl{10.1007/978-3-319-75396-6\_16}
\end{bchapter}
\endbibitem

\bibitem[\protect\citeauthoryear{Pietsch et~al.}{2011}]{PietschYK11}
\begin{bchapter}
\bauthor{\bsnm{Pietsch}, \binits{P.}},
\bauthor{\bsnm{Yazdi}, \binits{H.S.}},
\bauthor{\bsnm{Kelter}, \binits{U.}}:
\bctitle{Generating realistic test models for model processing tools}.
In: \beditor{\bsnm{Alexander}, \binits{P.}},
\beditor{\bsnm{Pasareanu}, \binits{C.S.}},
\beditor{\bsnm{Hosking}, \binits{J.G.}} (eds.)
\bbtitle{26th {IEEE/ACM} International Conference on Automated Software
  Engineering {(ASE} 2011), Lawrence, KS, USA, November 6-10, 2011},
pp. \bfpage{620}--\blpage{623}.
\bpublisher{{IEEE}},
\blocation{Piscataway, NJ}
(\byear{2011}).
\doiurl{10.1109/ASE.2011.6100140}
\end{bchapter}
\endbibitem

\bibitem[\protect\citeauthoryear{Ali et~al.}{2024}]{AliRB24}
\begin{bchapter}
\bauthor{\bsnm{Ali}, \binits{S.J.}},
\bauthor{\bsnm{Reinhartz{-}Berger}, \binits{I.}},
\bauthor{\bsnm{Bork}, \binits{D.}}:
\bctitle{How are {LLMs} used for conceptual modeling? {An} exploratory study on
  interaction behavior and user perception}.
In: \beditor{\bsnm{Maass}, \binits{W.}},
\beditor{\bsnm{Han}, \binits{H.}},
\beditor{\bsnm{Yasar}, \binits{H.}},
\beditor{\bsnm{Multari}, \binits{N.J.}} (eds.)
\bbtitle{Conceptual Modeling - 43rd International Conference, {ER} 2024,
  Pittsburgh, PA, USA, October 28-31, 2024, Proceedings}.
\bsertitle{Lecture Notes in Computer Science},
vol. \bseriesno{15238},
pp. \bfpage{257}--\blpage{275}.
\bpublisher{Springer},
\blocation{Cham}
(\byear{2024}).
\doiurl{10.1007/978-3-031-75872-0\_14}
\end{bchapter}
\endbibitem

\bibitem[\protect\citeauthoryear{C{\'{a}}mara et~al.}{2023}]{CamaraTBV23}
\begin{barticle}
\bauthor{\bsnm{C{\'{a}}mara}, \binits{J.}},
\bauthor{\bsnm{Troya}, \binits{J.}},
\bauthor{\bsnm{Burgue{\~{n}}o}, \binits{L.}},
\bauthor{\bsnm{Vallecillo}, \binits{A.}}:
\batitle{{On the assessment of generative {AI} in modeling tasks: an experience
  report with ChatGPT and {UML}}}.
\bjtitle{Softw. Syst. Model.}
\bvolume{22}(\bissue{3}),
\bfpage{781}--\blpage{793}
(\byear{2023})
\doiurl{10.1007/s10270-023-01105-5}
\end{barticle}
\endbibitem

\bibitem[\protect\citeauthoryear{Silva et~al.}{2024}]{SilvaMCKP24}
\begin{bchapter}
\bauthor{\bsnm{Silva}, \binits{J.}},
\bauthor{\bsnm{Ma}, \binits{Q.}},
\bauthor{\bsnm{Cabot}, \binits{J.}},
\bauthor{\bsnm{Kelsen}, \binits{P.}},
\bauthor{\bsnm{Proper}, \binits{H.A.}}:
\bctitle{Application of the tree-of-thoughts framework to {LLM}-enabled domain
  modeling}.
In: \beditor{\bsnm{Maass}, \binits{W.}},
\beditor{\bsnm{Han}, \binits{H.}},
\beditor{\bsnm{Yasar}, \binits{H.}},
\beditor{\bsnm{Multari}, \binits{N.J.}} (eds.)
\bbtitle{Conceptual Modeling - 43rd International Conference, {ER} 2024,
  Pittsburgh, PA, USA, October 28-31, 2024, Proceedings}.
\bsertitle{Lecture Notes in Computer Science},
vol. \bseriesno{15238},
pp. \bfpage{94}--\blpage{111}.
\bpublisher{Springer},
\blocation{Cham}
(\byear{2024}).
\doiurl{10.1007/978-3-031-75872-0\_6}
\end{bchapter}
\endbibitem

\bibitem[\protect\citeauthoryear{Ciatto et~al.}{2025}]{CiattoAMO24}
\begin{barticle}
\bauthor{\bsnm{Ciatto}, \binits{G.}},
\bauthor{\bsnm{Agiollo}, \binits{A.}},
\bauthor{\bsnm{Magnini}, \binits{M.}},
\bauthor{\bsnm{Omicini}, \binits{A.}}:
\batitle{Large language models as oracles for instantiating ontologies with
  domain-specific knowledge}.
\bjtitle{Knowledge-Based Systems}
\bvolume{310},
\bfpage{112940}
(\byear{2025})
\doiurl{10.1016/j.knosys.2024.112940}
\end{barticle}
\endbibitem

\bibitem[\protect\citeauthoryear{Bougleux et~al.}{2017}]{BougleuxBCFGV17}
\begin{barticle}
\bauthor{\bsnm{Bougleux}, \binits{S.}},
\bauthor{\bsnm{Brun}, \binits{L.}},
\bauthor{\bsnm{Carletti}, \binits{V.}},
\bauthor{\bsnm{Foggia}, \binits{P.}},
\bauthor{\bsnm{Ga{\"u}z{\`e}re}, \binits{B.}},
\bauthor{\bsnm{Vento}, \binits{M.}}:
\batitle{Graph edit distance as a quadratic assignment problem}.
\bjtitle{Pattern Recognit. Lett.}
\bvolume{87},
\bfpage{38}--\blpage{46}
(\byear{2017})
\doiurl{10.1016/j.patrec.2016.10.001}
\end{barticle}
\endbibitem

\bibitem[\protect\citeauthoryear{Zeng et~al.}{2009}]{ZengTWFZ09}
\begin{barticle}
\bauthor{\bsnm{Zeng}, \binits{Z.}},
\bauthor{\bsnm{Tung}, \binits{A.K.H.}},
\bauthor{\bsnm{Wang}, \binits{J.}},
\bauthor{\bsnm{Feng}, \binits{J.}},
\bauthor{\bsnm{Zhou}, \binits{L.}}:
\batitle{Comparing stars: On approximating graph edit distance}.
\bjtitle{{PVLDB}}
\bvolume{2}(\bissue{1}),
\bfpage{25}--\blpage{36}
(\byear{2009})
\doiurl{10.14778/1687627.1687631}
\end{barticle}
\endbibitem

\bibitem[\protect\citeauthoryear{Boria et~al.}{2020}]{BoriaBBG20}
\begin{barticle}
\bauthor{\bsnm{Boria}, \binits{N.}},
\bauthor{\bsnm{Blumenthal}, \binits{D.B.}},
\bauthor{\bsnm{Bougleux}, \binits{S.}},
\bauthor{\bsnm{Brun}, \binits{L.}}:
\batitle{Improved local search for graph edit distance}.
\bjtitle{Pattern Recognit. Lett.}
\bvolume{129},
\bfpage{19}--\blpage{25}
(\byear{2020})
\doiurl{10.1016/j.patrec.2019.10.028}
\end{barticle}
\endbibitem

\bibitem[\protect\citeauthoryear{Blumenthal and
  Gamper}{2018}]{BlumenthalGamper18}
\begin{barticle}
\bauthor{\bsnm{Blumenthal}, \binits{D.B.}},
\bauthor{\bsnm{Gamper}, \binits{J.}}:
\batitle{Improved lower bounds for graph edit distance}.
\bjtitle{{IEEE} Trans. Knowl. Data Eng.}
\bvolume{30}(\bissue{3}),
\bfpage{503}--\blpage{516}
(\byear{2018})
\doiurl{10.1109/TKDE.2017.2772243}
\end{barticle}
\endbibitem

\end{thebibliography}
